\documentclass[nohyper]{JHEP3}
\usepackage{cite}
\usepackage{epsfig}

\newcommand{\mycomm}[1]{\hfill\break $\phantom{a}$\kern-3.5em{\tt===$>$ \bf #1}\hfill\break}
\newcommand{\mycommA}[1]{\hfill\break $\phantom{a}$\kern-3.5em{\tt   $>$ \bf #1}\hfill\break}

\newcommand{\be}{\begin{equation}}
\newcommand{\ee}{\end{equation}}
\newcommand{\ba}{\begin{eqnarray}}
\newcommand{\ea}{\end{eqnarray}}

\def\eq#1{Eq.~(\ref{#1})}

\def\MSbar{\hbox{\tiny ${\overline{\rm MS}}$}}

\def\tot{\hbox{\tiny total}}

\def\sing{\hbox{\tiny sing.}}
\def\reg{\hbox{\tiny reg.}}

\def\PV{\hbox{\tiny PV}}

\catcode`\@=11 
\def\lsim{\mathrel{\mathpalette\@versim<}}
\def\gsim{\mathrel{\mathpalette\@versim>}}
\def\@versim#1#2{\vcenter{\offinterlineskip
        \ialign{$\m@th#1\hfil##\hfil$\crcr#2\crcr\sim\crcr } }}
\catcode`\@=12 

\title{Inclusive spectra in charmless semileptonic B decays
by Dressed Gluon Exponentiation}

\author{Jeppe R. Andersen  and  Einan Gardi\\
Cavendish Laboratory, University of Cambridge\\
Madingley Road, Cambridge, CB3 0HE, UK
}

\abstract{The triple differential spectrum in
$\bar{B}\longrightarrow X_u l \bar{\nu}$ is computed by Dressed
Gluon Exponentiation (DGE). In this framework the on-shell
calculation, converted into hadronic variables, can be directly used as
an approximation to the meson decay spectrum, without involving a
leading--power non-perturbative function. Sudakov~resummation for
the fully differential $\bar{B}\longrightarrow X_u l \bar{\nu}$
width is formulated in moment space, where moments are defined
using the ratio between the lightcone momentum components of the
partonic jet $p^+/p^-$ and the hard scale is $p^-$. In these
variables the correspondence with the
 $\bar{B}\longrightarrow X_s \gamma$ case is transparent. The
Sudakov exponent is known to next--to--next--to--leading logarithmic
accuracy. Further constraints are put on its Borel sum using the
cancellation of the leading renormalon ambiguity and the absence of
the next--to--leading one, which was proven in the large--$\beta_0$
limit and assumed here to be general. Based on the resummed
spectrum, matched to the fully differential NLO result, we calculate
the event fraction associated with experimental cuts on the hadronic
mass (or the small lightcone component) as well as on the lepton
energy. Finally, we extract $|V_{ub}|$ from recent measurements by Belle
and analyze the theoretical uncertainty.}


\preprint{Cavendish-HEP-05/16}

\begin{document}

\section{Introduction}

A central issue on the agenda of the B factories is the precise
determination of the CKM matrix element $V_{ub}$. Without
undermining the significance of exclusive hadronic final--state
analysis, measurements of the inclusive branching fraction of
charmless semileptonic decays, $\bar{B}\longrightarrow X_u l
\bar{\nu}$, provide the safest and most accurate determination of
$\left\vert V_{ub}\right\vert$
~\cite{Bornheim:2002du,Limosani:2005pi,Aubert:2004bv,Aubert:2005im,Kakuno:2003fk,Aubert:2005hb,Bizjak:2005hn}.

Theoretically, because of the inclusive nature of the measurement,
the calculation of the width $\Gamma\left(\bar{B}\longrightarrow X_u
l \bar{\nu}\right)$ in units of $\left\vert V_{ub}\right\vert^2$
relies primarily on QCD perturbation theory. The final state, which
in reality is composed of light hadrons, can be safely described in
the course of the calculation by
light quarks and gluons. Moreover, the dependence of the width on
the internal dynamics of the decaying B meson is power suppressed.
The Operator Product Expansion (OPE) implies that the total decay
width of a B meson is equal to the total decay width of a
hypothetical on-shell b quark up to ${\cal O}(\Lambda^2/m_b^2)$
corrections, which can be expressed as forward matrix elements of
certain local operators whose numerical values are determined based
on other measurements. These corrections are very small.

The main obstacle in extracting $\left\vert V_{ub}\right\vert$ from
inclusive semileptonic decays is the need to eliminate the huge
background due to B decay into charm. The most effective way to
maximize the rate while discriminating the charm background is to
select events where the hadronic system has small mass, $M_X\lsim
1.7$ GeV, see e.g.
\cite{Falk:1997gj,Bigi:1997dn,Kakuno:2003fk,Bizjak:2005hn}.

Because of the cuts, detailed theoretical understanding of the
differential distribution is essential. In the small--$M_X$ region
the hadronic system is most likely jet-like. Working with lightcone
coordinates, $P^+\leq P^-\leq M_B$ with $M_X^2=P^+P^-$, the small
$M_X$ region directly corresponds\footnote{The region where both
lightcone components are small is power suppressed.} to the region
where the smaller lighcone component gets small,
\hbox{$P^+\longrightarrow 0$}, so it involves highly non-local
correlation functions with large lightcone separation,
\hbox{$y^-\longrightarrow \infty$}, where $y^-$ is the Fourier
conjugate to $P^+$. At leading power in ${\cal O}(\Lambda/m_b)$ the
differential width involves the full lightcone momentum distribution
function of the b quark inside the meson, the so-called
``shape function''~\cite{Neubert:1993um,Bigi:1993ex}.
On the non-perturbative level, not much is known about
this distribution: it is difficult to compute it on the lattice and
although in principle it can be measured using the photon spectrum
in rare radiative decays $\bar{B}\longrightarrow X_s \gamma$,
realistic measurements constrain just the first few moments of this
function~\cite{Abe:2005cv,Aubert:2005cu}.

As it stands, the theoretical uncertainty in extracting $\left\vert
V_{ub}\right\vert$ from inclusive semileptonic decays is dominated
by the uncertainty in estimating the effect of cuts. A great variety
of experimental cuts has been proposed, which either aim at reducing
the sensitivity to the unknown momentum distribution of the b quark
in the meson by cutting out the phase--space regions where it is
important~\cite{Bauer:2000xf,Bauer:2001rc}, or alternatively, aim at
making the relation with the $\bar{B}\longrightarrow X_s \gamma$
spectrum as direct as
possible~\cite{Bosch:2004bt,Lange:2005qn,Lange:2005yw,Lange:2005xz}.
While the comparison between results for $\left\vert
V_{ub}\right\vert$ obtained using different cuts is useful for
controlling experimental as well as theoretical uncertainties,
better measurements can be achieved if cuts would be chosen based on
purely experimental considerations. Precise theoretical predictions
for the fully differential width are therefore urgently needed.

\subsection{The applicability of perturbation theory}

In perturbation theory, the fully differential width is currently
known~\cite{DFN} to the next--to--leading order (NLO), which
includes tree level as well as one--loop, i.e. ${\cal O}(\alpha_s)$,
virtual corrections, where the hadronic system is represented by a
single on-shell $u$ quark, as well as bremsstrahlung contributions
where the hadronic system is composed of a u-quark and a gluon, and
therefore has a non-vanishing mass. It has long been recognized that
fixed--order calculations do not provide even a qualitative
description of the differential width in the small--$M_X$ region,
where multiple emission of soft and collinear gluons is
important~\cite{KS,Aglietti:2000te,Aglietti:2001br,BDK,RD,Aglietti:2005mb,Bosch:2003fc,Bauer:2003pi,Bauer:2000yr,Bauer:2000ew,Leibovich:2000ig}.
In terms of partonic\footnote{The relation with the hadronic
lightcone coordinates is $p_j^{\pm}=P^{\pm}-\bar{\Lambda}$ where
$\bar{\Lambda}=M_B-m_b$, see \eq{P^+}.}
lightcone coordinates, $p_j^+\leq p_j^-\leq m_b$, the small $M_X$
region is characterized by a large hierarchy, $p_j^+/p_j^-\ll 1$.
Higher--order corrections containing Sudakov logarithms of the ratio
$p_j^+/p_j^-$ are large, and they must be resummed to all orders to
recover the characteristic Sudakov peak at small $M_X$.

According to the factorization properties of the {fully
differential} width in inclusive
decays~\cite{KS,Aglietti:2000te,Aglietti:2001br,BDK,RD,Aglietti:2005mb,Bosch:2003fc,Bauer:2003pi,Bauer:2000yr,Bauer:2000ew,Leibovich:2000ig},
 Sudakov logarithms exponentiate in moment space.
We shall work here with moments $n$ defined\footnote{There are other
viable definitions for moments; an alternative is considered in
Appendix~\ref{sec:xi_resummarion}, see \eq{jet_mass_mom_def} there.}
with respect to powers of $1-p_j^+/p_j^-$, as in \eq{r_mom_def}
below. The region of interest, of small $p_j^+$, is probed by high moments,
$n\longrightarrow \infty $.
The logarithms, $\ln n$, which become large in this limit,
originate in two distinct subprocesses
of different characteristic scales, which factorize in this moment space.
The first is the final--state
jet having a constrained invariant mass squared of ${\cal
O}(m_b^2/n)$~\cite{GR,BDK,RD}, which coincides with the Sudakov factor
of deep inelastic structure functions. The second corresponds to
the quark distribution in an on-shell heavy quark, the perturbative analogue
of the quark distribution in the meson~\cite{BDK,QD,RD}, and describes
soft radiation, $|k|\sim {\cal O}(m_b/n)$, from the nearly on-shell
heavy quark prior to the decay.

A major stumbling block, which essentially prohibits the
straightforward application of Sudakov resummation to inclusive B
decay phenomenology, is that the scales involved are almost too low
for perturbation theory to apply. While the b quark mass is heavy
compared to the QCD scale $\Lambda$, in the Sudakov limit,
$n\longrightarrow \infty $, the jet mass scale is
much lower than $m_b$ and the soft scale $|k|\sim {\cal O}(m_b/n)$
characterizing the quark distribution is yet much lower. Already at
moderately high moments, $n\sim 10$, the latter is dangerously close
to $\Lambda$ where the physics in non-perturbative. On these grounds
it is often argued~\cite{HLL,Benson:2004sg} that Sudakov resummation
is useless in this case. Conversely, when Sudakov resummation is
applied it is usually supplemented by an external infrared cutoff
along with parametrization of the non-perturbative quark
distribution function below this
scale~\cite{Bosch:2004bt,Lange:2005yw}.

While there are many ways to partially bypass the need to resum
corrections on the scale $|k|\sim {\cal O}(m_b/n)$, we argue that
this resummation is in fact largely under control and that it is
very useful. Ref.~\cite{RD} and the present investigation show that
definite predictions emerge when the available information on the
large--order behavior of the perturbative expansion is taken into
account. The sensitivity to low momentum scales, and specifically to
the meson structure, turns out to be smaller than it superficially
appears to be, and consequently the predictive power of perturbation
theory is higher than what one might a priori expect.

Infrared sensitivity appears in the moment--space Sudakov exponent
through infrared
renormalons~\cite{DGE_thrust,GR,CG,Gardi:2001di,Gardi:2002bg,Gardi:2004gj,Gardi:2005mf,QD}.
These generate divergences and invalidate the logarithmic accuracy
criterion \cite{BDK,RD}, but they definitely do not prohibit using
Sudakov resummation altogether: the Sudakov exponent needs to be
regarded as an asymptotic series, and summed up to all orders by
regularizing the renormalons in a systematic way. This is the essence
of Dressed Gluon Exponentiation (DGE). The application of this
approach to inclusive decays is particularly advantageous since the
leading infrared renormalon ambiguity, which dictates the divergence
of the series in the Sudakov exponent, cancels exactly with the
pole--mass ambiguity that is carried by kinematic power corrections
associated with the conversion from partonic to hadronic variables.
In this way, the on-shell calculation becomes directly useful for
phenomenology, in spite of the divergence of the series (or the
ambiguity) that appeared at the intermediate stage.

In the present paper we provide precise theoretical predictions for
the fully differential width in semileptonic decays using DGE. Our
calculation makes maximal use of Sudakov resummation and the
underlying renormalon structure. In this way we effectively minimize
the role of the unknown non-perturbative component of the
quark--distribution function and the associated uncertainties.

The method of DGE was already proven successful in the application
to the photon energy spectrum in $\bar{B}\longrightarrow X_s \gamma$
decays~\cite{RD,Gardi:2005mf}. Specifically, the first two central
moments defined over the range $E_{\gamma} > E_0$, were computed in
\cite{RD} in an essentially perturbative fashion (see below) without
involving any parametrization of the leading--power quark
distribution function. These predictions were later
compared~\cite{Gardi:2005mf} with experimental measurements from
Babar~\cite{Walsh} over a wide range of cuts, $E_0=1.9$ to 2.26 GeV, finding
very good agreement.

\subsection{Comparison with alternative approaches}

It is worthwhile re-examining from the present perspective the
conventional approach to inclusive spectra, which is based on
parametrizing the leading--power non-perturbative ``shape
function''~\cite{Neubert:1993um,Bigi:1993ex} in analogy with deep
inelastic structure--function phenomenology. Let us summarize the
qualitative differences.

 In Refs.~\cite{Neubert:1993um,Bigi:1993ex,Lange:2005yw} one begins by
considering directly the kinematic region where the small lightcone
component of the hadronic system is comparable to the momentum of
the light degrees of freedom in the meson, ${\cal O}(\Lambda)$. In
the $m_b\longrightarrow \infty$ limit the long--distance interaction
is captured by a single lightcone distribution, the ``shape
function''. No attempt is made to compute this function, which is
assumed {\em non-perturbative} from the beginning.  The ``shape
function'' is parametrized and the parameters are fixed by a fit to
the measured $\bar{B}\longrightarrow X_s \gamma$ spectrum (see e.g.
Sec.~4 in Ref.~\cite{Lange:2005yw}). Owing to the universality of
this function, the result can be directly used in semi-leptonic
decays. There is of course some bias owing to the functional form
assumed. In practice this can be dealt with by varying the function,
or better, by relating directly the semileptonic distribution to the
$\bar{B}\longrightarrow X_s \gamma$ data by weighted
integrals~\cite{Lange:2005qn,Lange:2005xz}.

Our approach is more ambitious: we wish to {\em compute} the
spectrum.  Relying on the fact that the heavy quark in the meson is
rarely far off shell, we compute the {\em on-shell} decay spectrum
and use it as a first approximation to the meson decay spectrum. In
this approximation the quark distribution in the meson is replaced
by its perturbative analogue, the quark distribution in an on-shell
heavy quark --- see Ref.~\cite{QD} for the precise definition. A key
element in this approach is the systematic treatment of infrared
renormalons. Most importantly, this includes a systematic definition
of the on-shell heavy quark state at the level of resummed
perturbation theory. Having full control of the large--order
behavior of the expansion, such definition is provided by Principal
Value Borel summation. Upon consistently using this definition in
both the Sudakov exponent and the quark pole mass (which enters the
computed spectrum through the conversion to hadronic variables) any
${\cal O}(\Lambda/m_b)$ ambiguities are avoided.  In this way
resummed perturbation theory yields definite predictions for the
on-shell decay spectrum. The latter provides an approximation to the
meson decay spectrum that does not require any ${\cal
O}(\Lambda/m_b)$ non-perturbative power corrections. Having
established that, we study higher power corrections that constitute
the ratio between the moments of the physical spectrum and the
computed one, which originate in the dynamical structure of the
meson. While these corrections gradually increase with the moment
index $n$ they have but a small effect on the global properties of
the spectrum.

It should be emphasized that in reality the spectrum in the
immediate vicinity of the endpoint, $P^+\simeq 0$, cannot be
accurately described by any of these approaches for two reasons:
first, the number of relevant ``shape function'' parameters
increases, and second, the transition to the exclusive region,
namely the hadronic structure of the {\em final} state, is not
addressed.

Recent formulations of the ``shape function'' approach,
e.g.~\cite{Lange:2005yw}, employ the tools of
factorization~\cite{KS,Bosch:2003fc} and can therefore make some use
of Sudakov resummation. Nevertheless the resummation there
effectively concerns the jet function only, as the ``shape
function'' is defined at an intermediate scale of order of the jet
mass. Moreover, the need to separate ``jet'' logarithms from
``soft'' ones requires a complicated infrared cutoff to be
introduced at any order in perturbation theory. The final answer is
affected by residual scale and scheme dependence.

In contrast, in our formulation Sudakov resummation is applied also
to the quark--distribution function. We thus make full use of the
available perturbative information. Moreover, the on-shell decay
spectrum we compute is itself {\em renormalization group invariant}:
it is absolutely free of any factorization scale or scheme
dependence; at the level of the Sudakov factor, it is also free of
renormalization--scale dependence owing to the additional
resummation of running--coupling effects we perform.

The conclusion from Ref.~\cite{RD} and the successful comparison
with experimental data~\cite{Gardi:2005mf} is that when using DGE,
the on-shell calculation, $b\longrightarrow X_s \gamma$, converted
into hadronic variables, directly provides a good approximation of
the B meson decay spectrum. Notably, in contrast with any
fixed--order approximation, it has approximately correct physical
support properties at large $E_{\gamma}$. This implies, in
particular, that in this framework the non-perturbative component of
the lightcone momentum distribution function is small, and at the
present accuracy can even be neglected. This conclusion may be
surprising: as discussed above, in other approaches the
parametrization of this non-perturbative distribution is absolutely
essential. In order to understand the difference one should recall
that
\begin{description}
\item{(1)\,} The separation between the perturbative and
non-perturbative components of the lightcone momentum distribution
function depends on the convention adopted. In our approach, the
calculated perturbative component is nothing but the quark
distribution in an on-shell heavy quark. It requires renormalization
owing to logarithmic ultraviolet singularities, however, {\em no
infrared cutoff is needed}. This distribution is infrared and
collinear safe and therefore well--defined to all orders in
perturbation theory. It requires of course Sudakov resummation,
owing to the large hierarchy between the hard scale $m_b$ and the
soft scale $m_b/n$. In contrast, {\em at the power level} the
Sudakov exponent is infrared sensitive: its Borel sum has
ambiguities scaling as powers of $(n \Lambda/m_b)$. In our approach
these are regularized using the Principal Value prescription
--- this is how we {\em define} the perturbative component of the
 quark distribution function.
\item{(2)\,} Beyond the currently available {\em  logarithmic} accuracy at which
the quark distribution is known (next--to--next--to--leading
logarithmic accuracy~\cite{QD}), the DGE calculation makes use of
additional information, which is important in constraining the
corresponding Borel sum. Most importantly, the exact cancellation of
the leading ($u=1/2$) renormalon ambiguity with kinematic power
corrections involving the pole mass~\cite{BDK} is used to fix the
corresponding residue~\cite{RD}. In addition, the absence of the
next--to--leading renormalon ($u=1$), which was proven in the
large--$\beta_0$ limit~\cite{BDK}, is assumed here to be general.
Thus, the dependence on the (Principal Value) prescription is
eventually restricted to higher renormalons ($u\geq 3/2$), which
have limited influence on the spectrum --- see Ref.~\cite{RD} and below.
\end{description}

Following the results of Refs.~\cite{RD,Gardi:2005mf}, our working
assumption in this paper is that the properly--defined quark
distribution in an on-shell heavy quark provides a good
approximation to the quark distribution in the meson. In reality the
two differ of course, and it is important to quantify the power
corrections that distinguish between them, a task that will require
theoretical as well as experimental input. This, however, is beyond
the scope of the present paper.

\subsection{The task and the strategy}

The main task on which we embark here is to provide a reliable
calculation of the partial decay width with experimentally--driven
cuts. Priority is given to cuts that maximize the rate, in
particular~\cite{Bizjak:2005hn} the charm--discriminating cut on the
hadronic mass $P^+P^-<M_X^2= (1.7 \,{\rm GeV})^2$ with an additional, experimentally
unavoidable, mild cut on the charged lepton energy $E_l>1$ GeV. Our
general strategy to extract $\left\vert V_{ub}\right\vert$ from data
is based on separately computing:
\begin{itemize}
\item{} The total
charmless semileptonic width
$\Gamma_{\tot}\left(\bar{B}\longrightarrow X_u l \bar{\nu}\right)$
in units of $\left\vert V_{ub}\right\vert^2$, using the available
NNLO result~\cite{vanRitbergen:1999gs} and renormalon resummation.
\item{} The effect of experimental cuts as the event fraction
\begin{equation}
R_{\rm cut}\equiv \frac{\Gamma(\bar{B}\longrightarrow X_u l
\bar{\nu}\,\,{\rm restricted\,\, phase\,\, space})}
{\Gamma(\bar{B}\longrightarrow X_u l \bar{\nu}\,\,{\rm entire\,\,
phase\,\, space})}, \label{splitting_calculation}
\end{equation}
which we compute using DGE and match to the fully differential NLO
result in moment space.
\end{itemize}
The partial branching ratio measured by the B factories can then be
compared to
\begin{eqnarray}
\label{Delta_cal_B_th_general} \Delta {\cal
B}(\bar{B}\longrightarrow X_u l \bar{\nu}\, \,{\rm restricted\,\,
phase\,\, space}) = \tau_B\Gamma_{\tot}\left(\bar{B}\longrightarrow
X_u l \bar{\nu}\right) R_{\rm cut}.
\end{eqnarray}
The advantages in splitting the calculation of the partial width
this way are:
\begin{itemize}
\item{} The event fraction $R_{\rm cut}$ is free of the ${\cal
O}(\Lambda/m_b)$ renormalon ambiguity associated with the overall
factor $m_b^5$ where $m_b$ is the quark pole mass. Note that
 ${\cal O}(n \Lambda/m_b)$ cut--related renormalon ambiguities
{\em are} present in the perturbative result for $R_{\rm cut}$, but
their effect is restricted to the moment--space Sudakov exponent.
Therefore, the perturbative expansion of the hard matching
coefficient for $R_{\rm cut}$ is renormalon free and truly dominated
by hard scales at higher orders.
\item{} The total charmless semileptonic width can be computed with
NNLO accuracy and in a way that renormalon--related ${\cal
O}(\Lambda/m_b)$ power effects explicitly cancel. One then obtains
the total width with theoretical uncertainty as low as $\sim 6 \%$.
Such accuracy cannot be achieved in a direct calculation of the
partial width in the restricted phase--space since (1) the
differential width is known in full only to NLO accuracy; and (2)
while Sudakov resummation can be used to take into account
parametrically--enhanced corrections associated with the phase-space
restriction, some residual uncertainty of both perturbative and
non-perturbative nature remains. Quantitative estimates are provided
in Sec.~\ref{sec:Numerical_results_and_uncertainty}.
\end{itemize}

The paper is divided into three main sections. In
Sec.~\ref{sec:total} we compute the total semileptonic width using
the NNLO result~\cite{vanRitbergen:1999gs} and renormalon
resummation. In Sec.~\ref{sec:resummation} we present the results
for resummation of the triple--differential width and work out NLO
matching formulae in moment space. Finally, in
Sec.~\ref{sec:Hadronic_var} we use these results to compute the
differential and partially--integrated width in hadronic variables,
taking into account the mass difference between the B meson and the
b quark. In this section we study the theoretical
uncertainty in $R_{\rm cut}$ and extract $\left\vert
V_{ub}\right\vert$ from recent measurements by
Belle~\cite{Bizjak:2005hn}.

\section{The total charmless semileptonic decay width\label{sec:total}}

According to the OPE the total $\bar{B}\longrightarrow X_u l
\bar{\nu}$ decay width is given by
\begin{eqnarray}
\label{total_width} \Gamma_{\tot}\left(\bar{B}\longrightarrow X_u l
\bar{\nu}\right) &=& \frac{G_F^2\left\vert
V_{ub}\right\vert^2m_b^5}{192\pi^3} F(\alpha_s(m_b))\\ \nonumber &=&
\frac{G_F^2\left\vert V_{ub}\right\vert^2m_b^5}{192\pi^3}
\left[1+a_0\frac{\alpha_s(m_b)}{\pi}+a_1\left(\frac{\alpha_s(m_b)}{\pi}\right)^2+\cdots
\,+\,{\cal O}\left(\frac{\Lambda^2}{m_b^2}\right)\right],
\end{eqnarray}
where the perturbative expansion has been computed to
NNLO~\cite{vanRitbergen:1999gs} and the matrix elements controlling
the ${\cal O}\left({\Lambda^2}/{m_b^2}\right)$ terms are not large
and amount to $\sim 1\%$ correction.

It is well known \cite{Bigi:1994em,Beneke:1994bc,Beneke:1994sw} that
while the decay width receives just small ${\cal
O}(\Lambda^2/m_b^2)$ power corrections, the perturbative expansions
of (a) the ratio between the pole mass and short--distance masses
(e.g. $m_{b}^{\MSbar}$); and (b) the function $F(\alpha_s(m_b))$ in
\eq{total_width} have {\em large} ${\cal O}(\Lambda/m_b)$ renormalon
ambiguities owing to infrared sensitivity. Furthermore, the {\em
first few terms} in these expansions show poor convergence; the
asymptotic behavior sets in early. A practical problem one needs to
address is how to optimally use the known coefficients in these
expansions to get a reliable estimate of $\Gamma_{\tot}$ in
\eq{total_width}. Often (e.g. \cite{Lange:2005yw,HLL}) this problem
is solved by resorting to a specific mass scheme where the mass and
the perturbative expansion of the width are separately renormalon
free, and hopefully converge well. Our approach is different. We use
the pole mass itself and regularize the leading renormalon using
Borel summation\footnote{The possibility to determine the
normalization of leading renormalon residues using the structure of
the singularity and the first few orders in the perturbative
expansion was considered by T.~Lee (in a different context) already
in~\cite{Lee_residues}. It was further observed that this
information can be used to systematically construct a bi-local
expansion for the Borel transform~\cite{Cvetic:2001sn}. Then Pineda
found~\cite{Pineda:2001zq} that in the case of the pole mass the
residue can be accurately determined, and used it to subtract the
corresponding divergence. More recently, it was shown that the
Principal Value Borel--resummed pole mass can be directly computed
from the bi-local expansion~\cite{Lee:2002px,RD} and used as an
alternative to short--distance or renormalon--subtracted mass
definitions.}, explicitly using the cancellation in
\eq{total_width}.

In Ref.~\cite{RD} we studied the mass ratio and expressed it as a
Borel integral where the Borel function is written as a bi-local
expansion~\cite{Cvetic:2001sn,Pineda:2001zq,Lee:2002px}
\begin{equation}
\label{mass_ratio_Borel}
\frac{m_b}{m_b^{\MSbar}}\,=\,1+\frac{C_F}{\beta_0}\int_0^{\infty} dz
\sum_{i=0}^{i_{\max}}b_i\,z^i+\frac{q}{(1-2z)^{1+\frac12
\delta}}\left[ 1+\sum_{k=1}^{k_{\max}}c_k(1-2z)^k\right],
\end{equation}
where $\delta\equiv \beta_1/\beta_0^2$, $c_k$ depend on the
coefficients of the $\beta$ function (see Appendix B in \cite{RD}),
and the coefficients $b_i$ in the regular part of the expansion can
be determined through order $b_2$ knowing the N$^3$LO
result~\cite{Melnikov:2000qh} for the mass ratio. Here both the {\em
exact} analytic structure of the singularity~\cite{Beneke:1994rs}
and the value of the residue $q$ fixing the large--order asymptotic
behavior of the expansion are used. The latter can be determined
with $\lsim 3\%$
accuracy~\cite{Pineda:2001zq,Lee:2003hh,Cvetic:2003wk,RD} using the
N$^3$LO result for the mass ratio. Hence, given a definite
integration prescription in the complex $z$ plain to avoid the
$z=\frac12$ singularity, \eq{mass_ratio_Borel} facilitates an
accurate calculation of both the ${\cal O}(1)$ real and the ${\cal
O}(\Lambda/m_b)$ imaginary parts of the mass ratio. With the central
values of the residue from Ref.~\cite{RD},
\begin{equation}
C_F q(N_f=3)/\pi=0.560;\qquad C_F q(N_f=4)/\pi=0.536 \label{q_det},
\end{equation}
one obtains:
\begin{equation}
\label{mass_ratio_results}
\left.\frac{m_b}{m_b^{\MSbar}}\right\vert_{N_f=3}\,=\,
1.161\,-\,0.0934\,i\;\qquad
\left.\frac{m_b}{m_b^{\MSbar}}\right\vert_{N_f=4} \,=\,
1.164\,-\,0.0768\, i,
\end{equation}
respectively. The variation in the number of light fermions is used
to estimate charm mass effects. Here we chose
 to {\em define} the Borel sum in \eq{mass_ratio_Borel} with the integration
contour going {\em below} the real axis avoiding the singularity at
$z=\frac{1}{2}$. This fixes the imaginary part of the mass ratio
(the physical real part is prescription independent). We evaluated
the integrals using:
\begin{eqnarray}
\int_{0}^{\infty} dz \frac{{\rm
e}^{-z/A}}{\left(1-\frac{z}{z_0}\right)^{1+\nu}}&=&
\frac{A}{\Gamma(1+\nu)}\int_{0}^{\infty}dl \frac{{\rm
e}^{-l}l^{\nu}} {1-\frac{lA}{z_0}}=
\frac{z_0}{\Gamma(1+\nu)}\frac{\pi}{\sin(\pi\nu)}\, {\rm
e}^{-z_0/A}(-z_0/A)^{\nu}
\nonumber \\
&&\hspace*{-50pt} +A^{1-\frac{1}{2} \nu} z_0^{\frac{1}{2} \nu}
\frac{{\rm e}^{-\frac{1}{2} z_0/A}}{1-\nu} \bigg[M_{\frac{1}{2}
\nu,\frac{1}{2}-\frac{1}{2} \nu}\left(z_0/A\right)
 -\frac{1}{\nu}M_{1+\frac{1}{2}\nu,\frac{1}{2}-\frac{1}{2}\nu}
 \left(z_0/A\right)\bigg],
\end{eqnarray}
where we assumed that $z_0=\frac12+i\varepsilon$ where $\varepsilon$
is an infinitesimally small {\em positive} number, and
$M_{\mu,\nu}(a)$ is the Whittaker $M$ function,
\[
M_{\mu,\nu}(a) = {\rm e}^{-\frac12 a} a^{\frac12 +\nu}\,
_1F_1\left(\Big[\frac12+\nu-\mu\Big],\Big[1+2\nu\Big],a\right).
\]

The next crucial observation is that the cancellation of the leading
renormalon ambiguity in \eq{total_width} implies that
\begin{eqnarray}
\label{Im_cancellation} \frac{{\rm
Im}\left\{F^{1/5}(\alpha_s(m_b))\right\}} {{\rm
Re}\left\{F^{1/5}(\alpha_s(m_b))\right\}}= -\frac{{\rm
Im}\left\{\frac{m_b}{m_b^{\MSbar}}\right\}} {{\rm
Re}\left\{\frac{m_b}{m_b^{\MSbar}}\right\}} \,=\,{\cal
O}(\Lambda/m_b),
\end{eqnarray}
and that the observable itself can be computed with ${\cal
O}(\Lambda^2/m_b^2)$ accuracy using the real parts (or Principal
Value) of the mass ratio and $F^{1/5}$:
\begin{eqnarray}
m_b F^{1/5} (\alpha_s(m_b))= m_b^{\MSbar}\times {\rm
Re}\left\{\frac{m_b}{m_b^{\MSbar}}\right\}\times {\rm
Re}\left\{F^{1/5}(\alpha_s(m_b))\right\}.
\end{eqnarray}
\eq{Im_cancellation} implies that the ambiguity of $F^{1/5}
(\alpha_s(m_b))$, just like that of the mass
ratio~\cite{Beneke:1994rs}, is a pure power term (i.e. a number
times $\Lambda/m_b$), not modified by logarithms. This is an exact
result.

The structure of the Borel singularity is therefore:
\begin{equation}
\label{F_Borel} F^{1/5}
(\alpha_s(m_b))\,=\,1+\frac{C_F}{\beta_0}\int_0^{\infty} dz
\sum_{j=0}^{j_{\max}}\tilde{b}_j\,z^j
+\frac{\tilde{q}}{(1-2z)^{1+\frac12 \delta}}\left[
1+\sum_{k=1}^{k_{\max}}c_k(1-2z)^k\right],
\end{equation}
where the state-of-the-art knowledge of the expansion of
$F(\alpha_s(m_b))$ in~\eq{total_width} (NNLO accuracy) allows
determination of $\tilde{b}_0$ and $\tilde{b}_1$ (so $j_{\max}=1$).
Evaluating the Borel sum in \eq{F_Borel} and using
\eq{Im_cancellation} and \eq{mass_ratio_results} we obtain
\begin{equation}
\left.\tilde{q}\right\vert_{N_f=3} = -1.191\,;\qquad\qquad
\left.\tilde{q}\right\vert_{N_f=4} = -1.146\,,
\end{equation}
respectively. This yields:
\begin{equation}
\label{F_fifth_result} \left.F^{1/5}
(\alpha_s(m_b))\right\vert_{N_f=3}\,\,=\,0.933 +0.0750\,i\,; \qquad
\left.F^{1/5} (\alpha_s(m_b))\right\vert_{N_f=4}\,\,=\,0.928+
0.0612\,i\,,
\end{equation}
respectively. Using \eq{F_fifth_result} and \eq{mass_ratio_results}
with
\begin{equation}
m_b^{\MSbar}\equiv m_b^{\MSbar}(m_b^{\MSbar})=4.19\pm0.05\,{\rm GeV}
\label{mb_MSbar_value}
\end{equation}
we find
\begin{equation}
\left.m_b^5F(\alpha_s(m_b))\right\vert_{N_f=3}=1926\pm 117 \,{\rm
GeV}^5; \qquad
 \left.m_b^5F(\alpha_s(m_b))\right\vert_{N_f=4}=1901\pm 116 \,{\rm GeV}^5.
\end{equation}
Thus, in units of $\left\vert V_{ub}\right\vert^2$, the
total semileptonic decay width is
\begin{eqnarray}
\label{total_width_over_Vub_sq} \frac{1}{\left\vert
V_{ub}\right\vert^2} \Gamma_{\tot}\left(\bar{B}\longrightarrow X_u l
\bar{\nu}\right)= 66.5 \pm 4 \,{\rm ps}^{-1},
\end{eqnarray}
where the error is dominated by the uncertainty in the value of the
short--distance mass in \eq{mb_MSbar_value}. The central value as
well as the error estimate are consistent with previous studies, see
e.g.~Ref.~\cite{Uraltsev:1999rr}.

\section{Resummation of the triple differential width\label{sec:resummation}}

\subsection{Partonic kinematics and triple differential width at NLO\label{sec:NLO}}

In order to describe the triple differential width in $b (p_b)
\,\longrightarrow\, X_u (p_j)\, l (k_l)\,\bar{\nu} (k_\nu)$ we
define the following kinematic variables: the charged lepton energy
fraction,
\begin{eqnarray}
\label{x_def} x&\equiv& 2p_b\cdot k_l/m_b^2,
\end{eqnarray}
the lightcone variables defined by the partonic jet
\begin{eqnarray}
\label{lightcone_variables}
\rho    &\equiv& p_j^+/m_b=(E_j-\left\vert \vec{p}_j\right\vert)/m_b,\nonumber \\
\lambda &\equiv& p_j^-/m_b=(E_j+\left\vert
\vec{p}_j\right\vert)/m_b,
\end{eqnarray}
and their ratio
\begin{eqnarray}
r  &\equiv& \frac{\rho}{\lambda}
=\frac{p_j^+}{p_j^-}=\frac{E_j-\left\vert \vec{p}_j\right\vert}
{E_j+\left\vert \vec{p}_j\right\vert},
\end{eqnarray}
which is the exponent of the rapidity. As we shall see below the
structure of the Sudakov exponent and the matching is particularly
simple when expressed in moments of $1-r$. Using the variables
$(\lambda,r,x)$  the phase--space integration is:
\begin{equation}
\Gamma_{\tot}=\int_0^1 d\lambda \int_{1-\lambda}^1 dx
\int_{0}^{(1-x)/\lambda}\!\!dr\,\frac{d\Gamma(\lambda,r,x)}{d\lambda
dr d x}. \label{partonic_ps}
\end{equation}
The perturbative expansion of the triple differential width takes
the form:
\begin{eqnarray}
\label{triple_diff_r}
\frac{1}{\Gamma_0}\frac{d\Gamma(\lambda,r,x)}{d\lambda d r dx}
=V(\lambda,x) \delta(r)+ R(\lambda,r,x)
\end{eqnarray}
where
\begin{equation}
\Gamma_0=\frac{G_F^2\left|V_{ub}\right|^2m_b^5}{192 \pi^3}
\label{Gamma0}
\end{equation}
and
\begin{eqnarray}
\label{V_and_D_r} V(\lambda,x)&=&w_0(\lambda,x)
+\frac{C_F\alpha_s(m_b)}{\pi}w_1(\lambda,x)+\cdots,\\ \nonumber
R(\lambda,r,x)&=&\frac{C_F\alpha_s(m_b)}{\pi}
k_1(\lambda,r,x)+\cdots,
\end{eqnarray}
correspond to virtual and real corrections, respectively.

The ${\cal O}(\alpha_s)$ coefficients (NLO) are known since
long~\cite{DFN}, however, they were computed there in terms of the
hadronic mass and energy variables which are less suited for
resummation as compared to the lightcone variables introduced above.
Using $(\lambda,r,x)$ the virtual coefficients are:
\begin{eqnarray}
\label{V_r_coef} w_0&=& 12\,(2 - x - \lambda )\,(x + \lambda  - 1)\\
\nonumber w_1&=&
  12\,(2-x  -\lambda )\,( x+ \lambda  - 1)
\bigg[ - {\displaystyle \frac {5}{4}} - {\displaystyle \frac {\pi ^{
2}}{3}}  - {\rm Li}_2(1- \lambda)\bigg]\\ \nonumber && \,+ 6
(2\,\lambda  + 2\,x - 5)\,\mathrm{ ln}(\lambda )\,( x + \lambda  -
1)
\end{eqnarray}
and the real--emission coefficient is:
\begin{eqnarray}
\label{c_1} k_1(\lambda,r,x)&=&\frac{6\mathrm{ln}\,r} {r\,(r -
1)^{2}} (x - 2 + \lambda  + r\,\lambda ) \Big( r\,\lambda  (1-
\lambda) + x + \lambda - 1 \Big)\times \\ \nonumber && \bigg[2 +
2\,r^{3} + r^{3}\,\lambda ^{2} - 2\,r^{3}\,\lambda  - 6\,
r^{2}\,\lambda  + r^{2}\,\lambda ^{2} + 2\,r^{2} + 2\,r - 2\,r\,
\lambda \bigg]
\\ \nonumber &+&
 \frac{3}{r(1-r)}
\,( \lambda  + r\,\lambda  - r\,\lambda ^{2} + x-1)\,(x - 2
 + \lambda  + r\,\lambda )\,\times \\ \nonumber &&
 \bigg[4\,r^{2}\,\lambda ^{2} - 10\,r^{2}\,
\lambda  + 7\,r^{2} + 2\,r - 10\,r\,\lambda  + 7\bigg]
 \\ \nonumber
 &+& {\displaystyle \frac {6\,( - 2\,r^{2}\,\lambda  + 2\,r
^{2} + r^{2}\,\lambda ^{2} - 2\,r\,\lambda  + 4\,r + 2)\,( - 1 +
r\,\lambda )\,( - 1 + \lambda )\,\lambda \,\mathrm{ln}(r)}
{(1-r)^{2}}}  \\ \nonumber &+& {\displaystyle \frac
{3\,(r^{2}\,\lambda ^{2} - 8\,r\, \lambda  + 8\,r + r\,\lambda ^{2}
+ 8)\,( - 1 + r\,\lambda )\,(
 - 1 + \lambda )\,\lambda }{1-r}}\\\nonumber
 &-&
\frac{6 (x - 1 + r\,\lambda )(\lambda  +
x-1)\,\mathrm{ln}\,r}{(1-r)^{4}}
  \bigg[ - 14\,r + 42\,r^{2}\,\lambda
- 2\,r\,\lambda ^{2} - 2\,r^{4}\,\lambda ^{2}+ 16\,r^{3} \,\lambda
  \\ \nonumber
  &+&
  2\,r^{4}\,\lambda - 22\,r^{2}\,\lambda ^{2} + 16\,r
\,\lambda  - 4\,r^{3} + 2\,\lambda  - 22\,r^{3}\,\lambda ^{2}
\mbox{} - 14\,r^{2} + \lambda ^{3}\,r^{4} - 4
\\ \nonumber &+& 4\,r^{3}\,\lambda
 ^{3} + r^{2}\,\lambda ^{3}\bigg]
  - \frac{3(x - 1 + r\,\lambda
 )\,( - 1 + \lambda  + x)}{ (1-r)^{3}}
\bigg[6\,r^{3}\,\lambda ^{3} + 11\,r^{3}\,\lambda  - 16\,r^{3}\,
\lambda ^{2} \\ \nonumber &+& 67\,r^{2}\,\lambda  -
64\,r^{2}\,\lambda ^{2} -  22\,r^{2} + 6\,r^{2}\,\lambda ^{3} -
16\,r\,\lambda ^{2} - 28\,r
 + 67\,r\,\lambda  - 22 + 11\,\lambda \bigg].
\end{eqnarray}
The $r\longrightarrow 0$ singular (non integrable) terms in $k_1$,
namely
\begin{eqnarray}
\label{c_1_sing} k_1^{\sing}(\lambda,r,x) &=& -12\,(2- x - \lambda
)\, ( x + \lambda  - 1)\,\left[\left(\frac{\mathrm{ln}
\,r}{r}\right)_{*}
+ \frac{7}{4}\left(\frac{1}{r}\right)_{*}\right] \nonumber  \\
&=&-w_0\,\left[\left(\frac{\mathrm{ln}\, r}{r}\right)_{*} +
\frac{7}{4}\left(\frac{1}{r}\right)_{*}\right]
\end{eqnarray}
are regularized as $()_*$ distributions, i.e.
 \begin{eqnarray}
\label{star_dist} \int_0^{r_0} dr F(r) \left(\frac{1}{r}\right)_{*}
=F(0)\ln r_0+\int_0^{r_0}
dr \Big(F(r)-F(0)\Big) \frac{1}{r},\\
\int_0^{r_0} dr F(r) \left(\frac{\ln r}{r}\right)_{*} =F(0)\frac12
\ln^2 r_0 +\int_0^{r_0} dr \Big(F(r)-F(0)\Big)\frac{\ln r}{r},
\end{eqnarray}
where $F(r)$ is a smooth test function. The remaining, regular part
of $k_1$
\begin{eqnarray}
k_1^{\reg}(\lambda,r,x)= k_1(\lambda,r,x) - k_1^{\sing}(\lambda,r,x)
\end{eqnarray}
is integrable, so it does not require any regularization.

As usual, at ${\cal O}(\alpha_s)$ and beyond the separation into
real and virtual terms is not unique. In order to compare the result
quoted here with NLO expressions for the triple differential width
in terms of other kinematic variables (e.g.~in~\cite{DFN}) one must
take account of the fact that terms proportional to $\delta(r)$ are
contained both in the $()_*$ distributions and in the virtual terms
and, depending on the variables chosen, they may be split
differently between the two.

\subsection{The Sudakov exponent{\label{sec:Sud_exp}}}

Owing to multiple soft and collinear gluon emission, Sudakov
logarithms, such as the \hbox{$r\longrightarrow 0$} singular terms
of \eq{c_1_sing}, appear at any order in perturbation theory. Such
terms spoil the convergence of the perturbative expansion and must
be resummed. Sudakov logarithms in inclusive decay spectra are
associated with two independent
subprocesses~\cite{KS,BDK,RD,Gardi:2004gj} (see also
Refs.~\cite{Bosch:2003fc,Leibovich:2000ig,Bauer:2000ew,Bauer:2000yr,Bauer:2003pi,Aglietti:2000te,Aglietti:2001br,Aglietti:2005mb}):
 the {\em soft function}, which is the
Sudakov factor of the {\em heavy quark distribution
function}~\cite{QD} and the {\em jet function}, summing up radiation
that is associated with an unresolved final--state quark jet of a
given invariant mass. This function is directly related to the
large-$x$ limit of deep inelastic structure functions; see
Ref.~\cite{RD,GR} for further details.

\subsubsection*{Singular terms in the large--$\beta_0$ limit}

Ref.~\cite{BDK} has generalized the concept of Sudakov resummation
in inclusive decay spectra beyond the perturbative (logarithmic)
level. It has been shown that, when considered to all orders, the
moments corresponding to each of the above subprocesses contain
infrared renormalons and therefore certain power corrections
exponentiate together with the logarithms.

The calculation in Ref.~\cite{BDK} was based on evaluating the
real--emission diagrams with a single dressed gluon using the Borel
technique: the dimension of the gluon propagator is modified:
$1/(-k^2) \longrightarrow 1/(-k^2)^{1+u}$, and in this way one gains
control of the physical scale for the running coupling.
Ref.~\cite{BDK} used the leptonic and partonic invariant mass
variables $z\equiv q^2/m_b^2$ and $1-\xi\equiv p_{j}^2/m_b^2$, that
are related to the lightcone variables of \eq{lightcone_variables}
by:
\begin{eqnarray}
z=(1-\rho)(1-\lambda); \qquad \qquad \xi= 1-\lambda\rho,
\end{eqnarray}
and obtained the following result for the $\xi\longrightarrow 1$
singular terms in the triple differential width in the
large--$\beta_0$ limit:
\begin{eqnarray}
\label{sl_logs} &&\hspace*{-10pt}\left. \frac{1}{\Gamma_0}
\frac{d\Gamma(\xi,z,x)}{dz dx d\xi}\right\vert_{\xi\longrightarrow
1} =12\,(x-z)(1+z-x)\,\times \bigg\{ \delta(1-\xi)+
\\ \nonumber
&&\hspace*{00pt} +\,\frac{C_F}{\beta_0}\int_0^{\infty}\frac{du}{u}\,
T(u)\left(\frac{\Lambda^2}{m_b^2}\right)^u
  \left[B_{\cal
S}(u)(1-z)^{2u}\left(1-\xi\right)^{-1-2u} -B_{\cal
J}(u)\left(1-\xi\right)^{-1-u} \right] \bigg\}+\cdots
\end{eqnarray}
where the dots stand for terms that are subleading in $\beta_0$ or
are suppressed by powers of $(1-\xi)$, and the Borel functions
corresponding to the two Sudakov anomalous dimensions are:
\begin{eqnarray}
\label{B_DJ_large_beta0}
B_{\cal S}(u)&=&{\rm e}^{\frac53 u}(1-u)\,+\,{\cal O}(1/\beta_0 ),\nonumber \\
B_{\cal J}(u)&=&\frac{1}{2}\,{\rm e}^{\frac53 u}
\left(\frac{1}{1-u}+\frac{1}{1-u/2}\right) \frac{\sin\pi u}{\pi
u}\,+\,{\cal O}(1/\beta_0 ).
\end{eqnarray}
Here and below we use the scheme--invariant Borel
transform~\cite{Grunberg:1992hf} where $T(u)$ is the Laplace
conjugate of the coupling. In the large--$\beta_0$ limit (one--loop
running) $T(u)=1$. Below we work in the full theory and define
$T(u)$ as the Laplace conjugate of the 't~Hooft coupling, see Eq.
(2.18) in Ref.~\cite{RD}.

\eq{sl_logs} reveals the physical scales that control the running
coupling within the soft and the jet subprocess, respectively: the
soft scale is $|k|\sim m_b(1-\xi)/(1-z)$ while the invariant mass
squared of the jet is $k^2\sim m_b^2(1-\xi)$.

\subsubsection*{Exponentiation}

Sudakov resummation can be done using different variables.  One
natural possibility is to resum Sudakov logarithms of $1-\xi$,
corresponding to the limit where the hadronic invariant mass gets
small, for fixed leptonic invariant mass $z$. This avenue is
followed in Appendix~A. Here instead we choose to revert to the
lightcone variables of \eq{lightcone_variables}, where the exponent
and the matching procedure are simpler. The comparison with
Appendix~\ref{sec:xi_resummarion} provides a useful consistency
check.

The Sudakov limit $\xi\longrightarrow 1$ corresponds to the
phase--space limit where the smaller lightcone momentum component
$\rho$ tends to zero. The large component $\lambda$ can either be
small or of ${\cal O} (1)$; fortunately, the region where also
$\lambda$ is small is power suppressed in the Born--level weight
(see \eq{bar_w0}) and is therefore unimportant. Consequently,
Sudakov resummation can be readily applied to resum logarithms of
$\rho$ or of $r$. The latter possibility is particularly attractive.
The singular terms for $r\longrightarrow 0$ in the large--$\beta_0$
limit take the form:
\begin{eqnarray}
\label{sl_logs_r_lambda} && \left.\frac{1}{\Gamma_0}
\frac{d\Gamma(\lambda,r,x)}{d\lambda dr dx}
\right\vert_{r\longrightarrow 0} =w_0\, \bigg\{ \delta(r)+
\nonumber \\
&& \hspace*{20pt} +\frac{C_F}{\beta_0}\int_0^{\infty}\frac{du}{u}\,
T(u)\left(\frac{\Lambda^2}{m_b^2 \lambda^2}\right)^u \left[B_{\cal
S}(u) r^{-1-2u} -B_{\cal J}(u) r^{-1-u} \right] \bigg\}+\cdots,
\end{eqnarray}
so both the soft and jet functions depend just on $r$, except for an
overall dependence on $\lambda$ that scales the argument of the
coupling. It is this dependence that makes the difference with the
radiative decay case \cite{RD}, which simply corresponds to the substitution of
$\lambda=1$ in~\eq{sl_logs_r_lambda}.

Multiple emission of soft and collinear gluons is taken into account
to all orders by exponentiation of the {\em fully differential}
width in {\em moment space}. It is necessary to consider the fully
differential width, since only then the hard configuration, from
which soft and collinear gluons are radiated, is fixed\footnote{This
point has recently been pointed out in Ref.~\cite{Aglietti:2005mb}.
We note that the variable $u$ introduced in that paper is identical
to our variable $r$.}. The integration over the variables $\lambda$
and $x$ that control the hard configuration itself can only be
performed after exponentiation has been carried out. Exponentiation
takes place in moment space because the phase space of soft and
collinear radiation factorizes there. Defining moments by
\begin{eqnarray}
\label{r_mom_def} \frac{d\Gamma_{n} (\lambda,x)}{d\lambda dx}
 &\equiv & \frac{1}{\Gamma_0}\int_{0}^{(1-x)/\lambda} dr
 \left(1-r\right)^{n-1}
\frac{d\Gamma(\lambda,r,x)}{d\lambda dr dx}
 \\ \nonumber
&\simeq  & \frac{1}{\Gamma_0}\int_{0}^{1} dr \left(1-r \right)^{n-1}
\left.\frac {d\Gamma(\lambda,r,x)}{d\lambda dr
dx}\right\vert_{r\longrightarrow 0} +{\cal O}(1/n),
\end{eqnarray}
the singular (non-integrable) terms for $r \longrightarrow 0$ of the
form appearing in \eq{sl_logs_r_lambda} generate terms that contain
powers of $\ln n$, while integrable terms generate ${\cal O}(1/n)$
terms, that are neglected at this stage. These terms will eventually
be taken into account (to a given order) by matching the resummed
result to a fixed--order expression. The resummed, triple
differetial distribution can then be computed to all orders by the
following inverse Mellin transformation:
\begin{eqnarray}
\label{sl_logs_resummed_r} &&\hspace*{-10pt} \frac{1}{\Gamma_0}
\frac{d\Gamma(\lambda,r,x)}{d\lambda dr dx }
=\int_{c-i\infty}^{c+i\infty} \frac{d n}{2\pi i} \,
\frac{d\Gamma_{n} (\lambda,x)}{d\lambda dx}\,
\left(1-{r}\right)^{-n}.
\end{eqnarray}
Computing the moments in~\eq{r_mom_def} with
\eq{sl_logs_r_lambda} and exponentiating the terms
that diverge in the large--$n$ limit we get the
following DGE formula:
\begin{eqnarray}
\label{Gamma_n_large_n} && \left.\frac{d\Gamma_{n}
(\lambda,x)}{d\lambda dx}\right\vert_{{\rm large}\,\, n} =
\left[V(\lambda,x)+\Delta
R^{\infty}(\lambda,x)\right]\,\times\exp\bigg\{
\frac{C_F}{\beta_0}\int_0^{\infty}\frac{du}{u} \, T(u)
\left(\frac{\Lambda^2}{m_b^2\lambda^2}\right)^u\\\nonumber &&
 \hspace*{20pt} \left[B_{\cal
S}(u)\Gamma(-2u)\left({n}^{2u}-1\right)  -B_{\cal
J}(u)\Gamma(-u)\left({n}^u-1\right) \right] \bigg\}\,+\,{\cal
O}(1/n)
\end{eqnarray}
where the square brackets that multiply the Sudakov factor contain
the virtual terms $V(\lambda,x)$ of \eq{triple_diff_r} as well as
$\Delta R^{\infty}$ that is defined as the $n\longrightarrow \infty$
limit of the {\em difference} between the moments of the full
real--emission contribution $R(\lambda,r,x)$ and the terms that are
included in the exponent, at any given order. $\Delta R^{\infty}$
will be determined in the next section at ${\cal O}(\alpha_s)$ in
the process of matching \eq{Gamma_n_large_n} to the full NLO
expression; see Eqs. (\ref{Delta_R_infty}) and
(\ref{Delta_K_1_infty}).

\eq{Gamma_n_large_n} summarizes the exact, all--order structure of
the Sudakov factor.
Comparing \eq{Gamma_n_large_n} to~\eq{sl_logs_r_lambda}, which was
derived in the large--$\beta_0$ limit, one notes
 that terms that are subleading in $\beta_0$ arise from several sources:
(1) the exponentiation;
(2) the function $T(u)$ summarizing the dependence
 on the two--loop $\beta$ function coefficient $\beta_1$;
(3) the anomalous--dimension functions
$B_{\cal S}(u)$ and $B_{\cal J}(u)$ that receive ${\cal O}(1/\beta_0 )$
contributions starting at ${\cal O}(u)$.
The Sudakov factor in $\bar{B}\longrightarrow X_s \gamma$~\cite{RD} can be
recovered by substituting $\lambda=1$ in \eq{Gamma_n_large_n}.

The perturbative expansions of $B_{\cal S}(u)$ and $B_{\cal J}(u)$
have recently been determined~\cite{QD,RD} to ${\cal O}(u^2)$, i.e.
the NNLO; explicit expressions appear in Sec. 2.2 in~\cite{RD}, see
\eq{B_DJ} below. In the following we shall compute the Borel sum in
\eq{Gamma_n_large_n} directly, using the Principal Value
prescription. Before doing so, however, it is worthwhile looking at
the conventional approach to Sudakov resummation, where a
logarithmic--accuracy criterion is used.

\subsubsection*{Resummation with a fixed logarithmic accuracy }

Let us write the exponent in \eq{Gamma_n_large_n} as an integral of
the Sudakov anomalous dimensions over the range of scales:
\begin{eqnarray}
\label{Gamma_n_large_n_FLA} && \frac{d\Gamma_{n}
(\lambda,x)}{d\lambda dx} = \left[V(\lambda,x)+\Delta
R^{\infty}(\lambda,x)\right]\,\times\exp\Bigg\{
\int_0^{1}dr\,\frac{(1-r)^{n-1}-1}{r}\\\nonumber && \hspace*{30pt}
\left[ \int_{m_b^2\lambda^2r^2}^{m_b^2\lambda^2 r}
\frac{d\mu^2}{\mu^2}{\cal A}(\alpha_s(\mu^2)) +{\cal
B}(\alpha_s(m_b^2\lambda^2r)) -{\cal D}(\alpha_s(m_b^2\lambda^2r^2))
\right]
 \Bigg\}
\end{eqnarray}
where the anomalous dimension functions have the following
expansions in the ${\overline{\rm MS}}$ scheme:
\begin{eqnarray}
\label{A_cusp_expansion} {\cal A}\big(\alpha_s(\mu^2)\big)&=&
\sum_{n=1}^{\infty}
A_n\left(\frac{\alpha_s^{\MSbar}(\mu^2)}{\pi}\right)^n \,=\,
\frac{C_F}{\beta_0}\int_0^{\infty} du \,T(u)
\,\left(\frac{\Lambda^2}{\mu^2}\right)^u\, B_{\cal A}(u),\nonumber
\\
{\cal B}\big(\alpha_s(\mu^2)\big)&=& \sum_{n=1}^{\infty}
B_n\left(\frac{\alpha_s^{\MSbar}(\mu^2)}{\pi}\right)^n\,=\,\frac{C_F}{\beta_0}\int_0^{\infty}
du \,T(u) \,\left(\frac{\Lambda^2}{\mu^2}\right)^u\, B_{\cal
B}(u),\nonumber
\\
{\cal D}\big(\alpha_s(\mu^2)\big)&=& \sum_{n=1}^{\infty}
D_n\left(\frac{\alpha_s^{\MSbar}(\mu^2)}{\pi}\right)^n\,=\,\frac{C_F}{\beta_0}\int_0^{\infty}
du \,T(u) \,\left(\frac{\Lambda^2}{\mu^2}\right)^u\, B_{\cal D}(u),
\end{eqnarray}
where the known\footnote{The coefficient $A_3$
was recently computed in the course of the three--loop calculation
of the Altarelli--Parisi evolution kernel~\cite{MVV}. $B_2$ has been
known for some time~\cite{Vogt:2000ci,GR} based on two--loop calculations in deep--inelastic
scattering. $D_2$ was recently computed~\cite{QD} directly from the
renormalization of the corresponding Wilson--line operator~\cite{KM} at two--loops.
Independent confirmation of this result has been possible owing to an exact all--order
correspondence~\cite{QD} between the heavy quark distribution and
fragmentation functions in the Sudakov limit and a recent two-loop calculation of the
latter~\cite{MM}.
} coefficients $A_{1,2,3}$, $B_{1,2}$ and $D_{1,2}$
are detailed in Eqs. (2.5) through (2.7) in Ref.~\cite{RD} and the all--order
relation with the Borel functions $B_{\cal S}(u)$ and $B_{\cal
J}(u)$ of \eq{Gamma_n_large_n} above is
\begin{eqnarray}
\label{DB_to_SJ}
B_{\cal S}(u)&= & B_{\cal A}(u) - u B_{\cal D}(u),\nonumber \\
B_{\cal J}(u)&= & B_{\cal A}(u) - u B_{\cal B}(u).
\end{eqnarray}

The moment--space Sudakov factor of \eq{Gamma_n_large_n_FLA}
can be written as:
\begin{equation}
{\rm Sud}(m_b\lambda ,n)=\exp\left\{ \sum_{k=0}^{\infty}
g_k(\tau)\left(\frac{\alpha_s^{\MSbar}(\lambda^2
m_b^2)}{\pi}\right)^{k-1}\right\}
 \label{sum_FLA},
\end{equation}
where
\[
\tau\equiv \frac{\alpha_s^{\MSbar}(\lambda^2 m_b^2)}{\pi}\beta_0 \ln n.
\]
A fixed--logarithmic--accuracy approximation is obtained by a given truncation
of the sum over $k$ in the exponent. The first three coefficients
$g_{k}(\tau)$, which sum up the logarithms to NNLL accuracy, are:
\begin{eqnarray}
\label{g_i} {g_{0}}(\tau) &=&  \frac{A_{1}}{{\beta _{0}}^{2}}
{\left(\left( 1- \tau\right)\,\ln\left(1 - \tau  \right) -\frac12
\left( 1 -2 \tau \right) \,\ln\left( 1-2\tau\right)\right)}
\\ \nonumber
{g_{1}}(\tau) &=&  \frac{1}{\beta _{0}}\left[{{{A_{1} \gamma_E}}
\bigg( - \,\ln\left( 1-\tau\right) +  \,\ln\left(
1-2\tau\right)\bigg) + {B_{1}}\,\ln\left( 1-\tau\right) - { \frac
{1}{2}} \,{{D}_{1}}\,\ln\left( 1-2\tau\right)}\right]
\\ \nonumber &&  + {\frac {{A_{2}}}{{\beta _{0}}^{2}} \bigg( - \ln\left( 1-\tau\right)
 + { \frac {1}{2}} \,\ln\left( 1-2\tau\right)
\bigg)\,}  \\ \nonumber &&  + {\frac {{A_{1}}\,{\beta _{1}}}{{\beta
_{0}}^{3}} \left( - { \frac {1}{2}} \,\ln\left( 1-2\tau\right) - {
\frac {1}{4}} \,\ln\left( 1-2\tau\right)^{2} + { \frac {1}{ 2}}
\,\ln\left( 1-\tau\right)^{2} + \ln\left( 1-\tau\right)\right)\,}
 \\ \nonumber
{g_{2}} (\tau)&=& {A _{1}}\left({ \frac {1}{2}}  - { \frac {1}{
1-2\tau}}  + { \frac {1}{ - 2\, \tau + 2}} \right)\,\left(\gamma_E
^{2} + \frac{\pi^2}{6}\right)
\\\nonumber &&+
{B_{1}\,\gamma_E }\left(1 - { \frac {1}{ 1-\tau}} \right)\,
+ {{D}_{1}\,\gamma_E}\left({ \frac {1}{ 1-2\tau} }  - 1\right) \, \\
\nonumber && + { \frac{1}{{\beta _{0}}} \left\{{ \left( { \frac
{1}{1-\tau }} - { \frac {1}{1-2\,\tau }} \right)\,{A_{2}\gamma_E} +
\left({    1-\frac {1}{1-\tau  }} \right)\,{B_{2}} + \left( - {
\frac {1}{2}}  - { \frac {1}{4\,\tau  - 2}}
\right)\,{{D}_{2}}}\right\} }  \\ \nonumber && +
\frac{1}{\beta_{0}^{2 }}\,\bigg\{\bigg[ \bigg({ \frac { \,\ln\left(
1-\tau\right)}{ \tau - 1}} - { \frac { \,\ln\left(1 - 2 \,\tau
\right)}{2\,\tau - 1}} + \,{ \frac {1}{\tau - 1}}  - { \frac
{1}{2\,\tau
 - 1}} \bigg)\,{A_{1}\gamma_E} \\ \nonumber &&
 + \left( - { \frac {1}{\tau  - 1}}  - { \frac {\ln\left(
1-\tau\right)}{\tau  - 1} } - 1\right)\,{B_{1}} + \left({ \frac
{1}{2}}  + { \frac {1}{4\,\tau - 2}}  + { \frac {1}{2}} \,{ \frac
{\ln\left( - 2\,\tau
 + 1\right)}{2\,\tau  - 1}} \right)\,{{D}_{1}}\bigg]{\beta _{1}} \\ \nonumber &&
 + \left( - { \frac {1}{4}}  + { \frac {1}{8\,\tau - 4}} -
{ \frac {1}{2\, \tau  - 2}} \right)\,{A_{3}}\bigg\}
 \\
\nonumber && +\,\frac{1}{\beta _{0}^{3}}\,\bigg\{\left({ \frac
{\ln\left( 1-\tau\right)}{\tau  - 1 }} + { \frac {3}{2\,\left(\tau
- 1\right)}}  + { \frac {3}{4}}  - { \frac {1}{2}} \, { \frac
{\ln\left( 1-2\tau\right)}{2\,\tau
  - 1}}  - { \frac {3}{4\,\left(2\,\tau  - 1\right)}} \right)\,{A
_{2}}\,{\beta _{1}} \\ \nonumber &&  + \left(\ln\left( 1-\tau\right)
- { \frac {1 }{2}} \,\ln\left( 1-2\tau\right)
 - { \frac {1
}{4}}  - { \frac {1}{2\,\tau  - 2}}  + { \frac {1}{8\,\tau  -
4}} \right)\,{A_{1}}\,{\beta _{2 }}\bigg\}  \\
\nonumber &&  + \frac{{A_{1}}\,{\beta _{1}}^{2}}{{\beta
_{0}}^{4}}\,\bigg\{ - { \frac {1}{2}} \,{ \frac {\ln^{2}\left(
1-\tau\right)}{\tau  - 1}}  - { \frac {\ln\left( 1-\tau\right)\,\tau
}{ \tau  - 1}}  - { \frac {1}{2\,\tau - 2}}  + { \frac {1}{4}} \,{
\frac {\ln^{2}\left(
 1-2\tau\right)}{2\,\tau  - 1}}    \\
\nonumber && + { \frac {\ln\left( 1-2\tau\right)\,\tau }{2\,\tau  -
1 }} + { \frac {1}{8\,\tau  - 4}}  - { \frac {1}{4}} \bigg\},
\end{eqnarray}
where the coefficients of the anomalous dimensions defined in
\eq{A_cusp_expansion} and the $\beta$ function are in the
${\overline {\rm MS}}$ scheme; they are given in Sec. 2.1 in
Ref.~\cite{RD}.

Next, we apply the inverse Mellin transform of
\eq{sl_logs_resummed_r} in order to convert the resummed result to
momentum space at NNLL accuracy (see Ref.~\cite{RD} and Sec. 3.4 in
Ref.~\cite{DGE_thrust}). Let us consider the integrated width with
$r<r_0$ for some $r_0<(1-x)/\lambda$. The NNLL resummed result,
matched to the full NLO expression takes the form:
\begin{eqnarray}
\label{NNLL_x_space}
\frac{1}{\Gamma_0}\frac{d\Gamma(\lambda,r<r_0,x)}{d\lambda dx} &=&
\left[V(\lambda,x) +\Delta R^\infty(\lambda,x) \right]\times\\
\nonumber && \hspace*{-52pt}
\frac{\exp\bigg\{g_0(\omega)\left(\frac{\alpha_s^{\MSbar}(\lambda^2
m_b^2)}{\pi}\right)^{-1}+g_1(\omega)+
g_2(\omega)\frac{\alpha_s^{\MSbar}(\lambda^2 m_b^2)}{\pi}\bigg\}}
{\Gamma\left(1-\beta_0\left(g_0^{\prime}
(\omega)+g_1^{\prime}(\omega)\frac{\alpha_s^{\MSbar}(\lambda^2
m_b^2)}{\pi}\right) \right)} \times\nonumber \\ \nonumber &&
\hspace*{-52pt}  \left[ 1+\frac12g_0^{\prime\prime}(\omega)\beta_0^2
\frac{\alpha_s^{\MSbar}(\lambda^2 m_b^2)}{\pi}\Big(
\Psi^2\left(1-\beta_0g_0^{\prime}(\omega)\right)
-\Psi^{\prime}\left(1-\beta_0g_0^{\prime}(\omega)\right) \Big)
\right] \\ \nonumber &+&\frac{C_F\alpha_s(m_b^2)}{\pi}\,\int_0^{r_0}
dr k_1^{\reg}(\lambda,r,x),
\end{eqnarray}
where
\[
\omega\equiv\frac{\alpha_s^{\MSbar}(\lambda^2 m_b^2)}{\pi}\beta_0
\ln\frac{1}{r_0},
\]
and where $\Delta R^\infty(\lambda,x)$ is given explicitly in Eqs.
(\ref{Delta_R_infty}) and (\ref{Delta_K_1_infty}) below. The regular
term at ${\cal O}(\alpha_s)$ is included here directly in momentum
space.
\begin{figure}[t]
\begin{center}
\epsfig{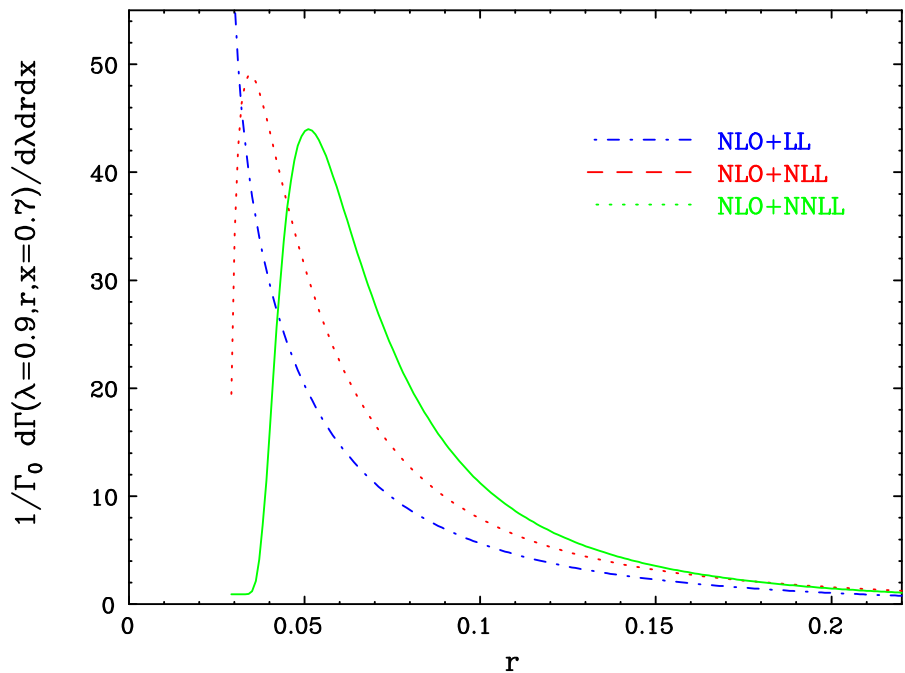}
\caption{\label{fig:matched_NNLL_spectrum} The differential spectrum
based on the fixed--logarithmic--accuracy formula
of~\eq{NNLL_x_space}, which is matched to~NLO. The LL, NLL and NNLL
accuracy results are plotted as dotdashed, dashed and full line,
respectively. The three curves end at $r\simeq 0.028$, where the
resummed results become complex owing to the Landau singularity at
$\omega=1/2$.}
\end{center}
\end{figure}

Fig.~\ref{fig:matched_NNLL_spectrum} shows the differential spectrum
as a function of $r$, for the chosen values of $\lambda=0.9$ and
$x=0.7$, computed as a derivative of the resummed spectrum in
\eq{NNLL_x_space}. The large differences between curves
corresponding to increasing logarithmic accuracy demonstrate the
problematic nature of this expansion, in which infrared renormalons
(in particular the one at $u=\frac12$) are unregulated. This problem
is solved in what follows by computing the moment--space Sudakov
exponent using Borel summation.

\subsubsection*{Calculation of the Sudakov factor by Borel summation}

Direct evaluation of the Borel integral in the Sudakov exponent
of~\eq{Gamma_n_large_n} using the Principal--Value prescription
makes optimal use of the known all--order structure of the exponent.
In the context of DGE for inclusive spectra, this regularization has
been proven effective in several respects~\cite{RD}:
\begin{itemize}
\item{} No Landau singularities are present.
\item{} It provides a systematic definition of the perturbative sum.
Using this definition, cancellation of ambiguities can be realized
--- this has been explicitly used for the leading renormalon
ambiguity associated with the pole mass.
\item{} The Sudakov factor is a real--valued
function of the moment variable.
\item{} The resummed spectra has modified
support properties. With the appropriate $B_{\cal S}(u)$ (see below), the support is
 close to that of the physical non-perturbative distribution.
\end{itemize}

The calculation of the Borel sum in~\eq{Gamma_n_large_n} in a given
regularization for the renormalons, requires, in principle, the knowledge
of $B_{\cal S}(u)$ and $B_{\cal J}(u)$ for any positive $u$. In
contrast to the large--$\beta_0$ limit (\ref{B_DJ_large_beta0}),
analytic expressions for these functions in QCD are not known. The
fixed--logarithmic--accuracy approach described above uses only the
{\em expansion} of these functions, as well as factors $\Gamma(-2u)$
and $\Gamma(-u)$ that multiply them, around the origin ($u=0$). As
discussed in detail in Sec. 2.3 in Ref.~\cite{RD}, one may use
additional information on the Borel functions $B_{\cal S}(u)$ and
$B_{\cal J}(u)$ to constrain further the Sudakov factor. Most
importantly, $B_{\cal S}(u)$ at $u=\frac12$  was determined in
Ref.~\cite{RD} with a good accuracy using the
cancellation~\cite{BDK} of the leading infrared renormalon ambiguity
in the Sudakov factor with that of the pole mass.

To understand the origin of renormalons in the Sudakov factor note
that the integration over $r$ near the $r\longrightarrow 0$ limit
necessarily involves some contribution from the infrared region.
>From \eq{Gamma_n_large_n_FLA} it is obvious that for small $r$ the
coupling is evaluated at small momentum scales, where it is out of
perturbative control. In the Borel formulation, when computing the
moments in \eq{r_mom_def} where the integrand is given by
\eq{sl_logs_r_lambda}, the integration gives rise to the factors
$\Gamma(-2u)$ and $\Gamma(-u)$ in the soft and jet parts of the
exponent, respectively. These factors contain infrared renormalons
poles at integer and half integer values of $u$. This way infrared
sensitivity translates into power--like ambiguity of the Borel
sum~\cite{BDK,RD,CG,DGE_thrust,Gardi:2002bg,Gardi:2001di},
corresponding to powers of $n\Lambda/(\lambda m_b)$ and
$n\Lambda^2/(\lambda m_b)^2$ in
the soft and jet factors, respectively; the latter are obviously
less important than the former. As shown in Refs.~\cite{BDK,RD}, the
leading power ambiguity on the soft scale proportional to
$(n-1)\Lambda/(\lambda m_b)$, cancels exactly with kinematic power
corrections. In the present context such corrections arise when
expressing $r$ in terms of the hadronic variables; see
Sec.~\ref{sec:Hadronic_var}.

Subleading power ambiguities on the soft scale, corresponding to
powers of $(n\Lambda/(\lambda m_b))^j$ with\footnote{The absence of ambiguity of
${\cal O}((n\Lambda/(\lambda m_b))^2)$ is discussed below.}
$j\geq 3$, are related to the
momentum distribution of the $b$ quark in the meson. In principle,
these ambiguities are removed only upon including a non-perturbative
{\em function}, namely {\em an infinite set} of power corrections,
that makes for the difference between the quark distribution in an
on-shell heavy quark and that in a meson. The closer to the endpoint
one cuts the distribution, the larger is the weight of high moments,
and with it the significance of these power corrections. Our analysis
of the ${\bar B}\longrightarrow X_s \gamma$ decay~\cite{RD} has
shown that if $B_{\cal S}(u)$ is sufficiently constrained (see below)
and kinematic power corrections are included,
one can determine the photon--energy spectrum to a good
accuracy neglecting any additional non-perturbative power corrections.

In principle, the theoretical uncertainty associated with the
unknown form of the anomalous dimension $B_{\cal S}(u)$ away from the origin, and the unknown
non-perturbative power terms that characterize the quark
distribution in the {\em meson} are two completely distinct issues.
In reality, however, both result in similar,
parametrically--enhanced power effects, and are therefore impossible
to distinguish phenomenologically.

The approach we take in this paper follows closely what was done in
Ref.~\cite{RD}: starting with the large--$\beta_0$ result for
$B_{\cal S}(u)$ and $B_{\cal J}(u)$, we include ${\cal O}(u)$ and
${\cal O}(u^2)$ terms to obtain the exact exponent to NNLO. For
$B_{\cal S}(u)$ we further include ${\cal }O(u^3)$ and higher--order
terms so as to match the computed residue of  at $u=\frac12$,
getting\footnote{This specific ansatz was called `model $(c)$' in Ref.~\cite{RD}.}
\begin{eqnarray}
\label{B_DJ} B_{\cal S}(u)&=& {\rm e}^{\frac53
u}(1-u) \exp\left\{c_2 u+\frac12 \left[
c_3-c_2^2+\frac{C_A}{\beta_0}\left(\frac{5}{18}\pi^2+\frac{7}{9}
-\frac{9}{2}\zeta_3\right)\right]u^2\right\}\times W(u),\nonumber\\
B_{\cal J}(u)&=&\frac{1}{2}\,{\rm e}^{\frac53
u} \left(\frac{1}{1-u}+\frac{1}{1-u/2}\right) \frac{\sin\pi u}{\pi
u}\,\times  \\ \nonumber&& \hspace*{-39pt}\exp{\bigg\{c_2 u+
\left[c_3-c_2^2+\frac{C_A}{\beta_0}\left(\frac{29}{72}\pi^2
-\frac{43}{72}-5\zeta_3\right)
+\frac{C_F}{\beta_0}\left(-\frac{\pi^2}{4}+\frac{3}{16}
+3\zeta_3\right)\right]\frac{u^2}{2!}+{\cal O}(u^3)\bigg\}},
\end{eqnarray}
where $c_{2,3}$ are given in Eq. (2.22) in Ref.~\cite{RD} and where
\begin{equation}
W(u)\equiv {\rm e}^{t_1 u +\frac12 t_2 u^2}\, \left(1-t_1 u+\frac12
(t_1^2-t_2)u^2\right)= 1+{\cal O}(u^3). \label{W}
\end{equation}
Here $t_{1,2}$ are fixed requiring:
\begin{eqnarray}
\label{C_def}
 B_{\cal S}(u=1/2) \,=\,0.91423\,\pm3\%\,({\rm
computed});\quad B_{\cal S}(u=3/2) \,=\,- 0.23366\,  \times\, C,
\end{eqnarray}
where we used the computed normalization of the leading renormalon
residue of the pole mass at $u=\frac 12$ (see eq. (2.30), (2.36) and
(4.8) in \cite{RD} with $N_f=4$) and an arbitrary normalization constant $C$ at
$u=\frac 32$.

We stress that the ansatz of \eq{B_DJ} {\em assumes} that
$B_{\cal S}(u)$ vanishes at $u=1$ and has no other zeros at positive $u$,
as in the large--$\beta_0$ limit~(\ref{B_DJ_large_beta0}).
The constraint $B_{\cal S}(u=1)=0$  was found to be quite important~\cite{RD}.
Here we explicitly assume that it holds in the full theory and
do not investigate the possibility of it being violated. The vanishing of
$B_{\cal S}(u=1)$ implies in particular that the perturbative Sudakov factor of
~\eq{Gamma_n_large_n} is free of ${\cal O}((n\Lambda/(\lambda m_b))^2)$
ambiguities\footnote{A well-known analogous situation is the absence of the
{\em leading} renormalon ambiguity~\cite{KS_DY} in Drell--Yan production~\cite{BB_DY}.
Also there parametrically--enhanced power corrections appear, which are related to renormalon
ambiguities in the Sudakov exponent~\cite{Gardi:2001di}. However, based on the analytic structure
of the soft Sudakov anomlaous dimension computed in the large--$\beta_0$ limit, the first power
correction is assumed not to appear.}.
Thus, after the cancellation of the $u=\frac 12$ ambiguity, the leading ambiguity
is  ${\cal O}((n\Lambda/(\lambda m_b))^3)$, as already mentioned. To the extent that
renormalons do indeed give good indication of which non-perturbative effects are
important, the absence of the $u=1$ renormalon is well
supported by the successful comparison of the prediction for the
first two cut moments in ${\bar B}\longrightarrow X_s \gamma$ with experimental data.

Note also that \eq{B_DJ} tends to zero at asymptotically large $u$.
In this respect it differs from the
large--$\beta_0$ limit. It is important to stress though that convergence
of the Borel integrals at large $u$ is guaranteed independently of
this assumption, owing to the factors $\Gamma(-2u)$ and
$\Gamma(-u)$ in \eq{Gamma_n_large_n}.

\begin{figure}
\begin{center}
\epsfig{file=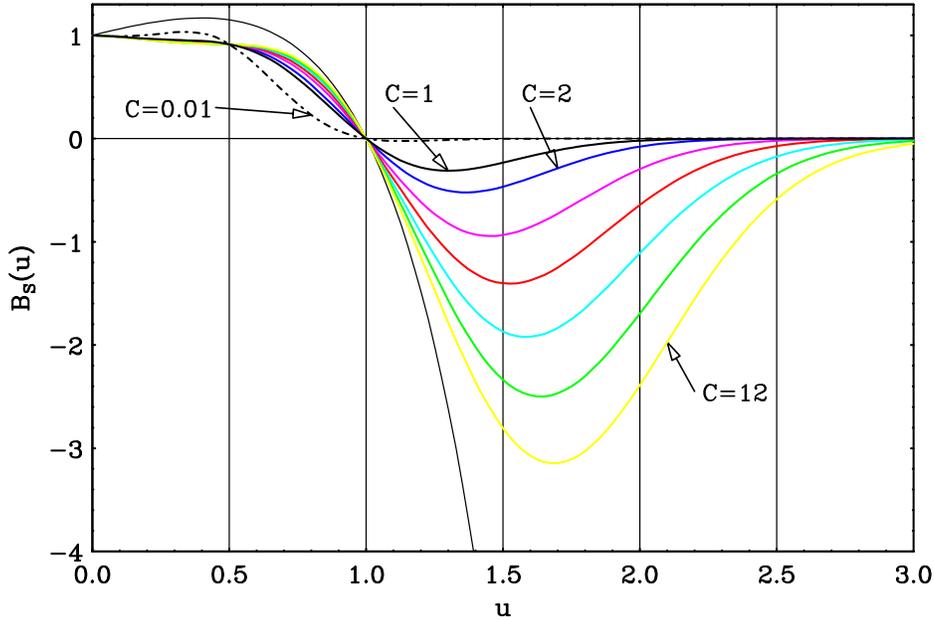,angle=90,width=12.4cm}
\caption{The function $B_{\cal S}(u)$ with vertical lines indicating
the positions of possible renormalons owing to the poles of $\Gamma(-2u)$ in
\eq{Gamma_n_large_n}. The thin line shows the
large--$\beta_0$ limit of \eq{B_DJ_large_beta0} while the thick lines show
$B_{\cal S}(u)$ according to \eq{B_DJ} with $N_f=4$ and
with different assignments of the parameter $C$ that controls the
residue of the Borel singularity in the
Sudakov exponent at $u=\frac32$: $C=0.01$ is shown as a dotdashed line while the default value
$C=1$ and $C=2,4,6,8,10,12$ are shown as full lines.
\label{fig:C_dependence}}
\end{center}
\end{figure}

In \eq{B_DJ} we parametrize
$B_{\cal S}(u)$ at intermediate $u$ values by a single parameter
$C$. When $B_{\cal S}(u)$ is inserted into \eq{Gamma_n_large_n} this
number determines the renormalon residue at $u=\frac32$. The effect of
$C$ on the function $B_{\cal S}(u)$ is shown if
Fig.~\ref{fig:C_dependence}.
While the actual residue in the full theory is not known, it is expected to be
smaller than in the large--$\beta_0$ limit, where $B_{\cal
S}(u=3/2)\simeq -6.09$. Such a large residue can be obtained in the
ansatz of \eq{B_DJ} with $N_f=4$ by setting $C\simeq 26$
($W(3/2)\simeq 20$). In model $(c)$ in Sec. 2.3 of \cite{RD}, we had
$B_{\cal S}(u=3/2)\simeq - 0.23366$ corresponding to $C=1$
($W(3/2)\simeq 0.776$). Here we keep $C$ as a free parameter by
which we gauge the theoretical uncertainty in the Sudakov factor.
The possible range of values for $C$ is discussed further in
Sec.~\ref{sec:Numerical_results_and_uncertainty}, where we also study the
impact of this parameter on the spectrum and on the measurable partial width,
see Fig.~\ref{fig:renormalon_at_3halves_influence} and
Fig.~\ref{fig:Mx_C_dep}, respectively.

\subsection{Matching and predictions for the logarithmic terms at NNLO\label{sec:matching}}

Using the definition of the moments in \eq{r_mom_def} with the NLO
expansion of \eq{V_and_D_r} we have
\begin{eqnarray}
\label{r_mom_def_NLO} \frac{d\Gamma_{n} (\lambda,x)}{d\lambda dx}
 &\equiv & \frac{1}{\Gamma_0}\int_{0}^{r_m} dr
 \left(1-r\right)^{n-1}
\frac{d\Gamma(\lambda,r,x)}{d\lambda dr dx}\\ \nonumber
&=&\int_{0}^{r_m} dr \left(1-r\right)^{n-1} \bigg[V(\lambda,x)
\delta(r)+ R(\lambda,r,x)\bigg]
\\ \nonumber
&=& w_0(\lambda,x) +\frac{C_F\alpha_s(m_b)}{\pi}\bigg[w_1(\lambda,x)
+ K_1^n(\lambda,x)\bigg]+{\cal O}(\alpha_s^2),
\end{eqnarray}
where $r_m\equiv (1-x)/\lambda$ denotes the maximal value of $r$ and
\begin{eqnarray}
\label{C1_moments} K_1^n(\lambda,x)&\equiv &\int_{0}^{r_m} dr
\left(1-r\right)^{n-1} k_1(\lambda,r,x)\\ \nonumber &=&
K_1^{n\,\sing} (\lambda,x) + K_1^{n\,\reg} (\lambda,x).
\end{eqnarray}
where we split the integration into singular and regular terms for
$r\longrightarrow 0$:
\begin{eqnarray}
\label{sing_part} K_1^{n\,\sing} (\lambda,x)&\equiv & \int_{0}^{r_m}
dr \left(1-r\right)^{n-1} k_1^{\sing}(\lambda,r,x)
\end{eqnarray}
and
\begin{eqnarray}
\label{C1_reg_def} K_1^{n\,\reg} (\lambda,x)&\equiv & \int_{0}^{r_m}
dr \left(1-r\right)^{n-1} k_1^{\reg}(\lambda,r,x) \\ \nonumber &= &
\int_{0}^{r_m} dr \left(1-r\right)^{n-1} \bigg[k_1(\lambda,r,x) -
k_1^{\sing}(\lambda,r,x)\bigg]
\end{eqnarray}
where $k_1^{\sing}(\lambda,r,x)$ and $k_1(\lambda,r,x)$ is given by
\eq{c_1_sing} and \eq{c_1}, respectively. In Appendix~\ref{sec:matching_coeff}
we evaluate
$K_1^{n\,\sing} (\lambda,x)$ and $K_1^{n\,\reg} (\lambda,x)$
explicitly. The result for the singular part reads
\begin{eqnarray}
\label{C_1_sing_very_explicit}
K_1^{n\,\sing} (\lambda,x)
&=& -w_0(\lambda,x)\bigg\{
\frac12\left[ \frac{\pi^2}{6}-\Psi_1(n)+\left(\Psi(n)+\gamma_E\right)^2 \right]-\frac74(\Psi(n)+\gamma_E)\\ \nonumber &&
-\sum_{k=0}^{\infty} \frac{(1-r_m)^{n+k}}{n+k}\bigg[\ln(r_m) +\Psi(n+k)-\Psi(n)\bigg]
-\frac74\sum_{k=0}^{\infty}\frac{(1-r_m)^{n+k}}{n+k}
\bigg\},
\end{eqnarray}
while the regular part is given in \eq{C_1_reg}.

In order to determine the $n$--independent coefficient that
multiplies the Sudakov factor in \eq{Gamma_n_large_n} at ${\cal
O}(\alpha_s)$,
\begin{equation}
\label{Delta_R_infty} \Delta R^{\infty}(\lambda,x)= \Delta
K_1^{\infty}(\lambda,x)\frac{C_F\alpha_s(m_b)}{\pi}\,+\,{\cal
O}(\alpha_s^2),
\end{equation}
we compare the expansion of \eq{Gamma_n_large_n} with the NLO result
of~\eq{r_mom_def_NLO}. We find that the ${\cal O}(\alpha_s)$ terms
that are not generated by the expansion of the exponent are:
\begin{equation}
\label{Delta_K_1} \Delta K_1^n(\lambda,x) =K_1^n(\lambda,x) - \bigg[
- {\displaystyle \frac {L ^{2}}{2}}  + \bigg( - \gamma_E  +
{\displaystyle \frac {7}{ 4}} \bigg)\,L\bigg]  w_0,
\end{equation}
where $L=\ln n$, so
\begin{equation}
\label{Delta_K_1_infty} \Delta K_1^{\infty}(\lambda,x) \equiv
\lim_{n\to \infty}\Delta K_1^n(\lambda,x)=-w_0(\lambda,x)
\left[\frac{\pi^2}{12}+\frac{\gamma_E^2}{2}-\frac74 \gamma_E\right].
\end{equation}

Using this coefficient \eq{Gamma_n_large_n} yields the following
perturbative expansion:
\begin{eqnarray}
\label{expansion_of_exponent_mom} \left.\frac{d\Gamma_{n}
(\lambda,x)}{d\lambda dx}\right\vert_{{\rm large}\,\, n}
&=&w_0(\lambda,x)+\bigg\{\bigg[ - {\displaystyle \frac {L ^{2}}{2}}
+ \bigg( - \gamma_E  + {\displaystyle \frac {7}{ 4}} \bigg)\,L +
-\frac{\gamma_E^2}{2}-\frac{\pi^2}{12}+\frac{7}{4}\gamma_E\bigg]
\,w_0(\lambda,x) \nonumber \\
&& \hspace*{40pt} + w_1(\lambda,x)\bigg\}\,\frac{C_F\alpha_s
(m_b)}{\pi}\nonumber \\\nonumber & +& \bigg\{\bigg[{\displaystyle
\frac {C_F^{2}\,L^{4}}{8}}  + \bigg(\bigg( - {\displaystyle \frac
{7}{8}}  + {\displaystyle \frac {\gamma_E }{2} } \bigg)\,C_F^{2} +
\bigg( - {\displaystyle \frac {11\,C_A }{24}}  + {\displaystyle
\frac {N_f}{12}} \bigg)\,C_F\bigg) \,L^{3} \nonumber \\  & +&
\bigg(\bigg({\displaystyle \frac {11\,C_A}{12}}  - {\displaystyle
\frac {N_f}{6}} \bigg)\,C_F\,\ln \lambda
 + \bigg( - {\displaystyle \frac {21}{8}} \,\gamma_E  +
{\displaystyle \frac {3}{4}} \,\gamma_E^{2} + {\displaystyle
\frac {49}{32}} +\frac{\pi^2}{24}\bigg)\,C_F^{2}  \nonumber \\
 &+& \bigg(\bigg({\displaystyle \frac {
\gamma_E }{4}}  - {\displaystyle \frac {13}{144}} \bigg)\,N_f +
\bigg({\displaystyle \frac {95}{288}}  - {\displaystyle \frac {11\,
\gamma_E }{8}}  + {\displaystyle \frac {\pi ^{2}}{24}}
\bigg)\,C_A\bigg)\,C_F\bigg)L^{2}\\ \nonumber &+& \bigg(
\bigg(\bigg({\displaystyle \frac {7}{12}}  - {\displaystyle \frac
{\gamma_E }{3}} \bigg)\,N_f + \bigg( - {\displaystyle \frac
{77}{24}}  + {\displaystyle \frac {11\,\gamma_E }{6}}
\bigg)\,C_A\bigg)\, C_F\,\ln \lambda  \\\nonumber & +&
\bigg({\displaystyle \frac {3}{32}}  + {\displaystyle \frac {3}{2}}
\,\zeta_3 - {\displaystyle \frac { 13\pi ^{2}}{48}}
+\frac{49\gamma_E}{16}+\frac{\pi^2\gamma_E}{12}
-\frac{21\gamma_E^2}{8}+ \frac{\gamma_E^3}{2}
 \bigg)\,C_F^{2}  \\\nonumber &
 +& \bigg(
\bigg( - {\displaystyle \frac {85}{144}}  + {\displaystyle \frac
{5}{ 72}} \,\pi ^{2} - {\displaystyle \frac {13}{72}} \,\gamma_E  +
{\displaystyle \frac {1}{4}} \,\gamma_E^{2}\bigg)\,N_f \\ \nonumber
& +& \bigg({\displaystyle \frac {905}{288}}  + {\displaystyle \frac
{\pi ^{2}\,\gamma_E }{12}}  + {\displaystyle \frac {95\, \gamma_E
}{144}}  - {\displaystyle \frac {67\,\pi ^{2}}{144}}  -
{\displaystyle \frac {11\,\gamma_E^{2}}{8}}  - {\displaystyle \frac
{1}{4}} \,\zeta_3\bigg)\,C_A\bigg)C_F\bigg)L\bigg]w_0(\lambda,x)
 \\ \nonumber &
+& \bigg[ - {\displaystyle \frac {L^{2}\,C_F}{2}}  + \bigg(
 - \gamma_E  + {\displaystyle \frac {7}{4}} \bigg)\,C_F\,L\bigg]\,
C_F w_1(\lambda,x)\bigg\} \left(\frac{\alpha_s(m_b)}{\pi}
\right)^{2} +\cdots.
\end{eqnarray}
All the coefficients of the logs determined here
are exact. It is important to emphasize that virtual corrections at
one--loop order, $w_1(\lambda,x)$, contribute to log terms ($C_F^2L$ and
$C_F^2 L^2$) at ${\cal O}(\alpha_s^2)$. Similarly, two-loop virtual corrections,
$w_2(\lambda,x)$, which are yet unknown, contribute to
subleading logarithms at ${\cal O}(\alpha_s^3)$ and beyond.

Complete matching at ${\cal O}(\alpha_s)$ can be achieved by
including ${\cal O}(1/n)$ terms additively, using ``R matching'':
\begin{eqnarray}
\label{Gamma_n_R_matched} && \left.\frac{d\Gamma_{n}
(\lambda,x)}{d\lambda dx}\right\vert_{\rm matched} =
\left[V(\lambda,x)+ \Delta
K_1^{\infty}(\lambda,x)\frac{C_F\alpha_s(m_b)}{\pi}+ {\cal
O}(\alpha_s^2)\right] \, \times \\ \nonumber && \exp\bigg\{
\frac{C_F}{\beta_0}\int_0^{\infty}\frac{du}{u} \, T(u)
\left(\frac{\Lambda^2}{m_b^2\lambda^2}\right)^u \left[B_{\cal
S}(u)\Gamma(-2u)\left({n}^{2u}-1\right)  -B_{\cal
J}(u)\Gamma(-u)\left({n}^u-1\right) \right] \bigg\} \\ \nonumber &&+
\left[\Delta K_1^n(\lambda,x)-\Delta K_1^{\infty}(\lambda,x)\right]
\frac{C_F\alpha_s(m_b)}{\pi} + {\cal O}(\alpha_s^2/n)
\end{eqnarray}
or, alternatively in front of the Sudakov factor, using ``log-R
matching''
\begin{eqnarray}
\label{Gamma_n_lnR_matched} && \left.\frac{d\Gamma_{n}
(\lambda,x)}{d\lambda dx}\right\vert_{\rm matched} =
\left[V(\lambda,x)\,+ \Delta
K_1^n(\lambda,x)\frac{C_F\alpha_s(m_b)}{\pi} + {\cal O}(\alpha_s^2)
\right] \times \\ \nonumber &&\exp\bigg\{
\frac{C_F}{\beta_0}\int_0^{\infty}\frac{du}{u} \, T(u)
\left(\frac{\Lambda^2}{m_b^2\lambda^2}\right)^u
 \, \left[B_{\cal
S}(u)\Gamma(-2u)\left({n}^{2u}-1\right)  -B_{\cal
J}(u)\Gamma(-u)\left({n}^u-1\right) \right] \bigg\}
\end{eqnarray}
where the terms missing at  ${\cal O}(\alpha_s^2)$ include
$n$--independent as well as ${\cal O}(1/n)$ terms.

Converting the expansion in \eq{expansion_of_exponent_mom} to
momentum space we obtain for the log-enhanced contribution to the
triple differential width (see \eq{V_and_D_r}) at this order:
\begin{eqnarray}
\label{D_r_NNLO}
\left.R\left(\lambda,r,x\right)\right\vert_{r\longrightarrow 0}&=&
\left( - C_F\,\left(\frac{\ln r}{r}\right)_{*} - {\displaystyle
\frac {7 }{4}}
\left(\frac{1}{r}\right)_{*}\,C_F\right)\,w_0\left(\lambda,x\right)\,
\frac{\alpha_s\left(m_b\right)}{\pi}\\\nonumber
&+&\Bigg\{\bigg\{{\displaystyle \frac
{C_F^{2}}{2}}\,\left(\frac{\ln^3 r}{r}\right)_{*}
  + \bigg[
{\displaystyle \frac {21\,C_F^{2}}{8}}  + \left( {\displaystyle
\frac {11\,C_A}{8}}  - {\displaystyle \frac {N_f}{4}}
\right)\,C_F\bigg]\,\left(\frac{\ln^2 r}{r}\right)_{*}
 \\ \nonumber
&+& \bigg[\left({\displaystyle \frac {11\,C_A}{6}}  - {\displaystyle
\frac {N_f}{3}} \right)\,C_F\,\mathrm{ln }\left(\lambda \right) +
\left({\displaystyle \frac {49}{16}}  - {\displaystyle \frac {\pi
^{2}}{6}} \right)\,C_F^{2} \\ \nonumber &+& \left( - {\displaystyle
\frac {13\,N_f}{72}}  + \left({\displaystyle \frac {\pi ^{2}}{ 12}}
+ {\displaystyle \frac {95}{144}} \right)\,C_A\right)\,C_F\bigg]\,
\left(\frac{\ln r}{r}\right)_{*} \\ \nonumber &+&
\bigg[\left({\displaystyle \frac {77\,C_A}{24}}  - {\displaystyle
\frac {7\,N_f}{12}} \right)\,C_F\, \mathrm{ln}\left(\lambda \right)
+ \left( - {\displaystyle \frac {\pi ^{2}}{6}}
 - {\displaystyle \frac {3}{32}}  - {\displaystyle \frac {1}{2}}
\,\zeta_3\right)\,C_F^{2} \\ \nonumber &+&
\left(\left({\displaystyle \frac {85}{144}}  - {\displaystyle \frac
{\pi ^{2}}{36}} \right)\,N_f + \left({\displaystyle \frac {17 \,\pi
^{2}}{72}}  + {\displaystyle \frac {1}{4}} \,\zeta_3 -
{\displaystyle \frac {905}{288}} \right)\,C_A\right)\,C_F \bigg]
\left(\frac{1}{r}\right)_{*}\bigg\}
w_0\left(\lambda,x\right)\mbox{} \\
&+& \nonumber
 \left[ - {\displaystyle \frac {7}{4}} \,
C_F^{2}\left(\frac{1}{r}\right)_{*} - \,C_F^{2}\left(\frac{\ln
r}{r}\right)_{*} \right]\,w_1\left(\lambda,x\right)\Bigg\}
\left(\frac{\alpha_s\left(m_b\right)}{\pi}\right)^{2} \!+\! {\cal
O}\left(\frac{\alpha_s\left(m_b\right)}{\pi}\right)^{3}.
\end{eqnarray}
This expression is useful for checking NNLO calculations. An
immediate application is the calculation of the log--enhanced part
of partially integrated and single differential distributions. For
example, the single differential distribution with respect to
$\rho$:
\begin{eqnarray}
\label{single_diff_rho_prediction}
\frac{1}{\Gamma_0}\frac{d\Gamma(\rho)}{d\rho}&=&\delta(\rho)\bigg[1
+\frac{C_F\alpha_s(m_b)}{\pi}\left(-\frac{13}{144}
-\frac{\pi^2}{2}\right)+{\cal O}(\alpha_s)^2\bigg]+
\bigg\{-\left(\frac{\ln(\rho)}{\rho}\right)_{*}-\frac{13}{6}\left(\frac{1}{\rho}\right)_{*}\nonumber
\\ \nonumber &-& \rho^2 (3-2 \rho) \ln^2(\rho) -\frac{1}{6}
\left(2-23 \rho-9 \rho^2+8 \rho^3\right) \ln(\rho)+
\frac{79}{18}+\frac{407}{72} \rho-\frac{367}{24} \rho^2 \\ \nonumber
&+&\frac{59}{6} \rho^3-\frac{25}{9} \rho^4+\frac{11}{24} \rho^5
-\frac{7}{72} \rho^6\bigg\} \frac{C_F\alpha_s(m_b)}{\pi} \\
\nonumber &+&\bigg\{\frac{1}{2} C_F
\left(\frac{\ln^3(\rho)}{\rho}\right)_{*} +\bigg[\frac{13}{4}
C_F+\frac{11}{8} C_A-\frac{1}{4} N_f\bigg]
\left(\frac{\ln^2(\rho)}{\rho}\right)_{*}
\\ \nonumber
&+&\bigg[\bigg(\frac{355}{72}+\frac{1}{3} \pi^2\bigg)
C_F+\bigg(\frac{1}{12} \pi^2+\frac{25}{24}\bigg) C_A -\frac{1}{4}
N_f\bigg] \left(\frac{\ln(\rho)}{\rho}\right)_{*}\\ \nonumber
&+&\bigg[\bigg(-\frac{3571}{1728}+\frac{61}{72} \pi^2+\frac{3}{2}
\zeta_3\bigg) C_F+\bigg(-\frac{1253}{288}
+\frac{1}{4} \zeta_3+\frac{13}{48} \pi^2\bigg) C_A \\
&+&\bigg(\frac{113}{144}-\frac{1}{36} \pi^2\bigg)
N_f\bigg]\left(\frac{1}{\rho}\right)_{*}+{\cal O}(\rho^0)
\bigg\}\,C_F\,\left(\frac{\alpha_s(m_b)}{\pi}\right)^2+\cdots.
\end{eqnarray}
The ${\cal O}(\alpha_s^2)$ coefficient that is leading in $N_f$ was
recently computed~\cite{HLL} and it is consistent with
\eq{single_diff_rho_prediction}. The other coefficients are new.

\subsection{Exponentiation beyond logarithms\label{sec:Exponentiation_beyond_logs}}

Let us note that the moment--space function obtained by
integrating the $r\longrightarrow 0$ singular terms of the form
$r^{-1-2u}$ or $r^{-1-u}$ in \eq{sl_logs_r_lambda}, which are regularized as
$()_{*}$ distributions, is not purely logarithmic. Performing the
calculation leading to~\eq{Gamma_n_large_n} above but
{\em avoiding any further large--$n$ approximation}
one obtains the following Sudakov factor:
\begin{eqnarray}
\label{alt_Sud} {\rm Sud}(m_b\lambda ,n) &=&\exp\bigg\{
\frac{C_F}{\beta_0}\int_0^{\infty}\frac{du}{u} \, T(u)
\left(\frac{\Lambda^2}{m_b^2\lambda^2}\right)^u
  \\ \nonumber &&
\bigg[B_{\cal
S}(u)\Gamma(-2u)\left(\frac{\Gamma(n)}{\Gamma(n-2u)}-\frac{1}{\Gamma(1-2u)}\right)
\\ \nonumber &&\hspace*{100pt}
 -B_{\cal
J}(u)\Gamma(-u)\left(\frac{\Gamma(n)}{\Gamma(n-u)}-\frac{1}{\Gamma(1-u)}\right)
\bigg] \bigg\}.
\end{eqnarray}
Note that to any order in $u$:
$
{\Gamma(n)}/{\Gamma(n-u)}\simeq n^u
$
up to constant terms. Thus, upon
applying this approximation one returns to~\eq{Gamma_n_large_n}. There,  {\em only}
powers of $\ln n$ appear at any given order in $u$, while constants and ${\cal O}(1/n)$ terms are
entirely excluded from the exponent and appear only in the matching coefficient.
In~\eq{alt_Sud} such terms appear also in the exponent.

It is straightforward to match the Sudakov factor of \eq{alt_Sud} to NLO.
The result is more elegant than when using a ``purely logarithmic''
Sudakov factor (\eq{Gamma_n_lnR_matched} or
\eq{Gamma_n_R_matched}) because the constants for
$n\longrightarrow \infty$, which where included there in the process
of matching, appear in \eq{alt_Sud} as part the exponent. This is
true at higher orders too: when using \eq{alt_Sud}, $\Delta
R^{\infty}=0$. Consequently, it is just the purely virtual terms
$V(\lambda,x)$ (terms proportional to $\delta(r)$, after the
$()_{+}$ distributions have been defined) that must be multiplied by
the Sudakov factor in order to generate correctly the log--enhanced
terms at higher orders. Using \eq{alt_Sud} with ``R matching'' we
get
\begin{eqnarray}
\label{Gamma_n_R_matched_w_constant} \left.\frac{d\Gamma_{n}
(\lambda,x)}{d\lambda dx}\right\vert_{\rm matched} &=&
V(\lambda,x)\times {\rm Sud}(m_b\lambda ,n) +\widetilde{\Delta
K_1}^n(\lambda,x)\frac{C_F\alpha_s(m_b)}{\pi} + {\cal
O}(\alpha_s^2),
\end{eqnarray}
while with ``log-R matching'' we get
\begin{eqnarray}
\label{Gamma_n_lnR_matched_w_constant} \hspace*{-20pt}
\left.\frac{d\Gamma_{n} (\lambda,x)}{d\lambda dx}\right\vert_{\rm
matched} = \left[V(\lambda,x)\,+ \widetilde{\Delta
K_1}^n(\lambda,x)\frac{C_F\alpha_s(m_b)}{\pi} + {\cal O}(\alpha_s^2)
\right] \times {\rm Sud}(m_b\lambda ,n).
\end{eqnarray}
In both cases:
\begin{eqnarray}
\label{tilde_C_1_explicit} \widetilde{\Delta K}_1^{n} (\lambda,x)
&=& w_0(\lambda,x)\bigg\{ \bigg[\ln(r_m) +\frac74\bigg]\times
\sum_{k=0}^{\infty} \frac{(1-r_m)^{n+k}}{n+k}\\ \nonumber &&
+\sum_{k=0}^{\infty}
\frac{(1-r_m)^{n+k}}{n+k}\bigg[\Psi(n+k)-\Psi(n)\bigg]\bigg\}
\\ \nonumber &&+
 \sum_{i=-4}^{2} f_i(\lambda,x) I_{n+i}(r_m)
 +\sum_{i=-3}^{2} \tilde{f}_i(\lambda,x)J_{n+i}(r_m).
\end{eqnarray}
where the last line corresponds to $K_1^{\reg}$ obtained in \eq{C_1_reg}.
The definitions of the coefficients $f_i(\lambda,x)$ and $\tilde{f}_i(\lambda,x)$ and
the integrals $I_{n+i}(r_m)$ and $J_{n+i}(r_m)$ are summarized in
Appendix~\ref{sec:matching_coeff}.

Recall that all the terms in $\widetilde{\Delta K}_1^{n}
(\lambda,x)$ are suppressed by at least one power of $1/n$. Note, on
the other hand, that for {\em fixed} $n$ this matching term contains
logarithms of $r_m$, which get large for $r_m\longrightarrow 0$. In
particular, the first moment ($n=1$) is:
\begin{eqnarray}
\label{tilde_C_1_explicit_n_eq_1} \widetilde{\Delta K}_1^{n=1}
(\lambda,x) &=&
w_0(\lambda,x)\bigg\{-\frac{1}{2}\ln^2(r_m)-\frac{7}{4}\ln
(r_m)\bigg\}\,+\, K_1^{n=1\,\,\reg} (\lambda,x),
\end{eqnarray}
where $K_1^{n=1\,\,\reg}$ is given in \eq{C1_reg_n_eq_1}.

It is interesting to study the renormalon structure of the Sudakov
factor in \eq{alt_Sud} and compare it to the simplified version of
\eq{Gamma_n_large_n}. Let us examine in some detail the soft
function (a similar structure appears on the jet side). As discussed
in Sec. 4.1 in Ref.~\cite{RD}, power corrections associated with
renormalon ambiguities in the soft function modify the Sudakov
factor of \eq{Gamma_n_large_n} or \eq{alt_Sud} as
follows\footnote{As discussed in Sec. 4 below, the $j=1$ ambiguity
cancels upon converting to hadronic variables and the $j=2$
renormalon is absent since $B_{\cal S}(1)=0$. Thus, dynamical
non-perturbative power corrections actually appears only for $j\geq
3$.}:
\begin{eqnarray}
\label{NP_power_sum} &&\left.{\rm Sud}(m_b\lambda
,n)\right\vert_{\PV} \,\longrightarrow \,\left. {\rm Sud}(m_b\lambda
,n)\right\vert_{\PV}\times \\ \nonumber
&&\hspace*{40pt}\exp\left\{\sum_{j=1}^{\infty} \left(\frac{\Lambda
}{m_b\lambda}\right)^j f_j^{\PV}\pi \frac{C_F}{\beta_0}
\frac{T(j/2)}{j/2}
 B_{\cal S}(j/2) R(n,j/2)
\right\},
\end{eqnarray}
where the $n$--dependent residue function in the two cases is
\begin{eqnarray}
R(n,j/2)=\left\{
\begin{array}{ll}
{\displaystyle
{\rm Res}\, \Gamma(-2u)\left(n^{2u}-1\right){\bigg \vert}_{u=j/2}}
&{\rm for\,\, Eq.}\,\, (\ref{Gamma_n_large_n})
\\
{\displaystyle
{\rm Res}\, \Gamma(-2u)\left(\frac{\Gamma(n)}{\Gamma(n-2u)}-\frac{1}{\Gamma(1-2u)}\right)
{\bigg \vert}_{u=j/2}}
&{\rm for\,\, Eq.}\,\,  (\ref{alt_Sud})
\end{array}
\right.
\end{eqnarray}
and
$f_j^{\PV}$ are dimensionless non-perturbative parameters of order $1$.
PV stands for the Principal Values prescription, which is used to define
the perturbative sum on the one hand and the non-perturbative parameters on the other.
In \eq{NP_power_sum} the l.h.s. is regularization prescription dependent, while the
r.h.s. is not.

It is important to note that subleading renormalons ${\cal
O}\left(\left({\Lambda}/({\lambda m_b})\right)^j\right)$
corresponding to increasing $j$ have an additional {\em numerical
suppression} owing to the structure of the Sudakov exponent, namely
the fact that
\begin{equation}
\label{factorial_suppression}
 \left.{\rm Res}\,
\Gamma(-2u)\right\vert_{u=j/2} = \frac12 \frac{(-1)^{j+1}}{j!}.
\end{equation}
Considering only powers of $n \Lambda/(\lambda m_b)$ the residue
structure of the two exponents is the same. In order to analyze the
differences between them at smaller $n$ let us now compare between
the residue function $R(n,j/2)$ in the two cases for the first few
renormalons at $u=j/2$:
\begin{eqnarray}
\label{soft_residues}
\begin{array}{lll}
  u        & R(n,j/2)\,\,{\rm for\,\, Eq.}\,\, (\ref{Gamma_n_large_n})&
   R(n,j/2)\,\, {\rm for\,\, Eq.}\,\,  (\ref{alt_Sud}) \\
  \frac12 \hspace*{20pt} & +\frac12 (n-1)& +\frac12 (n-1) \\
  1        & -\frac14 (n-1)(n+1) & -\frac14 (n-1) (n-2) \\
  \frac32  & +\frac{1}{12} (n-1) (n^2+n+1)& +\frac{1}{12}(n-1) (n-2) (n-3) \\
  2        & -\frac{1}{48} (n-1) (n+1) (n^2+1) & -\frac{1}{48}(n-1) (n-2) (n-3) (n-4) \\
  \frac52  & +\frac{1}{240} (n-1) (n^4+n^3+n^2+n+1) \hspace*{20pt} & +\frac{1}{240}
  (n-1) (n-2) (n-3) (n-4)(n-5).\\
\end{array}
\end{eqnarray}
Obviously, using \eq{alt_Sud} the parametric enhancement of power
corrections is not as dramatic as it looks based
on~\eq{Gamma_n_large_n}. In \eq{alt_Sud} integer moments are
entirely free of certain power--like ambiguities which do show up in
\eq{Gamma_n_large_n}: according to the residue structure of
\eq{alt_Sud}, for $n=N$, $N$ being a positive integer, only
powers $(\Lambda/(\lambda m_b))^j$ with $j < N$ appear.

A priori, one might expect that all power corrections would become
of order one at $n\sim \lambda m_b/\Lambda$. Assuming that
renormalon ambiguities do indeed give good indication on
non-perturbative effects, the factorial suppression of
\eq{factorial_suppression} as well as the $n$ dependence of the
residues of \eq{alt_Sud} imply that power terms are altogether less
important and, in contrast with the naive expectation, the power
expansion corresponding to higher renormalons in \eq{NP_power_sum}
does not break down.

\subsection{Resummed double--differential width with a lepton--energy cut\label{sec:Analytic_lepton_energy_cut}}

Since the dependence on the lepton energy fraction $x$ appears in
the resummation formula (\ref{Gamma_n_lnR_matched_w_constant}) only
though the phase--space limits and the matching coefficients but
{\em not through the Sudakov factor}, it is possible to integrate
over $x$ analytically to obtain the double differential width with a
cut $x>x_0$:
\begin{eqnarray}
\label{double_diff_r_lambda} &&\hspace*{-10pt} \frac{1}{\Gamma_0}
\frac{d\Gamma(\lambda,r,x>x_0)}{d\lambda dr}
=\int_{c-i\infty}^{c+i\infty} \frac{d n}{2\pi i} \,\frac{d\Gamma_{n}
(\lambda,x>x_0)}{d\lambda} \, \left(1-{r}\right)^{-n}
\end{eqnarray}
where, based on \eq{Gamma_n_lnR_matched_w_constant},
\begin{eqnarray}
\label{double_diff_matched} &&\left.\frac{d\Gamma_{n}
(\lambda,x>x_0)}{d\lambda}\right\vert_{\rm matched} \equiv
\int_{\max\{x_0,1-\lambda\}}^1dx \,\left.\frac{d\Gamma_{n}
(\lambda,x)}{d\lambda dx}\right\vert_{\rm matched}\\\nonumber
&&\hspace*{30pt}= \left[\overline{V}_r(\lambda,x>x_0)\,+
\overline{\Delta K_1}^n(\lambda,x>x_0)\frac{C_F\alpha_s(m_b)}{\pi} +
{\cal O}(\alpha_s^2) \right] \times \\ \nonumber &&\hspace*{50pt}
\exp\bigg\{ \frac{C_F}{\beta_0}\int_0^{\infty}\frac{du}{u} \, T(u)
\left(\frac{\Lambda^2}{m_b^2\lambda^2}\right)^u
 \, \bigg[B_{\cal
S}(u)\Gamma(-2u)\left(\frac{\Gamma(n)}{\Gamma(n-2u)}-\frac{1}{\Gamma(1-2u)}\right)
\\ \nonumber &&\hspace*{100pt}
 -B_{\cal
J}(u)\Gamma(-u)\left(\frac{\Gamma(n)}{\Gamma(n-u)}-\frac{1}{\Gamma(1-u)}\right)
\bigg] \bigg\}.
\end{eqnarray}
In the Appendix~\ref{sec:partially_integrated_matching} we explicitly compute the
matching coefficients
at ${\cal O}(\alpha_s)$, namely, $\overline{V}_r(\lambda,x>x_0)$ and
$\overline{\Delta K_1}^n(\lambda,x>x_0)$, by
integrating the virtual coefficients of \eq{V_and_D_r} and the
real--emission moments of \eq{tilde_C_1_explicit}, respectively.

\section{Partial width with experimentally relevant cuts in hadronic
variables\label{sec:Hadronic_var}}

\subsection{Conversion to hadronic variables and power corrections\label{sec:conversion}}

So far we examined the resummed distribution in kinematic variables
associated with the partonic jet initiated by the u-quark.
Measurements do not distinguish between the u-quark jet and the soft
partons originating in the light degrees of freedom in the meson.
Working still in the approximation where the $b$ quark is on shell,
one defines $v\equiv p_B/M_B$ (where $v^2=1$) so $p_b=m_b v$ and the
momentum of the light degrees of freedom is $\bar{\Lambda} v =
(M_B-m_b)v$. The hadronic variables (in the B rest frame) are then:
$E_H=E_j+\bar{\Lambda}$, where $\bar{\Lambda}=M_B-m_b$, and
$\vec{P}_H=\vec{p}_j$. The relations with the corresponding hadronic
lightcone variables are ({\em cf}. \eq{lightcone_variables}):
\begin{eqnarray}
\label{P^+} P^+&=&E_H-|\vec{P}_H|=\bar{\Lambda}+E_j-|\vec{p}_j| =
m_b \rho +\bar{\Lambda},\\ \nonumber
P^-&=&E_H+|\vec{P}_H|=\bar{\Lambda}+E_j+|\vec{p}_j| = m_b \lambda +
\bar{\Lambda}.
\end{eqnarray}
In the small hadronic--mass--squared ($P^+P^-$) region, $\rho$ is
small, and the $\bar{\Lambda}$ term in $P^+$ is absolutely
essential. On the other hand, the subregion where also $\lambda$ is
small ($P^-$ is of order $\Lambda$) is unimportant because it is
power suppressed by the Born--level weight (see~\eq{bar_w0}).
Nevertheless, as we carefully treated all other ${\cal
O}(\Lambda/m_b)$ effects, we do not neglect $\bar{\Lambda}$ compared
to $m_b \lambda$ when converting the decay width to hadronic
variables.

\begin{figure}[t]
\begin{center}
\epsfig{file=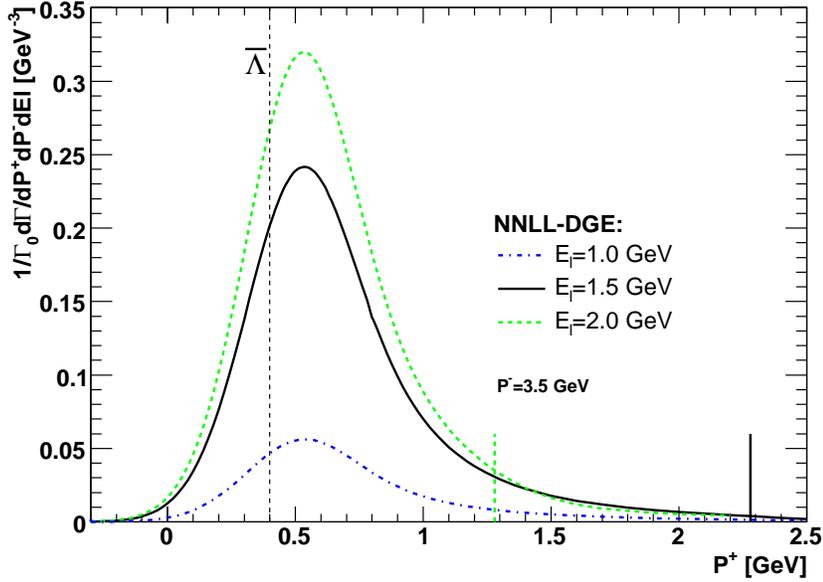,angle=0,width=11.9cm} \caption{The
fully differential width (in units of $\Gamma_0$) computed based
on~\eq{Gamma_n_lnR_matched_w_constant}, plotted as a function of the
small lightcone component $P^+$ at $P^-=3.5$ GeV and at
three different values of $E_l=1,\,1.5$ and $2$ GeV.
The maximal $P^+$ endpoint ($M_B-2E_l$), denoted by a vertical line,
should be: $3.3$, $2.3$ and $1.3$
GeV, respectively.
The $u=3/2$ renormalon residue parameter in \eq{B_DJ}
(see Fig.~\ref{fig:C_dependence}) is $C=1$. Note that the perturbative support properties, namely
$\bar{\Lambda} <P^+<M_B-2E_l$ are violated at both ends.
At the low $P^+$ end this is of course a desired consequence of the
resummation: the physical support properties $P^+>0$ are approximately
recovered. At the high $P^+$ end this is an ${\cal O}(\alpha_s^2/n)$
artifact, which is non-negligible only for large $E_l$.
 \label{fig:fully_diff_El_scan} }
\end{center}
\end{figure}

In terms of hadronic variables the total phase space is:
\begin{equation}
\label{hadronic_ps} \Gamma_{\tot}=\int_0^{M_B/2} dE_l
\int_{M_B-2E_l}^{M_B} dP^- \int_0^{M_B-2E_l} dP^+ \,\,
\frac{d\Gamma(P^+,\,P^-,\,E_l)}{dP^+\, dP^-\,dE_l }.
\end{equation}
A perturbative calculation of the differential width obviously
cannot fill the entire hadronic phase space. The fixed--order
perturbative result simply has a smaller phase space, {\em cf.}
\eq{partonic_ps}. However, resummation can, and indeed does modify
the support properties, violating energy and momentum conservation
which hold order by order. In this respect it is essential to
distinguish between the region targeted by the resummation and other
phase--space regions:
\begin{itemize}
\item{} In the small $P^+$ region (or, equivalently, the small hadronic mass region)
 the resummation\footnote{Note that we refer here
specifically to DGE where renormalons are regulated by the Principal
Value prescription. This improvement is not achieved in
fixed--logarithmic--accuracy Sudakov resummation, see e.g.
Fig.~\ref{fig:matched_NNLL_spectrum}, which suffers from Landau
singularities.} dramatically improves the description of the
differential width, and with it the support properties. The resummed
perturbative distribution of~\eq{sl_logs_resummed_r} extends to
negative $r$ values. In terms of hadronic variables, the
differential width extends to the region $P^+<\bar{\Lambda}$
 and approximately vanishes for $P^+<0$
(see Fig.~\ref{fig:fully_diff_El_scan}), consistently
with the physical support properties~\cite{RD}.
\item{} Away from the region targeted by the resummation, partonic phase--space
boundaries are also violated\footnote{Violation of the hard phase--space
boundary by resummed distributions is a familiar phenomenon, see e.g.~Ref.~\cite{Gardi:2002bg}}.
This, however, is an {\em artifact} of
uncontrolled ${\cal O}(1/n)$ higher--order perturbative corrections, which
depend on the specific scheme by which matching to the fixed--order
result is done.
As shown in Fig.~\ref{fig:fully_diff_El_scan} the phase--space limit
$r<(1-x)/\lambda$ (see \eq{partonic_ps}) is violated.
Fortunately, this makes just a small impact on
the observables we consider. This issue is discussed further in
Sec.~\ref{sec:Numerical_results_and_uncertainty}.
\end{itemize}

\begin{figure}
\begin{center}
\epsfig{file=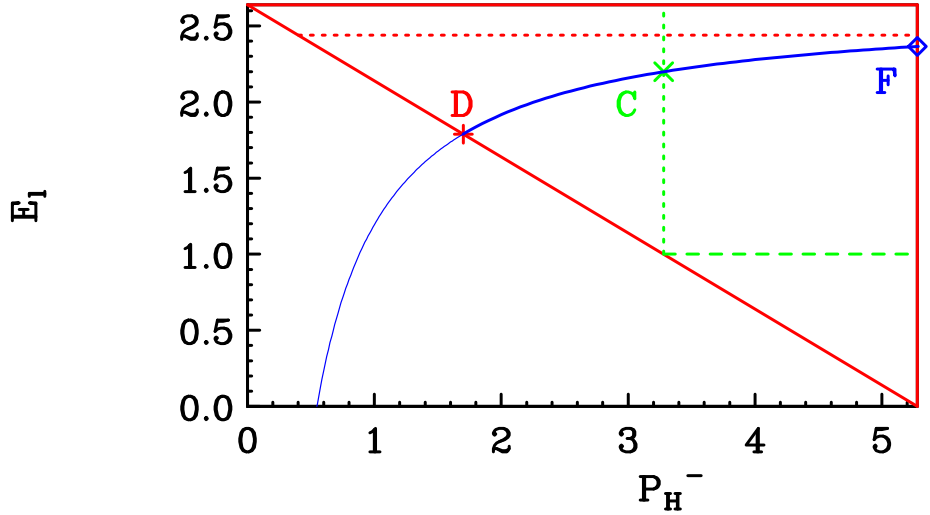,angle=90,width=12.4cm}
\caption{The full phase space in the ($P^-$, $E_l$) plain (upper
left triangle) after the $P^+$ integration has been performed. The
dotted horizontal line $E_l=m_b/2$ starting at $P_H^-=\bar{\Lambda}$
is the upper boundary of the {\em partonic} phase space.
A lower cut on the
charged lepton energy (here $E_l>E_0=1 \,{\rm GeV}$) removes the
small triangular region below the horizontal dashed line. An upper
cut on the hadronic invariant mass (here $P^+P^-<M_X^2=(1.7 \,{\rm
GeV})^2$) affects the distribution for $E_l<\frac12(M_B-M_X^2/P^-)$.
A full line separates between this region and the rest of the phase
space where this cut is irrelevant. On this line we denoted three
special points by $D$ $(P^-=M_X)$, $C$ ($P^-=M_B-2E_0$) and $F$
($P^-=M_B$), respectively.
      \label{fig:P_minus_El_ps}}
\end{center}
\end{figure}

Considering other phase--space limits of $x=2E_l/m_b$ or
$\lambda=(P^--\bar{\Lambda})/m_b$ that are not related to the
resummation variable $r$, the partonic phase--space limits of
\eq{partonic_ps} are unmodified by the resummation. The phase space
after the $P^+$ integration is shown in
Fig.~\ref{fig:P_minus_El_ps}, where we indicate both the hadronic
and the partonic upper limits\footnote{Note that although we shall
refer below to the hadronic phase space, $E_l\leq M_B/2$, the
numerical integration is actually done in the range $E_l\leq m_b/2$,
as dictated by the perturbative support properties.} on the charged
lepton energy.

As explained below, the transformation of the resummed distribution
in the small lightcone component from partonic to hadronic variables
has a crucial role in obtaining a correct, {\em unambiguous result}
(which is also characterized by approximate physical support
properties).

To this end, let us recall some of the observations of
Refs.~\cite{BDK,CG} and reformulate them in the context of the
semileptonic decay. The first observation, already discussed above,
is that renormalons in the
soft function of \eq{Gamma_n_large_n} have residues proportional to
inverse integer powers of the soft scale, indicating potential
non-perturbative power corrections of the form $(n\Lambda/(\lambda
m_b))^j$, where $j$ is a positive integer. It was also observed
there that the dominant part of these corrections should be
attributed to kinematic effects associated with the mass difference
between the meson and the quark, $\bar{\Lambda}=M_B-m_b$. In the
present context these correspond to identifying $r\equiv
p_j^+/p_j^-=\rho/\lambda$ in terms of hadronic variables using
\eq{P^+}:
\begin{equation}
r= \frac{P^+ - \bar{\Lambda}}{P^--\bar{\Lambda}},
\end{equation}
which effectively shifts the distribution in $r$ by
$\bar{\Lambda}/(\lambda m_b)$. Note that \eq{P^+} relies on the
assumption that the b quark is on-shell, so this transformation is
purely kinematic and the ``primordial'' motion of the b quark in the
meson is not taken into account. In order to account for
such non-perturbative dynamics, additional power corrections of the
form $(n\Lambda/(\lambda m_b))^j$ with $j\geq 2$ should be included,
as discussed below.

Still within the on-shell heavy--quark approximation, converting~\eq{sl_logs_resummed_r}
to hadronic variables using
\eq{P^+} one obtains:
\begin{eqnarray}
\label{Pplus_diff} &&\hspace*{-10pt} \frac{1}{\Gamma_0}
\frac{d\Gamma(P^+,\,P^-,\,E_l)}{dP^+\, dP^-\,dE_l}
=\frac{2}{m_b^2}\,\frac{1}{P^--\bar{\Lambda}}\,\int_{c-i\infty}^{c+i\infty}
\frac{d n}{2\pi i} \, \frac{d\Gamma_{n} (\lambda,x)}{d\lambda dx}\,
\,\left(1-\frac{P^+-\bar{\Lambda}}{P^--\bar{\Lambda}}\right)^{-n}\times
\nonumber
\\ &&\hspace*{180pt}
\frac{d\Gamma_n(\lambda=\frac{P^--\bar{\Lambda}}{m_b},x=\frac{2E_l}{m_b})}{d\lambda
dx} \\
\nonumber &&=\,\frac{2}{m_b^2}\,\frac{1}{P^-}
\int_{c-i\infty}^{c+i\infty}\frac{dn}{2\pi i}
\,\left(1-\frac{\bar{\Lambda}}{P^-}\right)^{n-1}
\left(1-\frac{P^+}{P^-}\right)^{-n}
 \frac{d\Gamma_n(\lambda=\frac{P^--\bar{\Lambda}}{m_b},x=\frac{2E_l}{m_b})}{d\lambda
dx},
\end{eqnarray}%
where no further approximation was made. The $\bar{\Lambda}$
dependence associated with the shift in $P^+$ factorizes {\em
exactly} in these variables. Note, in particular, that the kinematic
power correction factor $\left(1-{\bar{\Lambda}}/{P^-}\right)^{n-1}$
is valid for any $n$, not just for large $n$. Obviously, it makes no
impact of the first moment and it becomes increasingly important as
$n$ gets large.

It was further shown\footnote{The cancellation has been explicitly
checked by a calculation in the large--$\beta_0$ limit and argued to
be general.} in Ref.~\cite{BDK} that the product
\begin{eqnarray}
\label{Mellin_vs_Laplace} \frac{d\Gamma_{n} (\lambda,x)}{d\lambda
dx}\, \left(1-\frac{\bar{\Lambda}}{P^-} \right)^{n-1}\simeq
\frac{d\Gamma_{n} (\lambda,x)}{d\lambda dx}\, \exp
\left\{-\frac{(n-1)\bar{\Lambda}}{P^-}\right\}(1+{\cal
O}(\Lambda/P^-))
\end{eqnarray}
is free of the leading renormalon ambiguity, a linear ambiguity of
${\cal O}\left(\Lambda (n-1)/P^-\right)$ in the exponent, owing to
exact cancellation of the $u=\frac12$ ambiguities between the
Sudakov factor in the resummed moments and the one in
$\bar{\Lambda}$. The exponentiation of the corresponding power--like
ambiguity in \eq{Gamma_n_large_n} and the regularization of
$\bar{\Lambda}$ and of the Sudakov exponent using the same
prescription are crucial for this cancellation to take place. In the
Principal Value prescription the numerical result for
$\bar{\Lambda}$ is
\begin{eqnarray}
\bar{\Lambda}\equiv \bar{\Lambda}_{\PV}\equiv M_B-m_b^{\PV}
=5.28-4.88\pm 0.06 \,{\rm GeV} \,=\, 0.40 \pm 0.06\,{\rm GeV},
\end{eqnarray}
where we used the result for the mass ratio in
\eq{mass_ratio_results} with the short--distance quark mass value
of~\eq{mb_MSbar_value} and $N_f=4$.

Let us address now power corrections associated with the dynamical
structure of the meson. The perturbative calculation of the soft
Sudakov exponent accounts for radiation off the heavy quark that
puts it slightly off its mass shell. However, it cannot take into
account the way in which the virtuality of the heavy quark is
influenced by its non-perturbative interaction with the light
degrees of freedom in the meson. As discussed in
Refs.~\cite{BDK,RD}, in the present framework such non-perturbative
dynamics is reflected in additional power corrections of form
$(n\Lambda/(\lambda m_b))^j$ with $j\geq 2$. The detailed structure
of these power corrections can be read off from the renormalon
ambiguities of the soft Sudakov exponent. As done explicitly in
\eq{NP_power_sum} each power ambiguity is associated with a new
non-perturbative parameter $f_j^{\PV}$; upon including the power
term, the corresponding ambiguity is removed. The multiplicative
correction to the Sudakov factor in \eq{NP_power_sum} translates
directly into a multiplicative correction to partonic moments
computed, for example, in~\eq{Gamma_n_lnR_matched_w_constant}. Given
the kinematic cancellation of the leading ($u=1/2$) renormalon
ambiguity discussed above, and given that $B_{\cal S}(u=1)$ vanishes
(see \eq{B_DJ} and the discussion following it), the leading
dynamical power correction is associated with the $u=3/2$ ambiguity,
corresponding to $j=3$. This correction would modify the spectral
moments by
\begin{eqnarray}
\label{NP_power_leading} &&\left. \frac{d\Gamma_{n}
(\lambda,x)}{d\lambda dx}\right\vert_{\PV}\,\longrightarrow \,\left.
\frac{d\Gamma_{n} (\lambda,x)}{d\lambda dx}\right\vert_{\PV}\times
\\ \nonumber &&\hspace*{20pt}\exp\left\{ \left(\frac{\Lambda
}{m_b\lambda}\right)^3 f_3^{\PV} \pi \frac{C_F}{\beta_0}
\frac{T(3/2)}{3/2}
 B_{\cal S}(3/2) R(n,3/2)\,+\,{\cal O}\left(\left(\frac{\Lambda }{m_b\lambda}\right)^4\right)
\right\},
\end{eqnarray}
where $R(n,3/2)=\frac{1}{12} (n-3)(n-2)(n-1)$. Note that unless the
residue of the anomalous dimension $B_{\cal S}(3/2)$ and the
non-perturbative coefficient $f_3^{\PV}$ are much larger than one,
which we consider unlikely, \eq{NP_power_leading} amounts to a small
correction even for $n\sim \lambda m_b/\Lambda$.

The corrections of \eq{NP_power_leading} go beyond our minimal
model: here we regularize all the renormalons using the Principal
Value prescription and do not include any additional
non-perturbative power terms (i.e. we set $f_j^{\PV}=0$ for $j\geq
3$). The resulting, essentially perturbative, spectrum of
\eq{Pplus_diff} may therefore differ from the measured spectrum by
effects of the from of \eq{NP_power_leading}; because of the {\em
parametrically small} contribution to moments $n\lsim  \lambda
m_b/\Lambda$, these effects are restricted to a narrow region near
$P^+\simeq 0$. We shall revisit this issue in
Sec.~\ref{sec:Numerical_results_and_uncertainty} below when
estimating the related theoretical uncertainty.

In Appendix~\ref{sec:Phase_space} we explicitly show how the
resummed result for the
spectrum, computed in the previous section using moments of the
partonic lightcone momentum ratio $r=\rho/\lambda$, is used to
compute the partial branching fraction with experimentally--relevant
cuts. In Sec.~\ref{sec:diff_and_int} we use the results of Sections
\ref{sec:matching} and~\ref{sec:Exponentiation_beyond_logs} to
express the partially integrated width with a cut on $M_X$ (or on
$P^+$) and an additional mild cut on the lepton energy. In
Sec.~\ref{sec:int_lep_energy}  we derive expressions for the same
observables using the results of
Sec.~\ref{sec:Analytic_lepton_energy_cut} and
Appendix~\ref{sec:partially_integrated_matching}, where the cut on the
lepton energy is implemented analytically. Numerical results for
$R_{\rm cut}$ as a function on $M_X$ or $P^+$ are presented in
Sec.~\ref{sec:Numerical_results_and_uncertainty}, where we also
perform a detailed study of the theoretical uncertainty. Finally, in
Sec.~\ref{sec:extraction_Vub} we extract $|V_{ub}|$ from recent
measurements by Belle.

\subsection{Numerical results and theoretical uncertainty
estimates\label{sec:Numerical_results_and_uncertainty}}

In extracting $|V_{ub}|$ from experimental data according to
\eq{Delta_cal_B_th_general} the effect of kinematic cuts is taken
into account through the event fraction $R_{\rm cut}$ defined in
\eq{splitting_calculation}.
Here we shall use\footnote{Numerical analysis is done using a C program
that combines Borel integration,
Mellin inversion and phase--space integration~\cite{Web_page}.}
the tools of the previous
sections to evaluate $R_{\rm cut}$ for hadronic mass $M_X$
 or lightcone momentum $P^+$ cuts (with additional lower cut on
$E_l$) and estimate the theoretical uncertainties involved.

It should be emphasized that although the formulae in Sections~\ref{sec:diff_and_int}
and \ref{sec:int_lep_energy} are explicitly written for partial rates
and are normalized by $\Gamma_0$,
we are really dealing here with the perturbative expansion of $R_{\rm cut}$,
not the partial width and the full width separately.
A single calculation of $R_{\rm cut}$ in \eq{splitting_calculation} involves
using these formulae twice: with a cut
in the numerator, and with no cut in the denominator.
Clearly, $\Gamma_0$ cancels in the ratio.
This has far--reaching implications
in what concerns renormalization--scale dependence and renormalon ambiguities:
$R_{\rm cut}$, in contrast with the partial (or total) rate in units of $\Gamma_0$,
is not affected by the ${\cal O}(\Lambda/m_b)$ renormalon
ambiguity.
This ambiguity would have cancelled with the ambiguity of $m_b^5$ in $\Gamma_0$ had this
factor been included  and the series resummed. In $R_{\rm cut}$ it simply
cancels\footnote{Note, however, that cut--related renormalon ambiguities,
which are parametrically enhanced at large $n$,
are not canceled in this way; these are discussed below.}
between the numerator and the denominator.

As already mentioned, we focus on the cut that maximizes the rate,
as measured by Belle~\cite{Bizjak:2005hn}:
$P^+P^-<M_X^2=(1.7\, {\rm GeV})^2$ to discriminate the charm background
with an additional lepton--energy cut $E_l>E_0=1$ GeV.
Numerical integration of the differential distribution with respect to $P^-$ and
$E_l$, \eq{diff_M_X_cut}, over the available phase space yields
$R_{\rm cut} (M_X=1.7\, {\rm GeV}, E_0=1\, {\rm GeV}) = 0.615$.
This event fraction will be used in the next section to extract $|V_{ub}|$
from Belle data. In this calculation we used the matching
of~\eq{Gamma_n_lnR_matched_w_constant} where the renormalization scale for
the coupling was set as $\mu=P^-$. In the soft Sudakov exponent we chose $C=1$.
In the following we perform a detailed study of the
theoretical uncertainty in $R_{\rm cut}$.
We consider uncertainties  from three sources: (1) higher--order perturbative
corrections going beyond the resummation as well as beyond the NLO;
(2) renormalons and power corrections, where the main uncertainty
is associate with parametrically--enhanced power terms related
to the quark
distribution in the meson; (3) uncertainty in the values of
fundamental short--distance parameters: $\alpha_s^{\MSbar}$
and $m_b^{\MSbar}$.

\subsubsection*{Matching schemes, scale dependence and higher--order perturbative effects}

The resummation employed here focuses on improving the perturbative
expansion of the fully differential width in the particular
kinematic region where one lightcone component of the hadronic
system, $P^+$, is small. Matching has been performed to the full NLO
expression. Our purpose here is to estimate the effect of
higher--order perturbative corrections on~$R_{\rm cut}$. This is
done in two ways: first by comparing different matching schemes
introduced in sections~\ref{sec:matching}
through~\ref{sec:Analytic_lepton_energy_cut}, and then by
renormalization--scale variation in the matching coefficient.

\begin{figure}[htb]
\begin{center}
\epsfig{file=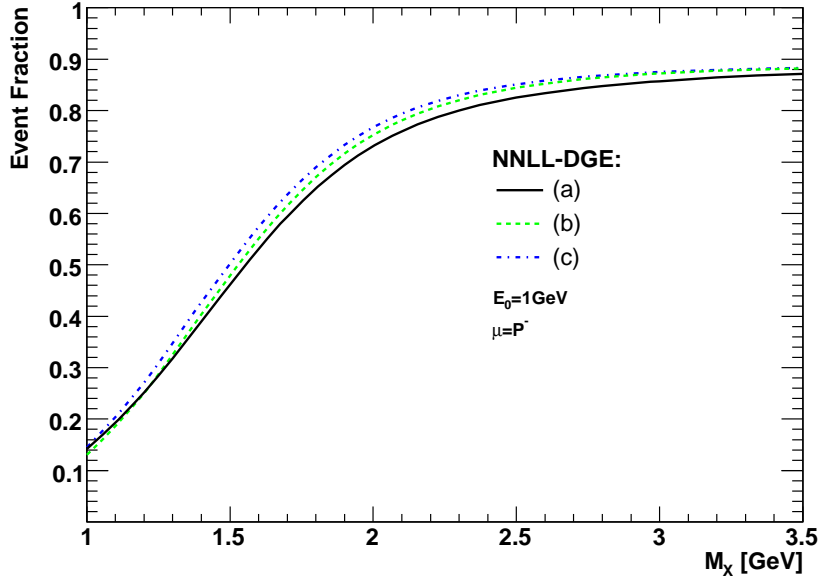,angle=0,width=11.9cm}
\caption{\label{fig:matching_schemes} The event fraction $R_{\rm
cut}$ within the range $P^+P^-<M_X^2$ and $E_l>1$ GeV, plotted as a
function of the cut value $M_X$. Three different matching schemes
are used: (a) the partially integrated distribution with respect to
$E_l$, namely \eq{int_M_X_cut} and \eq{double_diff_matched},
numerically integrated over $P^-$ (full line); (b) the fully
differential distribution computed by \eq{diff_M_X_cut} and the
matching of \eq{Gamma_n_lnR_matched_w_constant}, numerically
integrated over the
 $E_l$ and $P^-$ (dashed line); and (c) same as in (b) but with the matching of
\eq{Gamma_n_lnR_matched} (dotdashed line). }
\end{center}
\end{figure}

\begin{figure}[t]
\begin{center}
\epsfig{file=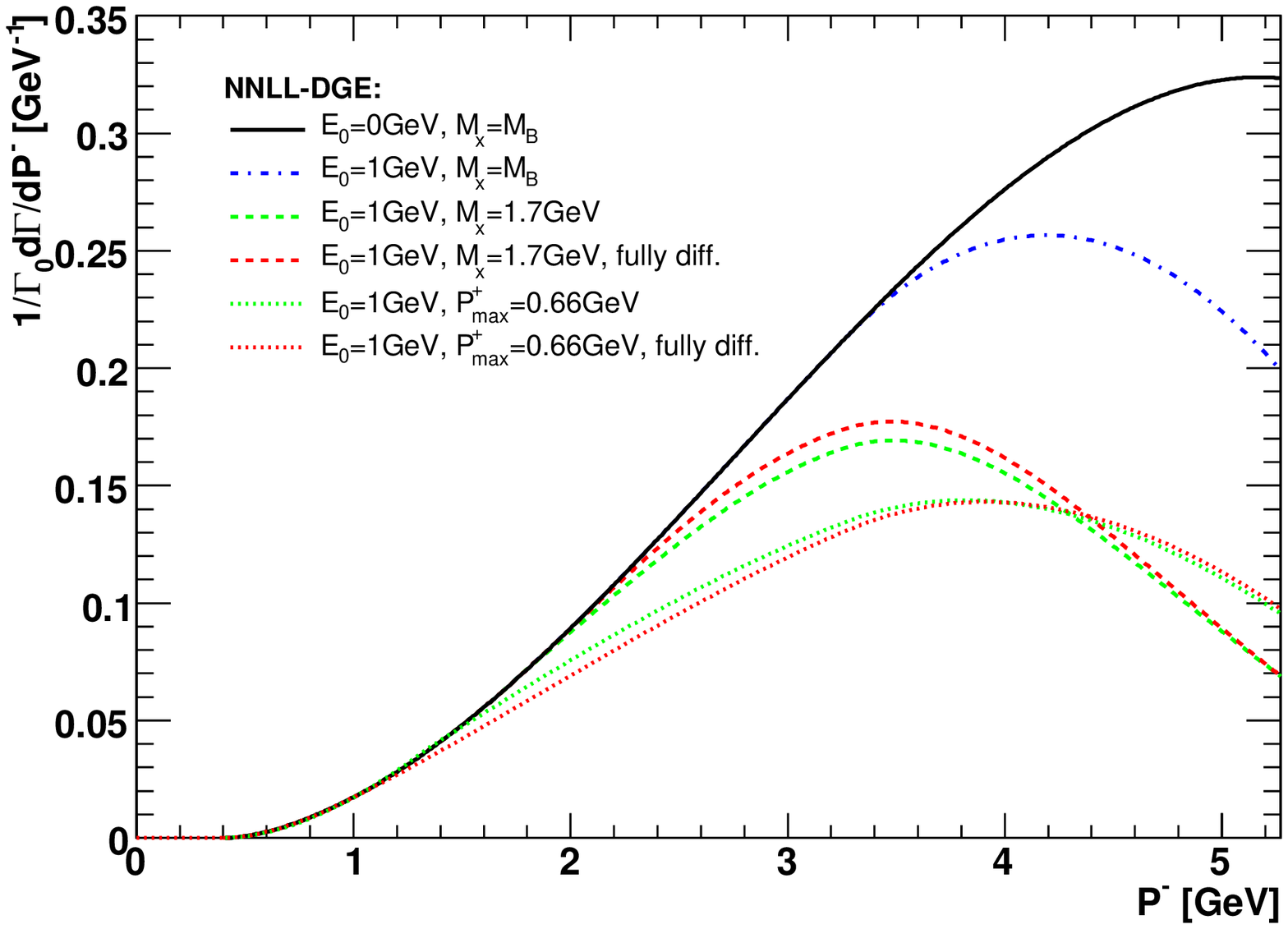,angle=0,width=11.9cm}
\caption{The $P^-$ distribution after integration over $P^+$ (using
the moment--space result)
and $E_l$ in four different situations:
no cuts (full line), $E_l>1$ GeV cut only
(dotdashed line),
$E_l>1$ GeV combined with $P^+P^-<M_X^2=(1.7\, {\rm GeV})^2$ cut (dashed), and finally
$E_l>1$ GeV combined with $P^+<P^+_{\max}=0.66$~GeV cut (dotted).
In each of the two last cases,
two different calculation procedures are used in order to gauge the sensitivity
to higher--order corrections (matching--scheme dependence):
(a) using the partially integrated distribution
with respect to $E_l$, namely
\eq{int_M_X_cut} and \eq{double_diff_matched} and (b)
numerically integrating over the differential distribution with respect to
$E_l$, \eq{diff_M_X_cut} and \eq{Gamma_n_lnR_matched_w_constant}.
As explained in the text, the difference between (a) and (b) exists
only when there is a $M_X$ or a $P^+_{\max}$ cut and only for
$P^->M_X$, or $P^->P^+_{\max}$, respectively.
\label{fig:diff_Pminus_different_cuts} }
\end{center}
\end{figure}

Let us first estimate the numerical significance of the difference
between different matching schemes, which are formally of ${\cal
O}(\alpha_s^2/n)$. Fig.~\ref{fig:matching_schemes} shows the result
of three different calculations of the event fraction $R_{\rm cut}$
with a hadronic mass cut $P^+P^-<M_X^2$ as a function of $M_X$, with
a fixed lepton--energy cut $E_l>1$ GeV, all with the same choice of
renormalization scale for the coupling in the matching coefficient,
$\mu=P^-$ (see below):
\begin{description}
\item{ (a)} Using the partially integrated distribution with respect to $E_l$, namely
\eq{int_M_X_cut} and \eq{double_diff_matched}.
\item{ (b)} Using the differential distribution with respect to
$E_l$, \eq{diff_M_X_cut} (or, equivalently, Eqs. (\ref{large_S_cut})
through (\ref{small_S_cut})) with the matching of
\eq{Gamma_n_lnR_matched_w_constant}.
\item{(c)} same as (b) with the matching of \eq{Gamma_n_lnR_matched}.
\end{description}
The origin of the difference between (c) and (b) is clear: the
constant terms, which were treated in \eq{Gamma_n_lnR_matched} as
part of the matching coefficient are exponentiated in
\eq{Gamma_n_lnR_matched_w_constant}. Less obvious is the difference
between (a) and (b), which is reflected also in the $P^-$
distributions plotted in Fig.~\ref{fig:diff_Pminus_different_cuts}.
Let us explain this difference for the example where an $M_X$ cut is
applied.

The first crucial observation is that the resummed distribution
(Fig.~\ref{fig:fully_diff_El_scan}) {\em does not respect} the phase
space limit $r<(1-x)/\lambda$, which holds order by order in
perturbation theory.  Instead, the distribution develops a tail that
extends up to $r=1$. By construction, the first moment, which is
determined by the matching coefficient (i.e. the fixed--order
result) alone, is unaffected by this tail. In contrast, whenever one
uses the distribution computed as an inverse--Mellin of the resummed
moment--space expression these details become important.

Considering next the way the phase space in
Fig.~\ref{fig:P_minus_El_ps} is covered we see that in case (a) only
the region $P^-<M_X$ is evaluated using the first moment whereas the
entire region $P^->M_X$ is reconstructed as an inverse Mellin
transform of the resummed expression. In case (b) the differential
distribution is used only where necessary: the entire region above
the line $E_l>\frac12(M_B-M_X^2/P^-)$ is computed using the first
moment and only the region below this line depends on the details of
the differential distribution. Consequently, (a) effectively uses
the distribution (computed as an inverse Mellin transformation) in
the region
\[
M_X<P^-<\min\left\{M_B,\,\frac{M_X^2}{\bar{\Lambda}}\right\},\qquad E_l>\frac12
\left(M_B-\frac{M_X^2}{P^-}\right),
\]
where (b) uses the $n=1$ moment instead.

Table~\ref{table:matching_scheme_dep} summarizes the results
corresponding to $P^+P^-<M_X^2=(1.7\, {\rm GeV})^2$.
\begin{table}
\begin{center}
\begin{tabular}{|c|c|c
|}
  \hline
matching scheme & Event Fraction $R_{\rm cut}$ ($M_X = 1.7$ GeV,
$E_0=1$ GeV) &
variation(\%)\\
\hline
  (a)   & 0.5941 & $-$3.5\\
  (b)   & 0.6154 & default\\
  (c)   & 0.6364 & +3.4\\
  \hline
\end{tabular}
\caption{Results of the event fraction $R_{\rm cut}$ within the
range $M_X < 1.7$ GeV and $E_l>1$ GeV using different matching
schemes (see text) with the scale of the coupling in the matching
coefficient set as $\mu=P^-$.\label{table:matching_scheme_dep}}
\end{center}
\end{table}
We see that for the $P^+P^-<M_X^2=(1.7\, {\rm GeV})^2$
cut, the matching--scheme dependence
amounts to less than $\pm 4\%$ uncertainty. Although the fully
differential calculation (b) is used to obtain the central value, we will estimate
other sources of uncertainty using scheme (a), in which numerical results are obtained
faster.

Next, let us consider the renormalization scale dependence. When
considering the partial width with stringent cuts, Sudakov
resummation is essential. As shown in Refs.~\cite{BDK,RD} and
Sec.~\ref{sec:Sud_exp} above, the logarithmic accuracy criterion
fails {\em because of large running--coupling effects}. By using DGE
we resum these effects as well. However, this resummation is
restricted to the Sudakov exponent. A priori, one might worry that
additional resummation of running--coupling effects should be
applied in the matching coefficient as well. Using scale variation we show
below that this resummation is probably not necessary for the
calculation of $R_{\rm cut}$ at the $\pm 5\%$ level.

When considering $R_{\rm cut}$ with very mild cuts
Sudakov resummation is irrelevant. But running--coupling effects,
which we have not resummed, may still be significant.
It is important to emphasize, that these effects
cannot be estimated using the expansion of the total
width\footnote{Indeed, the BLM scale of the total width is of order
$\Lambda$, see e.g.~\cite{Falk:1995me}.},~\eq{total_width}, which in
contrast with $R_{\rm cut}$, is affected by the infrared renormalon
corresponding to $m_b^5$. Determination of the BLM scale in $R_{\rm
cut}$ requires the knowledge of the matching coefficients of
Sec.~\ref{sec:matching} as a function of $\lambda$ and $x$ to ${\cal
O}(\alpha_s^2\beta_0)$. These calculations have not yet been
performed.

We therefore resort to estimating running--coupling effects in the
matching coefficients by means of scale variation. A natural scale
for the coupling in the matching coefficient is $P^-$, which is the
hard scale characterizing the hadronic system. $\mu=P^-$ will be our default value.
Clearly, as good an alternative would be the partonic $p_j^-$ ---
this scale appear as the hard scale in
Sudakov exponent, see e.g. \eq{Gamma_n_large_n} or \eq{sum_FLA}.

\begin{figure}[t]
\begin{center}
\epsfig{file=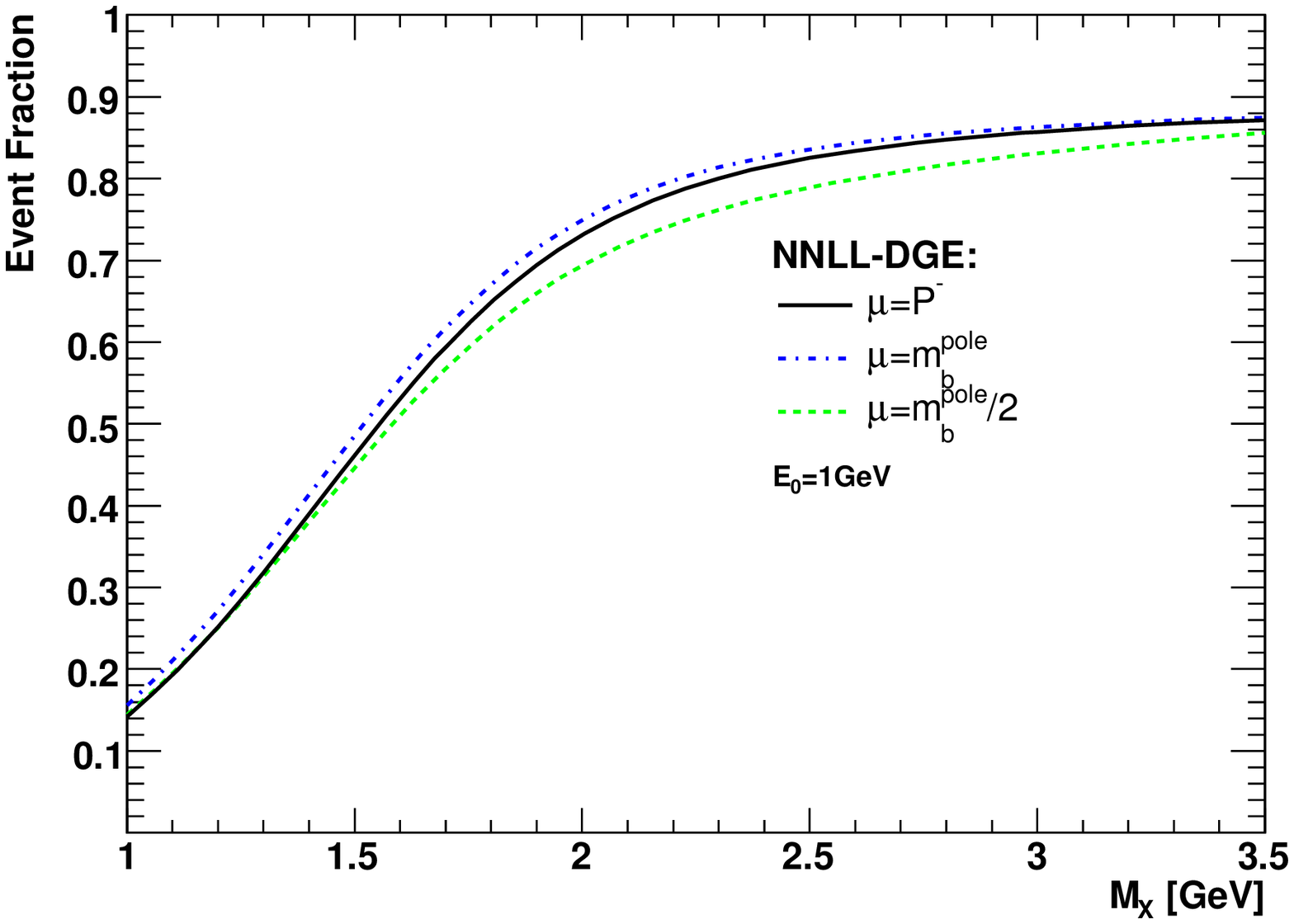,angle=0,width=11.9cm}
\caption{The event fraction $R_{\rm cut}$ within the range
$P^+P^-<M_X^2$ and $E_l>1$ GeV, computed using \eq{int_M_X_cut} and
\eq{double_diff_matched}, plotted as a function of the cut value
$M_X$. Different assignments for the argument of the coupling in the
matching coefficient are used. \label{fig:M_x_scale_dep} }
\end{center}
\end{figure}

Fig.~\ref{fig:M_x_scale_dep} presents the result for the event
fraction $R_{\rm cut}$ associated with a hadronic mass cut
$P^+P^-<M_X^2$ as a function of $M_X$ with a fixed lepton--energy
cut $E_l>1$ GeV, computed according to \eq{int_M_X_cut} and
\eq{double_diff_matched}, with three different assignments of the
renormalization scale $\mu$ in $\alpha_s^{\MSbar}(\mu)$ in both the
virtual and real--emission contributions to the matching
coefficient.
The results for $P^+P^-<M_X^2=(1.7\, {\rm GeV})^2$
(and $E_l>1$ GeV) are also summarized
in Table~\ref{table:matching_ren_scale_dep}.  We find that while the
coupling varies from $\alpha_s(m_b)\simeq 0.22$ to
$\alpha_s(m_b/2)\simeq 0.29$, i.e. by over $32\%$, $R_{\rm cut}$
varies only by about $8.3\%$.
\begin{table}
\begin{center}
\begin{tabular}{|l|c|c
|}
  \hline
scale of $\alpha_s$ & Event Fraction $R_{\rm cut}$ ($M_X=1.7$
GeV, $E_0=1$ GeV)& variation (\%)\\
\hline
  $\mu=P^-$   & 0.5941 &default\\
  $\mu=m_b$   & 0.6173 &+3.9\\
  $\mu=m_b/2$ & 0.5677 &$-$4.4\\
  \hline
\end{tabular}
\caption{Results of the event fraction within the range
$P^+P^-<M_X^2=(1.7\, {\rm GeV})^2$ and $E_l>E_0=1$ GeV using \eq{int_M_X_cut} and
\eq{double_diff_matched} with different assignments for the scale of
the coupling in the matching coefficient (the $n=1$
moment).\label{table:matching_ren_scale_dep}}
\end{center}
\end{table}

Let us consider now the default choice of $\mu=P^-$. As shown in
Fig.~\ref{fig:diff_Pminus_different_cuts} the $P^-$ distribution in
which no $M_X$ (or $P^+$) cuts are applied is peaked close to
$P^-\simeq M_B$. On the other hand the $M_X$ cut suppresses the
distribution at the hard--$P^-$ end, while it makes no impact on the
soft end. Therefore, the choice of matching--coefficient scale as
$\mu=P^-$ is a priori quite different from $\mu=m_b$. Nevertheless,
as shown in Table \ref{table:matching_ren_scale_dep} the overall
effect of this change on $R_{\rm cut}$ for $P^+P^-<M_X^2=(1.7\, {\rm GeV})^2$ (and
$E_l>1$ GeV) is just $\sim 3.9\%$.

Taking into account the scale uncertainty as $\pm 4.4\%$ and
combining it in quadrature\footnote{By regarding these two
uncertainties as independent we obtain a conservative uncertainty~estimate.} with the matching scheme dependence estimate of
$\pm 3.5\%$, we assign a total theoretical uncertainty of $5.6\%$ in
$R_{\rm cut}$ for $M_X = 1.7$ GeV owing to unknown higher--order
corrections. We expect these number would significantly reduce once
NNLO calculations of the fully differential width would be
completed.

\subsubsection*{Renormalons and power corrections}

Considering the leading, $u=1/2$ renormalon ambiguity,
it is essential to distinguish between
\begin{itemize}
\item{}
the ${\cal O}(\Lambda/m_b)$ renormalon in the
total width when written in units of $\Gamma_0$ (or the $n=1$ moment of the matching coefficient)
which cancels by defining $R_{\rm
cut}$ as we did, and
\item{}
the cut--related leading ${\cal O}((n-1) \Lambda/m_b)$ renormalon in
the Sudakov exponent which makes {\em all}  the perturbative moments
(but $n=1$) ambiguous and generates an ambiguous shift of the
normalized distribution.
\end{itemize}
While the former is canceled in $R_{\rm cut}$ by
definition, the latter affects only the partial width (i.e. the
numerator in $R_{\rm cut}$) and becomes more significant the deeper
the cut is. As discussed in detail in Sec. \ref{sec:conversion}, its cancellation~\cite{BDK,RD}
requires renormalon resummation in the Sudakov exponent and incorporating the corresponding
kinematic power corrections $\left(1-{\bar{\Lambda}}/{P^-}\right)^{n-1}$ according to
\eq{Pplus_diff}.
Having full control of the $u=1/2$ renormalon, including the normalization of its residue, and
having defined both the Sudakov factor and ${\bar{\Lambda}}$ using the Principal Value
prescription, the computed spectrum is free of any ${\cal O}(\Lambda/(\lambda m_b))$ artifact
of the non-physical on-shell heavy--quark state.

Beyond the $u=1/2$ renormalon the Sudakov factor has subleading
power corrections reflecting the non-perturbative dynamics of the b
quark in the B meson. As discussed in~Refs.\cite{BDK,RD} and in
Sections \ref{sec:Exponentiation_beyond_logs} and
\ref{sec:conversion} above, these corrections depend on the
definition of the perturbative sum applied in the Sudakov exponent.
Moreover, the perturbative Sudakov exponent
(\ref{Gamma_n_lnR_matched_w_constant}) as well as the corresponding
renormalon ambiguities (\ref{NP_power_leading}) depends on the
magnitude of the Borel function {\em away from the origin}, so the
assumptions made on $B_{\cal S}(u)$ are directly relevant. Let us
briefly summarize these assumptions:
\begin{description}
\item{(1)\,} We assumed that, as in the large--$\beta_0$
limit~(\ref{B_DJ_large_beta0}), the Borel function $B_{\cal
S}(u)$ vanishes at $u=1$ and has no additional zeros. The vanishing of $B_{\cal
S}(1)$ implies, in particular, that there is no ambiguity
${\cal O}((n \Lambda/(\lambda m_b))^2)$ in the exponent so the leading correction,
corresponding to the $u=3/2$ renormalon ambiguity,
takes the form of \eq{NP_power_leading};
\item{(2)\,} We assumed that $B_{\cal S}(u)$ gets small at
large $u$, although this is not required for the convergence of the
Borel integral. To control the contribution from intermediate values of $u$,
we included in the ansatz for $B_{\cal S}(u)$ in \eq{B_DJ} a single
free parameter $C$ which is proportional to
$B_{\cal S}(u=3/2)$; see Fig.~\ref{fig:C_dependence} and \eq{C_def}.
It is difficult to compute $B_{\cal S}(u=3/2)$, but numerically
large (as well as extremely small) values of $C$ can probably be
excluded by the following considerations: (a)
 for $C\simeq 26$ the residue at $u=3/2$ would
be as large as in the large--$\beta_0$, which is unlikely since
non-Abelian corrections tend to reduce renormalon residues; (b)
given the ansatz of \eq{C_def}, either large or small values of $C$
imply that $t_1$ and $t_2$  become unnaturally large, and so do higher--order
perturbative corrections to $B_{\cal S}(u)$. Eventually, these can be
computed explicitly to further constrain this function.
\end{description}
Having fixed the Borel transform of the anomalous dimensions in
\eq{B_DJ}, we compute the Sudakov exponent of \eq{alt_Sud} applying
the Principal Value prescription to all the Borel singularities. The
resulting DGE spectrum displays only small sensitivity to $C$, see
discussion below. Therefore, despite the formal dependence on the
regularization prescription for the $u=3/2$ and higher renormalons,
this resummed spectrum can be directly considered an approximation
to the meson decay spectrum. On the theoretical level, this is
supported by the following observations~\cite{RD}:
\begin{itemize}
\item{} According to the renormalon ambiguity pattern, the leading correction corresponds to the
{\em third} power of $n \Lambda/(\lambda m_b)$. \eq{NP_power_leading} implies that for
not--too--high moments ($n\lsim  \lambda m_b/\Lambda$) the non-perturbative effect is small,
and therefore only a narrow region near $P^+\simeq 0$ is affected.
\item{} Despite the finite gap between the physical support properties ($P^+>0$) and the
perturbative ones ($P^+>\bar{\Lambda}$), the resummed perturbative spectrum, quite
remarkably, has support that is close to the physical spectrum.
\end{itemize}
The final judgement, however, must be based on comparison with
experimental data. Comparing the predictions of Ref.~\cite{RD},
corresponding to $C=1$ in \eq{B_DJ} above, with experimental data
for $\bar{B}\longrightarrow X_s\gamma$, shows~\cite{Gardi:2005mf}
that
 this minimal model, which essentially has no free parameters,
 describes the spectrum well. In particular, the first two central
moments with cuts, namely the average energy and the width defined
with $E_\gamma>E_0$ (Eqs. (5.3) and (5.4) in Ref.~\cite{RD}) agree
with experiment within error, while for the third moment the data is
still statistically limited~\cite{Abe:2005cv,Aubert:2005cu}. It
should be emphasized that the theoretical error estimate
in~\cite{RD,Gardi:2005mf} did not include any variation of $B_{\cal
S}(u)$ nor any non-perturbative power corrections, but only
variation of the short--distance parameters $\alpha_s^{\MSbar}$ and
$m_b^{\MSbar}$ within their ranges of uncertainty.

Given the good description of the $\bar{B}\longrightarrow X_s\gamma$
spectrum~\cite{Gardi:2005mf} in the present framework, our default
choice here is not to include any parametrization of additional
non-perturbative corrections of ${\cal O}((n \Lambda/(\lambda
m_b))^j)$ with $j\geq 3$. As soon as theoretical predictions and
experimental results become sufficiently constraining,
parametrization of these corrections along the lines
of~\eq{NP_power_leading} would be in place.

We note that since the size of the leading correction in
\eq{NP_power_leading} is controlled by the product of $f_3^{\PV}$
and $B_{\cal S}(3/2)$ (or, equivalently, $C$) these two parameters
are strongly correlated. Based on data alone it would therefore not
be possible to distinguish between genuine non-perturbative effects
in the meson ($f_j^{\PV}$) and unknown higher--order corrections to
$B_{\cal S}(u)$ that modify its behavior away from the origin with
respect to \eq{B_DJ}. To determine $f_j^{\PV}$, one would need not
only very accurate data but also further theoretical constraints.

\begin{figure}[t]
\begin{center}
\epsfig{file=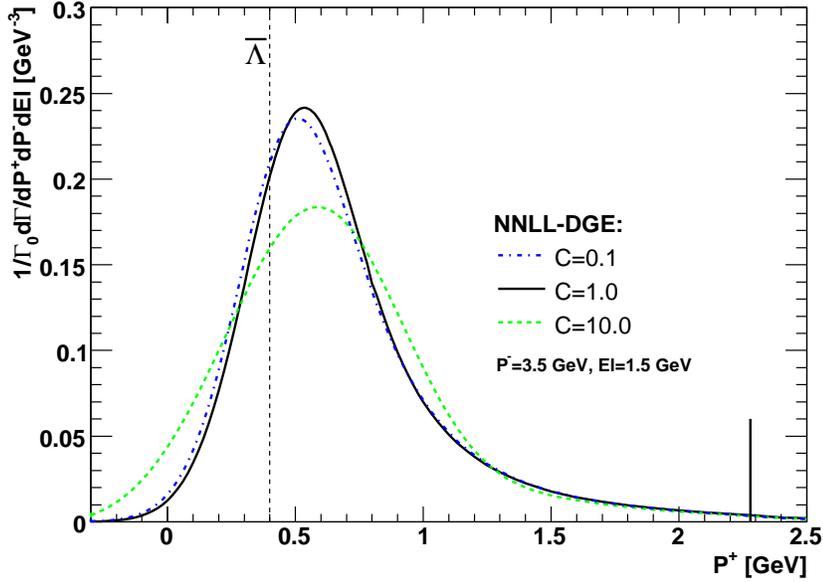,angle=0,width=11.9cm} \caption{The
fully differential width (in units of $\Gamma_0$) computed based
on~\eq{Gamma_n_lnR_matched_w_constant}, plotted as a function of the
small lightcone component $P^+$ at $E_l=1.5$ GeV and
$P^-=3.5$ GeV, for three different assignments of
the $u=3/2$ renormalon residue parameter $C$ in \eq{B_DJ}
(see Fig.~\ref{fig:C_dependence}). $C=1$ is
the value used in Ref.~\cite{RD}.
 \label{fig:renormalon_at_3halves_influence} }
\end{center}
\end{figure}

Let us therefore proceed by analyzing the numerical consequences of
varying $B_{\cal S}(u)$ in \eq{alt_Sud} by means of $C$, while
setting $f_j^{\PV}$ ($j\geq3$) in \eq{NP_power_leading} to zero.
Later we shall consider the uncertainty associated with a non-zero
$f_3^{\PV}$. $C$ influences the width of the spectrum as a function
of $P^+$, but even more, the left--right asymmetry in the shape.
This can be seen in Fig.~\ref{fig:renormalon_at_3halves_influence},
where it is also obvious that large values of $C$, such as $C=10$,
are disfavored as the distribution extends then further into the
negative $P^+$ domain, violating the physical support properties.

\begin{figure}[t]
\begin{center}
\epsfig{file=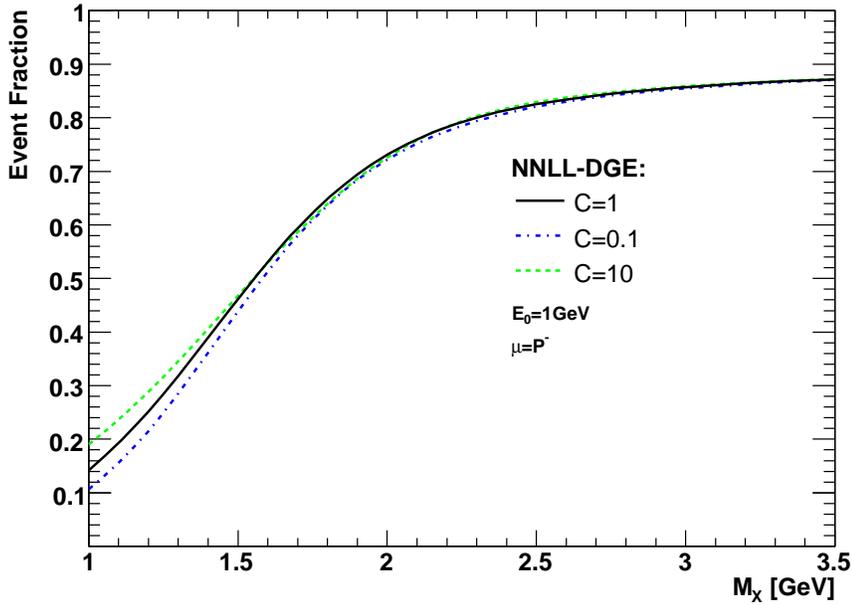,angle=0,width=12.4cm}
\caption{The event fraction $R_{\rm cut}$ within the range
$P^+P^-<M_X^2$ and $E_l>1$ GeV, computed using \eq{int_M_X_cut} and
\eq{double_diff_matched}, plotted as a function of the cut value
$M_X$. Different assignments for the unknown parameter $C$, which
controls the shape of the perturbative quark distribution function
(see \eq{C_def}) are used. In the matching coefficient the
renormalization scale $\mu=P^-$ is
chosen.\label{fig:Mx_C_dep}}
\end{center}
\end{figure}

The final effect of varying $C$ within the range $0.1$ to $10$ on
$R_{\rm cut}$ is shown in Fig.~\ref{fig:Mx_C_dep} as a function of
$M_X$. Results for the experimentally--relevant value of
$P^+P^-<M_X^2=(1.7\, {\rm GeV})^2$ (and $E_l>E_0=1$ GeV) are also summarized in
Table~\ref{table:C_dep}. We find that varying $C$ from $0.1$ to $10$
the event fraction varies by only $2.4 \%$. Note that the variation
is non-monotonous.
\begin{table}
\begin{center}
\begin{tabular}{|l|c|c
|}
  \hline
$C$ & Event Fraction $R_{\rm cut}$ ($M_X=1.7$
GeV, $E_0=1$ GeV) &variation (\%)\\
\hline
  $0.1$ & 0.5799 &$-$2.4  \\
  $1$   & 0.5941 &default\\
  $10$  & 0.5870 &$-$1.2   \\
  \hline
\end{tabular}
\caption{Results of the event fraction within the range
 $P^+P^-<M_X^2=(1.7\, {\rm GeV})^2$
GeV and $E_l>1$ GeV using \eq{int_M_X_cut} and
\eq{double_diff_matched} with different assignments of $C$ in
\eq{C_def}.\label{table:C_dep}}
\end{center}
\end{table}

Finally, let us return to the numerical effect of the leading
non-pertrbative correction in~\eq{NP_power_leading}. As discussed
above, one expects $f_j^{\PV}$ to be of ${\cal O}(1)$. For
$f_3^{\PV}=1$ and for our default value $C=1$ the effect is
\emph{very}  small. For example, for the experimentally--relevant
cut value of $P^+P^-<M_X^2=(1.7\, {\rm GeV})^2$ the correction to
$R_{\rm cut}$ is $\sim 0.2\%$; naturally, the effect increases
significantly when lowering the cut, for example, for
$P^+P^-<M_X^2=(1.4\, {\rm GeV})^2$ it amounts to $1.3 \%$. As
obvious from~\eq{NP_power_leading}, if one assumes $C=10$ instead,
the corrections are larger. In this case with $f_3^{\PV}=1$ one
obtains $0.8 \%$ and $4.0 \%$ for the two cuts, respectively. Given
how small the effect from varying $f_3^{\PV}$ is, and its strong
correlation with the variation of $C$, we do not include it in the
overall uncertainty estimate for $R_{\rm cut}$.

\subsubsection*{Parametric Uncertainty}

The numerical values reported above depend on two short distance parameters,
$m_b^{\MSbar}$ and
$\alpha_s^{\MSbar}$. Our default values are
$m_b^{\MSbar}=4.19$ GeV corresponding to a pole mass of
 $m_b\equiv m_b^{\PV}=4.88$ GeV and
$\alpha_s^{\MSbar}(m_Z)=0.119$ corresponding to $\alpha_s^{\MSbar}(m_b)=0.217$.
The number of light flavors was set as $N_f=4$.

To estimate the uncertainty in the computed values of $R_{\rm cut}$ we repeat
the calculation with different assignments of these parameters. The results
are summarized in Table~\ref{table:parametric_uncr}.
\begin{table}
\begin{center}
\begin{tabular}{|l|c|c
|}
  \hline
parameter & Event Fraction $R_{\rm cut}$ ($M_X=1.7$
GeV, $E_0=1$ GeV) &variation (\%)\\
\hline
  $\alpha_s^{\MSbar}(m_Z)=0.115$ & 0.5832 &$-1.8$  \\
  $\alpha_s^{\MSbar}(m_Z)=0.119$ & 0.5941 &default  \\
  $\alpha_s^{\MSbar}(m_Z)=0.122$ & 0.5986 & $0.8$    \\
\hline
  $N_f=3$  & 0.6008 & $+$1.1    \\
  $N_f=4$  & 0.5941 & default    \\
 \hline
$m_b^{\MSbar}=4.14$ & 0.5511      &$-7.2$    \\
$m_b^{\MSbar}=4.19$ & 0.5941      &default    \\
$m_b^{\MSbar}=4.24$ & 0.6325      &$+6.5$    \\
\hline
\end{tabular}
\caption{Results of the event fraction within the range
$P^+P^-<M_X^2=(1.7\, {\rm GeV})^2$ and $E_l>1$ GeV using \eq{int_M_X_cut} and
\eq{double_diff_matched} with different assignments of the short
distance parameters.\label{table:parametric_uncr}}
\end{center}
\end{table}
Clearly, the largest parametric uncertainty is related to the value of the
quark mass.

\subsubsection*{Uncertainty estimates for $P^+<P_{\max}^+$ cut}

Let us now perform the calculation of $R_{\rm cut}$ with a cut on
the small lightcone component $P^+$ in addition to a mild cut on the
lepton energy ($E_l>1$ GeV). As before, the central value is
obtained by numerical integration of the differential distribution
computed using \eq{Pplus_int} with the matching
of~\eq{Gamma_n_lnR_matched_w_constant} over the relevant $P^-$ and
$E_l$ phase space, where the renormalization scale in the matching
coefficient is set to be $\mu=P^-$ and $C=1$. The central value
computed this way for $P^+ < 0.66$ GeV and $E_l>1$ GeV (as in the
Belle measurement~\cite{Bizjak:2005hn}) is $R_{\rm cut}=0.535$.

\begin{figure}[htb]
\begin{center}
\epsfig{file=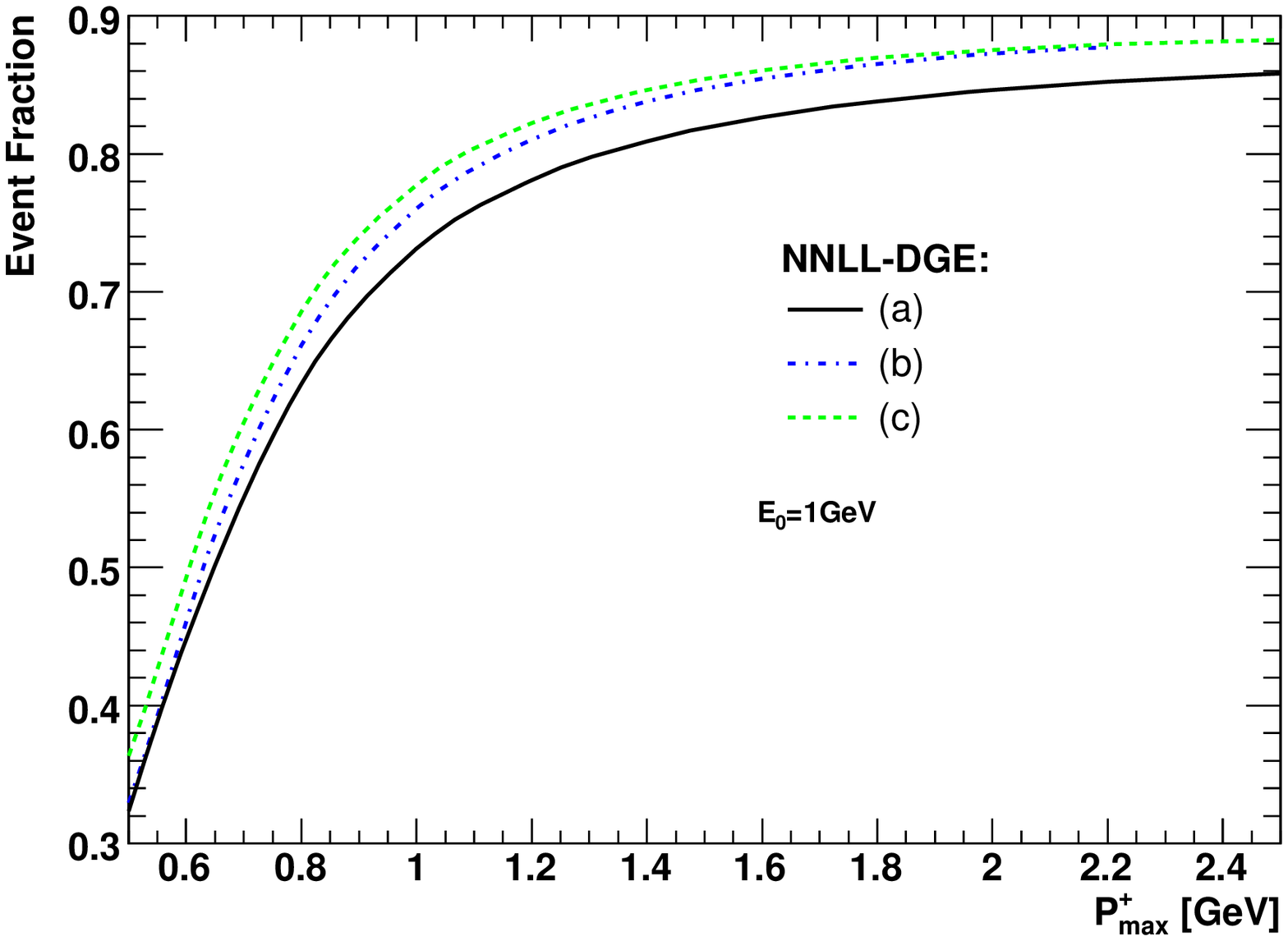,angle=0,width=12.4cm}
\caption{\label{fig:matching_schemes_Pplus} The event fraction $R_{\rm
cut}$ within the range $P^+<P^+_{\max}$ and $E_l>1$ GeV, plotted as a
function of the cut value $P^+_{\max}$. Three different matching schemes
are used: (a) the partially integrated distribution with respect to
$E_l$, namely Eqs. (\ref{int_r_H_lambda_P}) and  (\ref{int_M_X_cut_P_plus})
with \eq{double_diff_matched},
numerically integrated over $P^-$ (full line); (b) the fully
differential distribution computed by \eq{Pplus_int} and the
matching of \eq{Gamma_n_lnR_matched_w_constant}, numerically
integrated over the
 $E_l$ and $P^-$ (dashed line); and (c) same as in (b) but with the matching of
\eq{Gamma_n_lnR_matched} (dotdashed line). }
\end{center}
\end{figure}

The error analysis completely parallels the one applied for the $M_X$ cut.
We therefore briefly summarize the results. The matching scheme uncertainty
is summarized in Fig.~\ref{fig:matching_schemes_Pplus}. Numerical values
for $P^+ < 0.66$ GeV are given in Table~\ref{table:Pplus_matching_scheme_dep}.
\begin{table}
\begin{center}
\begin{tabular}{|c|c|c
|}
  \hline
matching scheme & Event Fraction $R_{\rm cut}$ ($P^+ < 0.66$ GeV,
$E_l>1$ GeV) &
variation(\%)\\
\hline
  (a)   & 0.5118 & $-$4.4\\
  (b)   & 0.5353 & default\\
  (c)   & 0.5659 & +5.7\\
  \hline
\end{tabular}
\caption{Results of the event fraction $R_{\rm cut}$ within the
range $P^+ < P^+_{\max}=0.66$ GeV and $E_l>1$ GeV using different matching
schemes (see Fig.~\ref{fig:matching_schemes_Pplus} and
discussion next to Table~\ref{table:matching_scheme_dep})
with the scale of the coupling in the matching
coefficient set as $\mu=P^-$.\label{table:Pplus_matching_scheme_dep}}
\end{center}
\end{table}

\begin{figure}[t]
\begin{center}
\epsfig{file=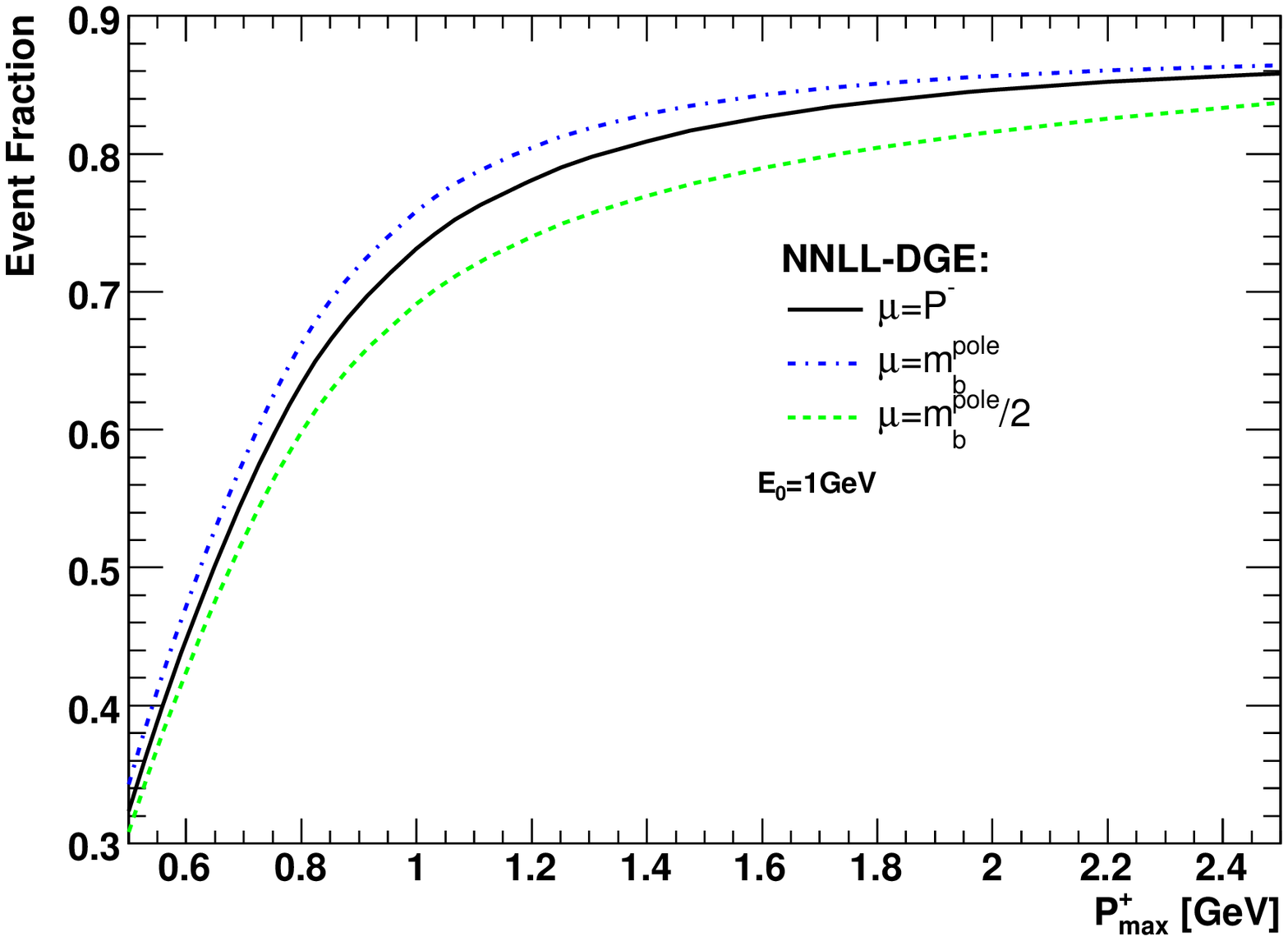,angle=0,width=12.4cm}
\caption{The event fraction $R_{\rm cut}$ within the range
$P^+<P^+_{\max}$ and $E_l>1$ GeV, computed using Eqs.
(\ref{int_r_H_lambda_P}) and  (\ref{int_M_X_cut_P_plus}) and
\eq{double_diff_matched}, plotted as a function of the cut value
$P^+_{\max}$. Different assignments for the argument of the coupling
in the matching coefficient are used. \label{fig:P_plus_scale_dep} }
\end{center}
\end{figure}

The renormalization scale dependence in the matching coefficient is
shown in Fig.~\ref{fig:P_plus_scale_dep}.
The result for $R_{\rm cut}$ of $P^+<P_{\max}^+=0.66$ GeV (and
$E_l>1$ GeV) is summarized
in Table~\ref{table:matching_ren_scale_dep_P_plus}.
\begin{table}
\begin{center}
\begin{tabular}{|l|c|c
|}
  \hline
scale of $\alpha_s$ & Event Fraction $R_{\rm cut}$ for $P^+<0.66$
GeV and $E_l>1$&variation(\%)\\
\hline
  $\mu=P^-$   & 0.5118 & default\\
  $\mu=m_b$   & 0.5385 &  +5.2\\
  $\mu=m_b/2$ & 0.4847 &  $-$5.3\\
  \hline
\end{tabular}
\caption{Results of the event fraction within the range $P^+<0.66$
GeV and $E_l>1$ GeV using \eq{int_M_X_cut_P_plus} and
\eq{double_diff_matched} with different assignments for the scale of
the coupling in the matching coefficient (the $n=1$
moment).\label{table:matching_ren_scale_dep_P_plus}}
\end{center}
\end{table}
The total uncertainty due to unknown higher--order corrections amounts to
$\pm 7.8\%$, somewhat higher than the uncertainty for the $M_X=1.7$ GeV cut, which is
5.6\%.

\begin{figure}[t]
\begin{center}
\epsfig{file=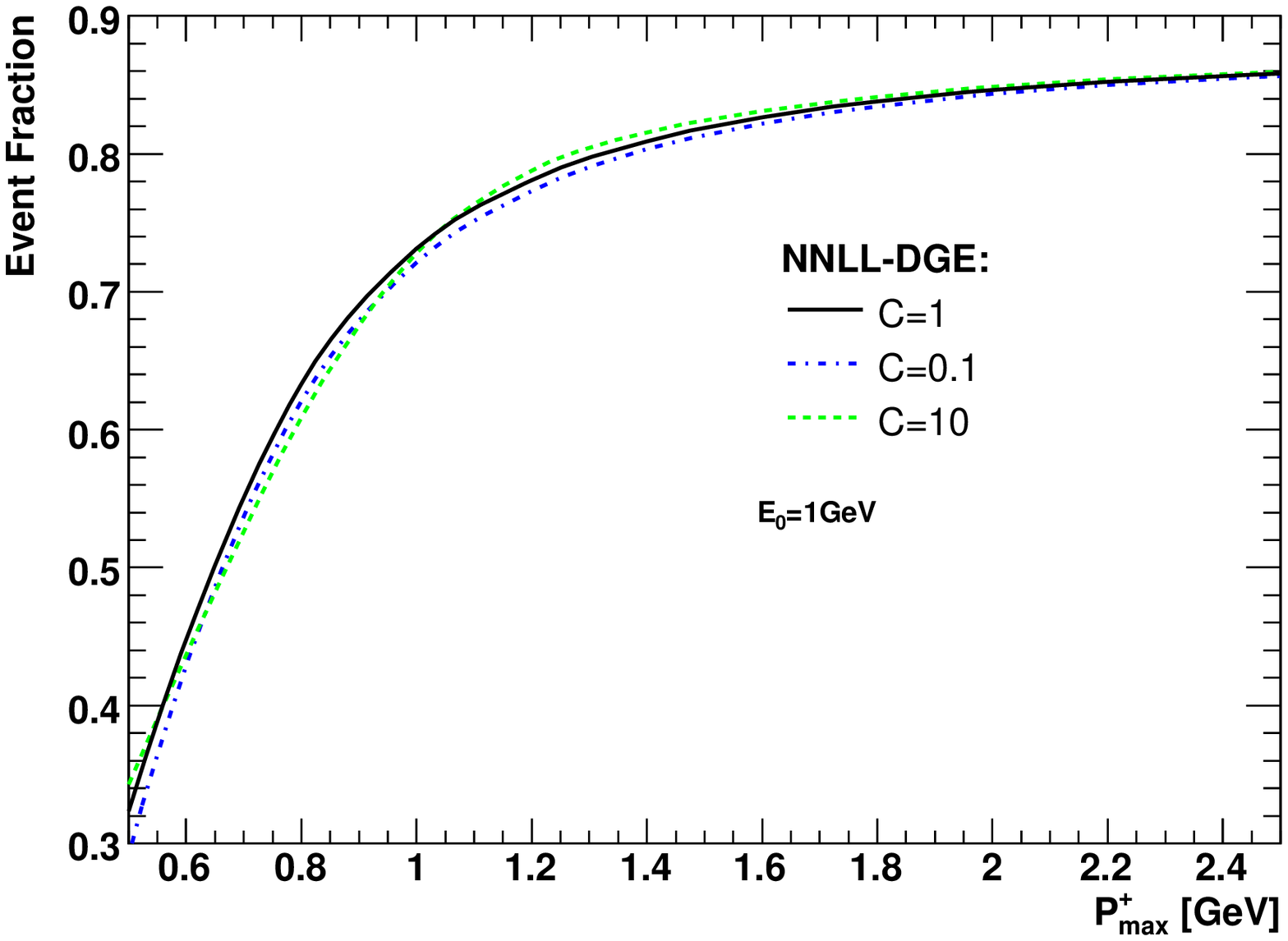,angle=0,width=12.4cm}
\caption{The event fraction $R_{\rm cut}$ within the range
$P^+<P^+_{\max}$ and $E_l>1$ GeV, computed using Eqs.
(\ref{int_r_H_lambda_P}) and  (\ref{int_M_X_cut_P_plus}) and
\eq{double_diff_matched}, plotted as a function of the cut value
$P^+_{\max}$. Different assignments for the parameter $C$, which
controls the shape of the perturbative quark distribution function
(see \eq{C_def}) are used. \label{fig:P_plus_C_dep} }
\end{center}
\end{figure}

Next, the uncertainty in the shape of the quark distribution function is obtained
by modifying $C$ between $C=0.1$ and $C=10$. The result is summarized in
Fig.~\ref{fig:P_plus_C_dep} and Table~\ref{table:C_dep_Pplus}.
\begin{table}
\begin{center}
\begin{tabular}{|l|c|c
|}
  \hline
$C$ & Event Fraction $R_{\rm cut}$ ($P^+<0.66$
GeV, $E_l>1$ GeV) &variation (\%)\\
\hline
  $0.1$ & 0.4971 &$-$2.9  \\
  $1$   & 0.5118 &default\\
  $10$  & 0.4909 &$-$4.1   \\
  \hline
\end{tabular}
\caption{Results of the event fraction within the range $P^+<P^+_{\max}=0.66$
GeV and $E_l>1$ GeV using \eq{int_M_X_cut_P_plus} and
\eq{double_diff_matched} with different assignments of $C$ in
\eq{C_def}.\label{table:C_dep_Pplus}}
\end{center}
\end{table}
Finally, the parametric uncertainty estimates are summarized in
Table~\ref{table:parametric_uncr_Pplus}. The total parametric uncertainty is
$\pm 12.5$.
\begin{table}
\begin{center}
\begin{tabular}{|l|c|c
|}
  \hline
parameter & Event Fraction $R_{\rm cut}$ ($P^+<0.66$
GeV, $E_l>1$ GeV) &variation (\%)\\
\hline
  $\alpha_s^{\MSbar}(m_Z)=0.115$ & 0.4922 & $-$3.8  \\
  $\alpha_s^{\MSbar}(m_Z)=0.119$ & 0.5118 & default  \\
  $\alpha_s^{\MSbar}(m_Z)=0.122$ & 0.5209 &  +1.8   \\
\hline
  $N_f=3$  & 0.5207     & 1.7    \\
  $N_f=4$  & 0.5118     & default    \\
 \hline
$m_b^{\MSbar}=4.14$ & 0.4513       & $-$11.8    \\
$m_b^{\MSbar}=4.19$ & 0.5118       & default    \\
$m_b^{\MSbar}=4.24$ & 0.5650       & $+$10.4    \\
\hline
\end{tabular}
\caption{Results of the event fraction within the range $P^+<0.66$
GeV and $E_l>1$ GeV using \eq{int_M_X_cut_P_plus} and
\eq{double_diff_matched} with different assignments of the short
distance parameters.\label{table:parametric_uncr_Pplus}}
\end{center}
\end{table}
In conclusion, when applying a $P^+<0.66$ cut instead of a
$P^+P^-<M_X^2=(1.7\, {\rm GeV})^2$ cut
the event fraction is reduced by $13\%-14\%$,
while the theoretical uncertainty increases.
This applies separately to all sources of uncertainty, namely the
unknown higher--order corrections (NNLO),
the shape of the quark distribution function ($C$) and the
values of $\alpha_s$ and $m_b$.

\subsection{Extracting $\left\vert V_{ub}\right\vert$\label{sec:extraction_Vub}}

Let us consider first a cut on the invariant mass of the hadronic
system, as well as a mild cut on the charged lepton energy, $E_l>E_0$,  as in
Ref.~\cite{Bizjak:2005hn}. The calculation of the partial branching
fraction takes the form:
\begin{eqnarray}
\label{Delta_cal_B_th}
&&\Delta {\cal B}(\bar{B}\longrightarrow X_u l \bar{\nu}, P^+P^-<M_X^2,\,E_l>E_0 )\nonumber \\
&&\hspace*{80pt}= \tau_B\Gamma_{\tot} \left(\bar{B}\longrightarrow
X_u l \bar{\nu}\right) R_{\rm cut}(P^+P^-<M_X^2,\,E_l>E_0 ).
\end{eqnarray}
The total effect of the cuts amounts to the following event
fraction:
\begin{eqnarray}
R_{\rm cut}&=& \frac{\Gamma(P^+P^-<(1.7\,{\rm GeV})^2\,,\,E_l>
1\,{\rm GeV})
/\Gamma_0}{\Gamma(P^+P^-<M_B^2,\,E_l>0)/\Gamma_0} \nonumber \\
&=&0.615 \pm 0.035_{[\rm  higher\, orders]}\pm 0.014_{[{\rm quark
\,distribution}]}\pm 0.044_{[{\rm parametric} \,(m_b^{\MSbar})]}. \label{E_cut_Mx_error}
\end{eqnarray}
Here the central value was computed using numerical integration of
the fully differential distribution, \eq{diff_M_X_cut} with the
matching of \eq{Gamma_n_lnR_matched_w_constant} where the
renormalization scale was set to $\mu=P^-$. The theoretical error
estimate was done as explained in
Sec.~\ref{sec:Numerical_results_and_uncertainty}: the uncertainty
associated with higher--order corrections is $ \pm 5.6\%$, the one
associated with the quark distribution function (renormalons on and
power correction on the soft scale) is~$ \pm 2.4\%$ and the
parametric uncertainty (dominated by the uncertainty in $m_b$) is~$
\pm 7.4\%$.

Using \eq{E_cut_Mx_error} together with the theoretical result for the total charmless
semileptonic width of \eq{total_width}
in \eq{Delta_cal_B_th} with
the experimental value of the $B$ lifetime $\tau_B=(1.604 \pm
0.011) \,{\rm ps}$, and comparing the result to the Belle measurement\footnote{
The experimental uncertainty is computed by adding the systematic and
statistical errors in quadrature.}
\cite{Bizjak:2005hn},
\begin{equation}
\Delta {\cal B}(\bar{B}\longrightarrow X_u l \bar{\nu}, P^+P^-<(1.7
\,{\rm GeV})^2,\,E_l>1\, {\rm GeV}) =1.24\cdot  10^{-3} \qquad (\pm 13.4\%)
\label{Delta_cal_B_exp}
\end{equation}
we get,
\begin{equation}
\label{V_ub_MX}
\left\vert V_{ub}\right\vert=\Big(4.35 \,\pm\, 0.28_{[{\rm
exp}]}\,\pm\,0.14_{[{\rm th-total} \,(m_b^{\MSbar})]}
 \,\pm\,0.22_{[{\rm th-cuts}]}\Big)\cdot 10^{-3},
\end{equation}
where the three sources of errors quoted separately are: \,(1) the
total experimental error
 \cite{Bizjak:2005hn}
on the measured BF in \eq{Delta_cal_B_exp}; \,\,(2) the error on the
total width, dominated by the uncertainty in $m_b^{\MSbar}$ in
\eq{mb_MSbar_value};\,\, (3) the theory error on the event fraction
associated with the hadronic invariant mass cut, computed
by adding the three sources of uncertainty in~\eq{E_cut_Mx_error} in quadrature.

Let us consider now the cut on the $P^+$ momentum as proposed
 in \cite{Bosch:2004bt} and measured in Ref.~\cite{Bizjak:2005hn} with $P^+<P_{\max}^+=0.66$ GeV.
The calculation of the partial branching fraction takes the form:
\begin{eqnarray}
\label{Delta_cal_B_th_P_plus}
&&\Delta {\cal B}(\bar{B}\longrightarrow X_u l \bar{\nu}, P^+<P_{\max}^+,\,E_l>E_0 )\nonumber \\
&&\hspace*{80pt}= \tau_B\Gamma_{\tot} \left(\bar{B}\longrightarrow
X_u l \bar{\nu}\right) R_{\rm cut}(P^+<P_{\max}^+,\,E_l>E_0 ).
\end{eqnarray}
The effect of the cuts amounts to the following event fraction:
\begin{eqnarray}
R_{\rm cut}&=& \frac{\Gamma(P^+<0.66\,{\rm GeV}\,,\,E_l> 1\,{\rm
GeV})
/\Gamma_0}{\Gamma(P^+<M_B,\,E_l>0)/\Gamma_0} \nonumber \\
&=&0.535 \pm 0.042_{[\rm  higher\, orders]}\pm 0.021_{[{\rm quark
\,distribution}]}\pm 0.063_{[{\rm parametric} \,(m_b^{\MSbar})]}. \label{E_cut_Mx_error_P_plus}
\end{eqnarray}
Comparing \eq{Delta_cal_B_th_P_plus} to the Belle measurement
\cite{Bizjak:2005hn},
\begin{equation}
\Delta {\cal B}(\bar{B}\longrightarrow X_u l \bar{\nu},
P^+<0.66\,{\rm GeV}\,,\,E_l> 1\,{\rm GeV}) =1.10\cdot  10^{-3} \qquad (\pm 17.2\%),
\label{Delta_cal_B_exp_P_plus}
\end{equation}
we get,
\begin{equation}
\label{V_ub_P_plus} \left\vert V_{ub}\right\vert=\Big(4.39 \,\pm\,
0.36_{[{\rm exp}]}\,\pm\,0.14_{[{\rm th-total} \,(m_b^{\MSbar})]}
 \,\pm\,0.38_{[{\rm th-cuts}]}\Big)\cdot 10^{-3}.
\end{equation}

Comparing $\left\vert V_{ub}\right\vert$ from the $P^+<0.66$ GeV cut in
(\ref{V_ub_P_plus}) to the one from $P^+P^-<M_X^2=(1.7\, {\rm GeV})^2$
in~(\ref{V_ub_MX})
and with Ref.~\cite{Bizjak:2005hn} we find:
\begin{itemize}
\item{} The central values of $\left\vert V_{ub}\right\vert$ we extract using
the two cuts agree very well.
\item{} Our results also agree very well with the value quoted by
Belle~\cite{Bizjak:2005hn} based on their own analysis of the $M_X$ cut, while
the central value obtained there for the $P^+$ cut is~$\sim 5\%$
higher\footnote{The values corresponding to the two cuts there
are still consistent within errors.}.
\item{} The $P^+$ cut is characterized by larger
 theoretical uncertainty; this is even more
pronounced than the increase in the experimental error.
The largest sources of theoretical uncertainty, for both cuts, are the value
of the quark mass and the unknown NNLO corrections.
\end{itemize}

\section{Conclusions\label{sec:conclusions}}

We have presented a calculation of the fully differential
distribution in $\bar{B}\longrightarrow X_u l\bar{\nu}$ decays that
relies only on resummed perturbation theory, and yet describes the
meson decay spectrum with approximately correct physical support in
the small--$P^+$ region. What makes this at all feasible is that the
leading renormalon divergence of the Sudakov exponent is fully under
control; the corresponding ambiguity of the resummed moments, which
amounts to a global ${\cal O}(\Lambda/P^-)$ shift, cancels out
exactly upon converting to hadronic variables, a conversion that
itself involves the a priori ambiguous pole-mass definition. While a
regularization (Principal Value) has been arbitrarily chosen at the
intermediate stage, the end result is unambiguous.

\begin{figure}[htb]
\begin{center}
\epsfig{file=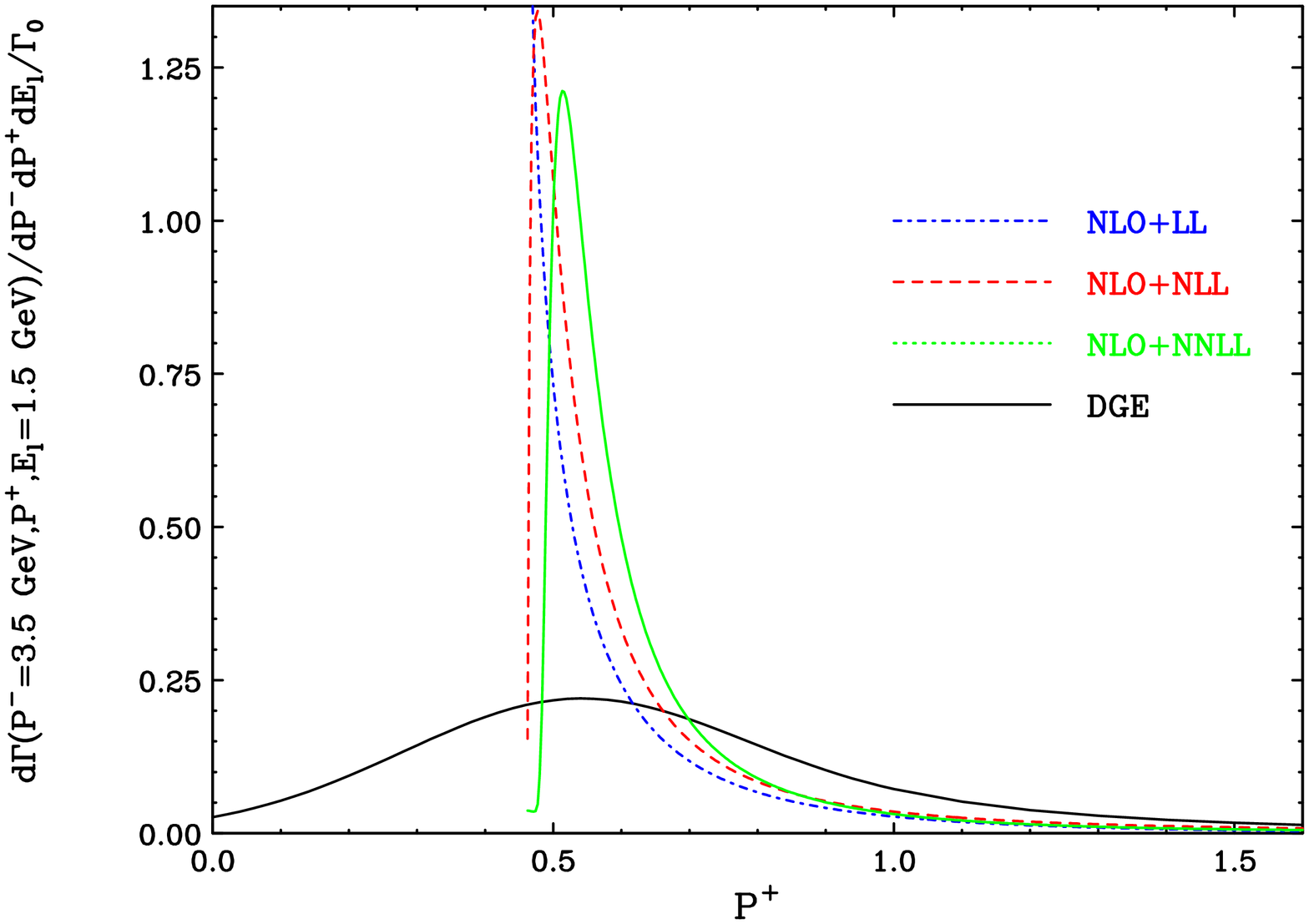,angle=90,width=7.3cm}
\epsfig{file=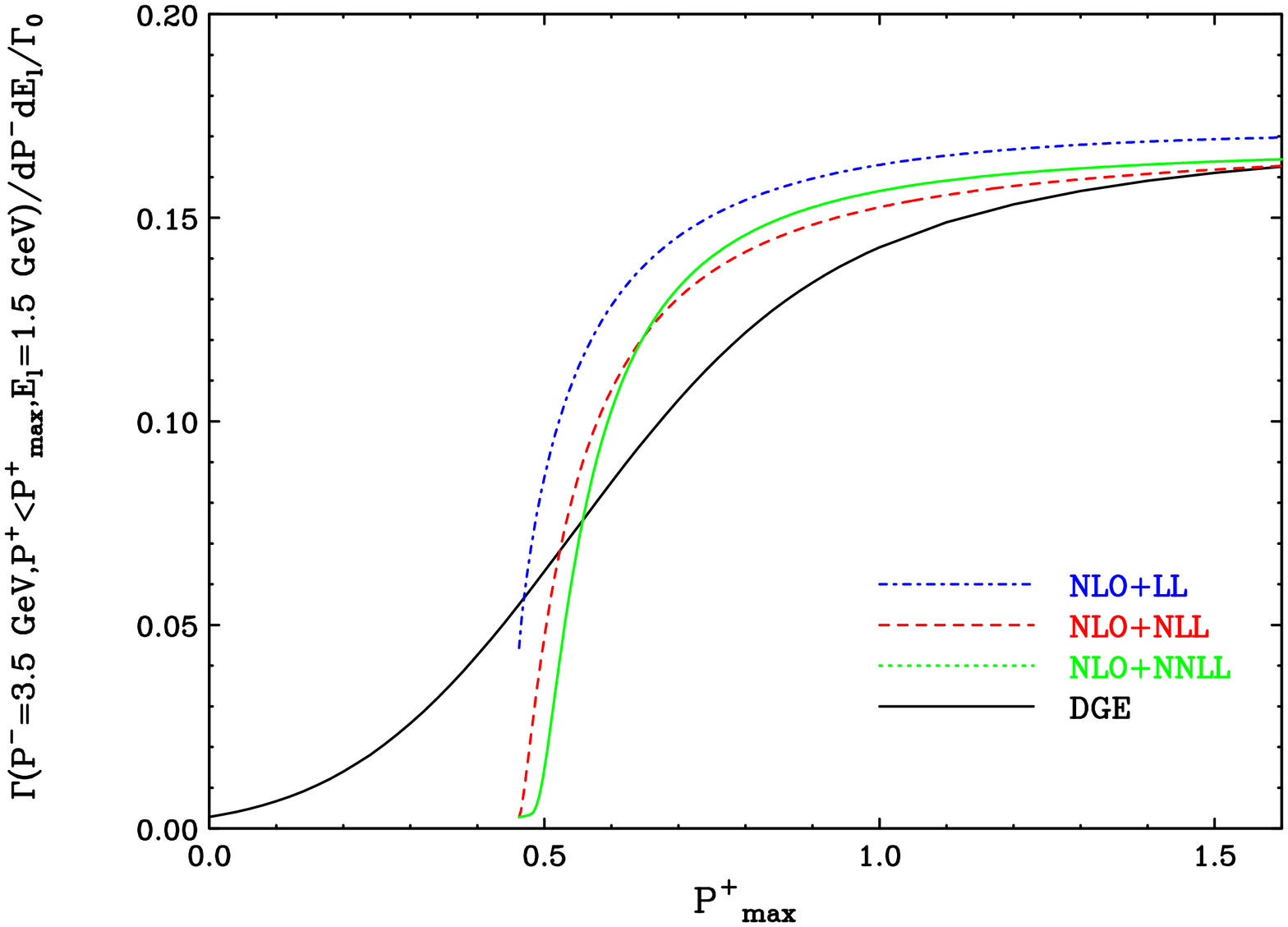,angle=90,width=7.3cm}
\caption{\label{fig:matched_NNLL_spectrum_compared_with_DGE} The
differential (left) and integrated (right) spectrum based on DGE and
on the fixed--logarithmic--accuracy formula of~\eq{NNLL_x_space},
plotted as a function of $P^+$. All calculations are matched to~NLO.
The LL, NLL and NNLL accuracy results are plotted as dotdashed,
dashed and full line, respectively. The DGE curve smoothly extends
below the partonic phase--space limit, filling up the region
$0<P^+<\bar{\Lambda}$. In contrast, the three
fixed--logarithmic--accuracy curves end above $\bar{\Lambda}$, at
the point
 where the resummed results become complex owing to the Landau singularity.}
\end{center}
\end{figure}

In sharp contrast, an attempt to use Sudakov resummation with fixed
logarithmic accuracy
--- no matter how high --- is bound to fail dramatically due to the
unregulated renormalon divergence\footnote{Moreover, in this case it
is not at all obvious which mass should be used in the conversion of
the partonic spectrum to hadronic variables.}. An example is
presented in
Figure~\ref{fig:matched_NNLL_spectrum_compared_with_DGE} that
compares the DGE spectrum at $E_l=1.5$ GeV and $P^-=3.5$  GeV to
that obtained by standard Sudakov resummation with LL, NLL and NNLL
accuracy. Obviously, DGE makes a qualitative difference.

It should be stressed that the detailed description of the spectrum
in the $P^+$--peak region, and especially near the $P^+=0$ endpoint,
is necessarily approximate
and it must rely on certain assumptions.
First of all, there are both perturbative and
non-perturbative contributions. Using moments the latter appear as powers of
${\cal O}(n\Lambda/P^-)$, so they are parametrically small at small~$n$, but they
gradually become important as $n$ increases. Uncertainties corresponding to
the first and hopefully the second powers of $n\Lambda/P^-$ can be avoided
--- but not the high powers.
Towards the small--$P^+$ endpoint ($P^+\longrightarrow 0$) high
moments become relevant and the parametric hierarchy is lost. What
dictates the details of the decay spectrum in this region is the
quark distribution in the meson. Physically one expects that this
distribution would differ from the one in an on-shell quark owing to
the interaction between the b quark and the light degrees of
freedom. It is our working assumption that these differences are
small and therefore better calculation of the latter amounts to
better description of the former. Eventually the difference between
the two should be quantified.

Non-perturbative contributions aside, computing the on-shell decay
spectrum itself within a given regularization scheme for the
renormalons is still an incredible task. In Ref.~\cite{RD} and in
the present work we made significant progress in this direction. As
we saw, the main difficulty is that the Borel sum in \eq{alt_Sud} is
not entirely dominated by the perturbative expansion in powers of
$u$; additional information on the behavior of the Borel integrand
away from the origin is needed. Since the general structure of the
exponent as a function of $n$ is known, the main uncertainty arises
from the soft anomalous dimension function $B_{\cal S}(u)$. The
cancellation of the leading renormalon ambiguity with the pole mass
allows us to determine $B_{\cal S}(u=1/2)$ accurately~\cite{RD}.
Based on the known analytic result in the large--$\beta_0$ limit, we
further assumed here that $B_{\cal S}(u=1)=0$ and that this function
has no other zeros. We then parametrized the contribution from the
region $B_{\cal S}(u\gsim 3/2)$ to facilitate uncertainty estimates.
We found that approximately correct non-perturbative support
properties are recovered in the Principal Value prescription if this
contribution is indeed small. This is demonstrated in
Fig.~\ref{fig:renormalon_at_3halves_influence}.

Under these assumptions we obtained definite predictions for the ${\bar
B}\longrightarrow X_s \gamma$~\cite{RD} and $\bar{B}\longrightarrow X_u
l\bar{\nu}$ spectra. So far comparison with experimental data has
been performed for the first two central moments in ${\bar
B}\longrightarrow X_s \gamma$ with varying cuts, finding very good
agreement~\cite{Gardi:2005mf}. More detailed comparison of both
distributions has a great potential in testing these
assumptions and, eventually, in quantifying power corrections.

The fully differential $\bar{B}\longrightarrow X_u
l\bar{\nu}$ spectrum computed here can be used to extract
$|V_{ub}|$ from any experimental measurement of the partial branching
fraction. The resummation employed is specifically designed to improve the
determination of the partial rate in the small hadronic mass (or small $P^+$)
region, which is the experimentally--favorable way of discriminating charm.
By numerically integrating the differential spectrum over the relevant
phase space we computed the fraction of events measured for a given experimental cut.
 Using recent measurements by Belle~\cite{Bizjak:2005hn}
for the branching fraction in the regions $P^+P^-<M_X^2=(1.7\,{\rm GeV})^2$ and
$P^+<P^+_{\max}=0.66$ GeV (with $E_l> 1\,{\rm GeV}$), we extracted $|V_{ub}|$
and analyzed the theoretical uncertainty. The final results are quoted in Eqs.
(\ref{V_ub_MX}) and (\ref{V_ub_P_plus}), respectively. We find that
\begin{itemize}
\item{} The most important
source of uncertainty is the value of the b quark mass.
It directly influences both the calculation of the total rate and
that of the event fraction associated with the cut.
Better determination of the mass will therefore translate directly
into better measurement of $|V_{ub}|$.
One natural possibility for the b quark mass measurement is
to make use of recent data on the $\bar{B}\longrightarrow X_s
\gamma$ spectrum~\cite{Abe:2005cv,Aubert:2005cu} together with calculated
spectrum~\cite{RD}.
\item{} Another important source of uncertainty is higher--order perturbative corrections.
NNLO calculation of the fully differential $\bar{B}\longrightarrow X_u
l\bar{\nu}$ is not beyond reach. When available, it can be readily matched
to the resummed spectrum. This will significantly
reduce the matching scheme dependence as well
as the renormalization scale dependence in the matching coefficient.
\item{} Finally, under the assumptions discussed above, the uncertainty associated with
the detailed shape of the quark distribution function is found to be quite
small. This is most striking in the $M_X$ cut, see Fig.~\ref{fig:Mx_C_dep}
and Table~\ref{table:C_dep}.
\end{itemize}

\section*{Acknowledgements}

We would like to thank Andrea Banfi for very useful discussions.
JRA acknowledges the support of PPARC (postdoctoral fellowship
PPA/P/S/2003/00281). The work of EG is supported by a Marie Curie
individual fellowship, contract number HPMF-CT-2002-02112.
EG wishes to thank the CERN theory unit and the INT at the
University of Washington for hospitality in the course of this work.

\appendix

\section{Resummation in terms of hadronic and leptonic invariant
masses\label{sec:xi_resummarion}}

Let us choose the following kinematic variables~\cite{BDK}, all
defined at the partonic (perturbative) level:
\begin{eqnarray}
x&\equiv& 2p_b\cdot k_l/m_b^2\\
y&\equiv& 2p_b\cdot q/m_b^2\\
z&\equiv& q^2/m_b^2 \label{z_def}\\
\xi&\equiv& 1-p_{j}^2/m_b^2=y-z \label{xi_def}
\end{eqnarray}
where momentum conservation implies $p_b=p_j+q$, where $p_j$ is the
total momentum of the jet and $q=k_l+k_{\nu}$ is the total momentum
of the leptons. The total semileptonic width into u-quark is:
\begin{equation}
\Gamma_{\tot}=\int_0^1 dz \int_z^1 dx
\int_{x-z(1-1/x)}^{1}\!\!\!\!d\xi\,\frac{d\Gamma(z,\xi,x)}{dzdxd\xi}
=\Gamma_0\,
\left[1+\frac{C_F\alpha_s(m)}{\pi}\left(\frac{25}{8}-\frac{\pi^2}{2}\right)+\cdots
\right]. \label{Gamma_tot_0}
\end{equation}
where $\Gamma_0$ is given by \eq{Gamma0}.

\subsection*{NLO result}

The perturbative expansion of the triple differential width take the
form:
\begin{eqnarray}
\label{triple_diff_xi_z_x}
\frac{1}{\Gamma_0}\frac{d\Gamma(\xi,z,x)}{d\xi d z dx} ={\cal
V}(z,x) \delta(\xi)+ {\cal R}(\xi,z,x)
\end{eqnarray}
where
\begin{eqnarray}
\label{V_and_D_xi} {\cal V}(z,x)&=&v_0(z,x)
+\frac{C_F\alpha_s(m_b)}{\pi}v_1(z,x)+\cdots,\\
{\cal R}(\xi,z,x)&=&\frac{C_F\alpha_s(m_b)}{\pi}
d_1(\xi,z,x)+\cdots.
\end{eqnarray}
The NLO result for the virtual coefficients is:
\begin{eqnarray}
\label{V_coef_xi} v_0(z,x) &=& 12\,(z + 1 -x )\,(x -z)\\ \nonumber
v_1(z,x) &=&-12\left(-2\ln(1-z)^2-\frac{1}{3}\pi^2-\frac{5}{4}-{\rm
Li}_2(z)\right)(x-z-1)(x-z)
\\ \nonumber && -6\ln(1-z)(5x-5z-4)(x-z)
\end{eqnarray}
while the real--emission coefficient is:
\begin{eqnarray}
\label{e1} d_1(\xi,z,x)&=&\bigg\{\frac{6\,(x - z - \xi )\,(  z -
x)\,t\,(  z + \xi - 2 )}{ ((z+\xi)^{2} - 4\,z)^{2}(  1 - \xi
)}\times  \\ \nonumber && (2\,z^{3} - 10\,z^{2} + 4\,z^{2}\,\xi  -
10\,z\,\xi  + 3\,\xi ^{2 }\,z + 13\,z + \xi ^{3} - 3\,\xi ) \\
\nonumber && \mbox{} + {\displaystyle \frac {6\,z\,t\,(  z + \xi - 2
)\,(5 - 6\,z + 2\,z\,\xi  + \xi ^{2} + 2\,z^{2} - 4\,\xi
)}{((z+\xi)^{2} - 4\,z) ^{2}}}  - 6\,(z - x\,z - x\,\xi  + x^{2})\,t
\\ \nonumber && (  z + \xi - 2)\bigg[2\,z^{4} + 6\,z^{3}\,\xi  -
10\,z^{3} + 7\,z^{2 }\,\xi ^{2} + 21\,z^{2} - 28\,z^{2}\,\xi  -
14\,\xi ^{2}\,z - 10 \,z \\ \nonumber && \mbox{} + 4\,z\,\xi ^{3} +
24\,z\,\xi  - 5\,\xi ^{2} + 2\,\xi ^{3 } + \xi ^{4}\bigg] \frac{1}{
((z+\xi)^{2} - 4\,z)^{3}}\bigg\}\mathrm{ln}\left(  {\displaystyle
\frac {1 + t}{  1 - t}} \right) \\ \nonumber && \mbox{} -
{\displaystyle \frac {3\,(7\,z^{2} + 4\,z\,\xi  - 18\,z
 + 6\,\xi  + \xi ^{2})\,(x - z - \xi )\,( - z + x)}{( - 1 + \xi )
\,((z+\xi)^{2} - 4\,z)}}  \\ \nonumber && \mbox{} + {\displaystyle
\frac {3\,z\,( - 9\,z + z\,\xi  + 10 + \xi ^{2} - 3\,\xi
)}{((z+\xi)^{2} - 4\,z)}}  + {\displaystyle \frac {3\,(z - x\,z -
x\,\xi  + x^{2})}{((z+\xi)^{2} - 4\,z)^{2}}}\times\\ \nonumber
&&\hspace*{30pt}\Big(11\,z^{3}
 + 17\,z^{2}\,\xi  - 28\,z^{2} + 7\,\xi ^{2}\,z + 20\,z - 38\,z\,
\xi  + 10\,\xi ^{2} + \xi ^{3}\Big)
\end{eqnarray}
with $t\equiv \sqrt{(z+\xi)^2-4z}/(2-z-\xi)$. The
$\xi\longrightarrow 1$ singular (non integrable) terms in $d_1$,
namely
\begin{equation}
\label{e1_sing} d_1^{\sing} = \left[\frac{3 (1+z-x) (x-z)}{1-\xi}
\left[-4\ln(1-\xi)+8\ln(1-z)-7\right]\right]_{*}
\end{equation}
are regularized as $()_*$ distributions, as in \eq{star_dist}.

\subsection*{Resummation}

The all--order structure of the $\xi\longrightarrow 1$ singular
terms in the triple differential width in the large--$\beta_0$ limit
has been determined in Sec.~3 of Ref.~\cite{BDK} using the Borel
technique. The result is summarized by \eq{sl_logs} above. Sudakov
resummation can now be performed by exponentiating these singular
terms in moment space. Defining moments by
\begin{eqnarray}
\label{jet_mass_mom_def} \frac{d\Gamma_N (z,x)}{dz dx}
 &\equiv &\frac{1}{\Gamma_0}\int_{x-z(1-1/x)}^{1} d\xi \xi^{N-1}
\frac{d\Gamma(\xi,z,x)}{d\xi dz dx}\\\nonumber & \simeq
&\frac{1}{\Gamma_0}\int_{0}^{1} d\xi \xi^{N-1}
\left.\frac{d\Gamma(\xi,z,x)}{d\xi dz
dx}\right\vert_{\xi\longrightarrow 1} +{\cal O}(1/N)
\end{eqnarray}
The singular terms for $\xi \longrightarrow 1$ generate terms that
contain powers of $\ln N$, while ${\cal O}(1/N)$ terms are
neglected. After matching the resummed distribution can terms can be
obtained by the following inverse Mellin formula:
\begin{eqnarray}
\label{sl_logs_resummed} &&\hspace*{-10pt} \frac{1}{\Gamma_0}
\frac{d\Gamma(\xi,z,x)}{dz dx d\xi} =\int_{c-i\infty}^{c+i\infty}
\frac{d N}{2\pi i} \, \left.\frac{d\Gamma_N (z,x)}{dz
dx}\right\vert_{\rm matched} \, \xi^{-N}.
\end{eqnarray}

The matched moment--space result takes the form:
\begin{eqnarray}
\label{matching_xi} && \left.\frac{d\Gamma_N (z,x)}{dz
dx}\right\vert_{\rm matched}
 =
 \left[{\cal V}(\lambda,x)\,+ \widetilde{\Delta D_1}^N(\lambda,x)\frac{C_F\alpha_s(m_b)}{\pi}
+ {\cal O}(\alpha_s^2) \right] \times \\ \nonumber && \exp\bigg\{
\frac{C_F}{\beta_0}\int_0^{\infty}\frac{du}{u} \, T(u)
\left(\frac{\Lambda^2}{m_b^2}\right)^u
\bigg[B_{\cal
S}(u)(1-z)^{2u}\Gamma(-2u)\left(\frac{\Gamma(N)}{\Gamma(N-2u)}-\frac{1}{\Gamma(1-2u)}\right)
\\\nonumber && \hspace*{160pt}
-B_{\cal
J}(u)\Gamma(-u)\left(\frac{\Gamma(N)}{\Gamma(N-u)}-\frac{1}{\Gamma(1-u)}\right)
\bigg] \bigg\}
\end{eqnarray}
To perform matching at ${\cal O}(\alpha_s)$ one needs to compute the
moments at this order
\begin{eqnarray}
D_1^N(\lambda,x) &\equiv &\frac{1}{\Gamma_0}\int_{x-z(1-1/x)}^{1}
d\xi \xi^{N-1} \frac{d\Gamma(\xi,z,x)}{d\xi dz dx}\,d_1(\xi,z,x),
\end{eqnarray}
where $d_1(\xi,z,x)$ is given by \eq{e1}. We find:
\begin{eqnarray}
\label{} D_1^N(\lambda,x)\!&=&\!
v_0(z,x)\bigg\{\frac12\bigg[\Psi_1(N)-\frac{1}{6}\pi^2-(\gamma_E+\Psi(N))^2\bigg]
  \\ \nonumber
&&\hspace*{30pt} +
\left(\frac{7}{4}-2\ln(1-z)\right)(\Psi(N)+\gamma_E)\bigg\}
\,+\,{\cal O}({1}/{N}).
\end{eqnarray}
These terms, including the constant for $N\longrightarrow \infty$,
are reproduced by the expansion of the exponent in \eq{matching_xi},
so $\widetilde{\Delta D_1}^N(\lambda,x)$ contributes only at the
level of ${\cal O}(1/N)$ corrections ($\widetilde{\Delta
D_1}^{\infty}(\lambda,x)=0$). Therefore, \eq{matching_xi} with the
coefficients of ${\cal V}(\lambda,x)$ given by \eq{V_coef_xi} can
readily be expanded to predict the log-enhanced terms at ${\cal
O}(\alpha_s^2)$. Converting this result back to $\xi$ space, we get:
\begin{eqnarray}
\label{logs_xi_z} \left. \frac{1}{\Gamma_0}
\frac{d\Gamma(\xi,z,x)}{dz dx d\xi}\right\vert_{\xi\longrightarrow
1}\!\!\!\!\! &=& v_0(z,x)+ \Bigg\{\bigg[ -
\,\left(\frac{\ln(1-\xi)}{1-\xi}\right)_{*} + \left( - {  \frac
{7}{4}}
 + 2\,\ln(1 - z)\right)\,\left(\frac{1}{1-\xi}\right)_{*}\bigg]\,\times
\nonumber \\
&&
 v_0(z,x) \,+\,
 v_1(z,x)\Bigg\}
\frac{C_F\alpha_s(m_b)}{\pi}+ \Bigg\{\bigg[
\frac{C_F^{2}}{2}\,\left(\frac{\ln^3(1-\xi)}{1-\xi}\right)_{*}
\\ \nonumber
&&
  + \left(\bigg( - 3\,\ln(1 - z) + {  \frac {21}{
8}} \bigg)\,C_F^{2} + \bigg({  \frac {11\,C_A}{8} - {  \frac
{N_f}{4}}  } \bigg)\,C_F\right)
\,\left(\frac{\ln^2(1-\xi)}{1-\xi}\right)_{*}\\ \nonumber &&
 + \bigg(
\bigg(4\,\ln(1 - z)^{2} - {  \frac {\pi ^{2}}{6}}
 + {  \frac {49}{16}}  - 7\,\ln(1 - z)\bigg)\,
C_F^{2} \mbox{} + \bigg(\bigg({  \frac {2}{3}} \,\ln(1 - z)
\\ \nonumber &&-
{  \frac {13}{72}} \bigg)\,N_f + \bigg({ \frac {\pi ^{2}}{12}}  - {
\frac {11}{3}} \,\mathrm{ ln}(1 - z) + {  \frac {95}{144}}
\bigg)\,C_A\bigg)\,
C_F\bigg)\left(\frac{\ln(1-\xi)}{1-\xi}\right)_{*} \mbox{} \\
\nonumber && + \bigg(\bigg( - {  \frac {\pi ^{2}}{6}}  + {  \frac
{1}{3}} \,\ln(1 - z)\,\pi ^{2} - {  \frac {3}{32}}  - {  \frac
{1}{2}} \, \zeta_3\bigg)\,C_F^{2}   \\ \nonumber && + \left( - {
\frac {\pi ^{2}}{36}}  - { \frac {2}{9}} \,\ln(1 - z) - {  \frac
{1}{3} } \,\ln(1 - z)^{2} + {  \frac {85}{144}} \right)\, C_F N_f
\\ \nonumber && \mbox{} + \bigg({  \frac {17}{9}} \,\ln(1 - z) + {
\frac {17\,\pi ^{2}}{72}}  + { \frac {1}{4}} \,\zeta_3 + {  \frac
{11}{6}} \, \ln(1 - z)^{2}- {  \frac {905}{288}}
\\ \nonumber &&
-{  \frac {1}{6}} \,\ln(1 - z)\,\pi ^{2}\bigg)\, C_F
C_A\bigg)\left(\frac{1}{1-\xi}\right)_{*}  \bigg]v_0(z,x) \mbox{} +
\bigg[ - \left(\frac{\ln(1-\xi)}{1-\xi}\right)_{*}
 \\ \nonumber &&
+ \bigg( - { \frac {7}{4}}  + 2\,\ln(1 - z)\bigg)\,
\left(\frac{1}{1-\xi}\right)_{*}\bigg]\,
\,C_F^{2}\,v_1(z,x)\Bigg\}\left(\frac{\alpha_s(m_b)}{\pi}\right)^{2}
\end{eqnarray}
Changing variable in \eq{logs_xi_z} to the lightcone momenta $\rho$
and $\lambda$ we get
\begin{eqnarray}
\label{logs_lambda_rho} \left. \frac{1}{\Gamma_0}
\frac{d\Gamma(\lambda,\rho,x)}{d\lambda d\rho
dx}\right\vert_{\rho\longrightarrow 0} &=&
v_0(z=1-\lambda,x)+\Bigg\{\bigg[ - \,\left(\frac{\ln
\rho}{\rho}\right)_{} + \left(\ln(\lambda ) - {  \frac
{7}{4}}\right)\left(\frac{1}{\rho}\right)_{} \bigg]\,\times
\nonumber \\ \nonumber && \hspace*{-50pt} v_0(z=1-\lambda,x)\,+\,
 v_1(z=1-\lambda,x)\Bigg\}
 \,\left(\frac{C_F\,\alpha_s(m_b)}{\pi}\right)
+ \bigg\{\bigg[{  \frac {1 }{2}} \,C_F^{2}\,\left(\frac{\ln^3
\rho}{\rho}\right)_{} \\  && \hspace*{-50pt} \mbox{} + \bigg( - {
\frac {3}{2}} \,C_F^{2}\, \ln(\lambda ) + {  \frac {21\,C_F^{2}
}{8}} + \left({  \frac {11\,C_A}{8}} - {  \frac {N_f}{4}}
\right)\,C_F\bigg)\, \left(\frac{\ln^2 \rho}{\rho}\right)_{}
\\ \nonumber && \hspace*{-50pt}
 + \bigg( - {  \frac {1}{2}} \,
C_F^{2}\,\ln(\lambda )^{2} \mbox{} + \bigg( - {  \frac
{7\,C_F^{2}}{4}}  + \bigg( {  \frac {N_f}{6}}  - {  \frac {
11\,C_A}{12}} \bigg)\,C_F\bigg)\,\ln(\lambda )
\\ \nonumber && \hspace*{-50pt}
+ \bigg(
 - {  \frac {\pi ^{2}}{6}}  + {  \frac {
49}{16}} \bigg)\,C_F^{2}
 + \bigg( - {  \frac {13\,N_f}{72}}  + \bigg(
{  \frac {95}{144}}  + {  \frac {\pi ^{2} }{12}}
\bigg)\,C_A\bigg)\,C_F\bigg)\left(\frac{\ln \rho}{\rho}\right)_{}
\\ \nonumber && \hspace*{-50pt}
 +\bigg(
{  \frac {3}{2}} \,C_F^{2}\,\ln( \lambda )^{3} \mbox{} + \bigg( - {
\frac {35\,C_F^{2}}{8}}  + \left(
 - {  \frac {11\,C_A}{24}}  + {
\frac {N_f}{12}} \right)\,C_F\bigg)\,\ln(\lambda )^{2}
\\ \nonumber && \hspace*{-50pt}
\mbox{} + \bigg(\bigg({  \frac {\pi ^{2}}{6}}  + {  \frac {49}{16}}
\bigg)\,C_F^{2} + ( - {  \frac {29\,N_f}{72}}  + \bigg({ \frac
{367}{144}}  - {  \frac {\pi ^{2}}{12}} \bigg)\,
C_A)\,C_F\bigg)\,\ln(\lambda )
\\ \nonumber && \hspace*{-50pt}
\mbox{} + \bigg( - {  \frac {\pi ^{2}}{6}}  - {  \frac {1}{2}}
\,\zeta_3 - {  \frac { 3}{32}} \bigg)\,C_F^{2} + \bigg(\bigg( - {
\frac {\pi ^{2} }{36}}  + {  \frac {85}{144}} \bigg)\,C_FN_f
\\ \nonumber
&& \hspace*{-50pt}+ \left( {  \frac {1}{4}} \,\zeta_3 + {  \frac {
17\,\pi ^{2}}{72}}  - {  \frac {905}{288}} \right)\,
C_FC_A\bigg)\,\bigg)\left(\frac{1}{\rho}\right)_{}\bigg]
v_0(z=1-\lambda,x)
\\ \nonumber && \hspace*{-50pt}
\mbox{} + \bigg[ - \left(\frac{\ln \rho}{\rho}\right)_{} +\left(
\ln(\lambda ) - {  \frac {7}{4}}\right)
\left(\frac{1}{\rho}\right)_{} \bigg]
\,C_F^{2}\,v_1(z=1-\lambda,x)\bigg\}\left(\frac{\alpha_s(m_b)}{\pi}\right)^{2}
\end{eqnarray}
These logarithms agree exactly with those predicted using the
variables $r$ and $\lambda$, \eq{triple_diff_r} with \eq{D_r_NNLO}.
Note, however, that the {\em separation} into virtual and real is
different when different variables are used to define the $()_{*}$
distributions.

\section{Calculation of the real-emission contribution at NLO in moment
space\label{sec:matching_coeff}}

\subsubsection*{Singular part}

For the singular part of \eq{sing_part} we get
\begin{eqnarray}
\label{C_1_sing} K_1^{n\,\sing} (\lambda,x)&= & -w_0(\lambda,x)
\int_{0}^{r_m} dr \left(1-r\right)^{n-1}\left[{\displaystyle \frac
{\ln (r) + \frac74}{r}}\right]_{*}
\\ \nonumber
&=& -w_0(\lambda,x)\bigg[\frac12 \ln^2(r_m)
+i_0^n(r_m)+\frac74\Big(\ln(r_m)+i_1^n(r_m)\Big)\bigg],
\end{eqnarray}
where we applied the definition of the $()_{*}$ distribution in
\eq{star_dist} and defined
\begin{eqnarray}
\label{i0} i_0^n(r_m) &\equiv& \int_0^{r_m}dr \bigg[(1-r)^{n-1} -1
\bigg]\frac {\ln r}{r}\\ \nonumber
           &=& (n-1)r_m \bigg[\, _4F_3([1, 1, 1, 2-n],[2, 2, 2],r_m)
           \,-\, \ln(r_m)\, _3F_2([1, 1, 2-n],[2, 2],r_m)\bigg]
           \\ \nonumber
           &=& \frac12\left[ \frac{\pi^2}{6}-\Psi_1(n)+\left(\Psi(n)+\gamma_E\right)^2-\ln^2 (r_m)
           \right]
           \!-\! \ln(r_m) s_1(n,1-r_m)
           \!-\!s_2(n,1-r_m)
\end{eqnarray}
and
\begin{eqnarray}
\label{i1} i_1^n(r_m) &\equiv& \int_0^{r_m}dr \bigg[(1-r)^{n-1} -1
\bigg]\frac {1}{r}\\ \nonumber
           &=& -r_m\,(n-1)\,\times\, _3F_2([1, 1, 2-n],[2, 2],r_m) \\ \nonumber
           &=& -\Big[\ln(r_m)+\Psi(n)+\gamma_E\Big]-s_1(n,1-r_m)\,,
\end{eqnarray}
where $s_{1,2}$ denote the infinite sums:
\begin{eqnarray}
\label{s12} s_1(n,1-r_m) &\equiv & \sum_{k=0}^{\infty}
\frac{(1-r_m)^{n+k}}{n+k}\\ \nonumber
&=&\frac{(1-r_m)^n}{n}\,_2F_1([1,n],[n+1],1-r_m)=(1-r_m)^n\,{\rm
LerchPhi}(1-r_m,1,n)
\\ \nonumber s_2(n,1-r_m) &\equiv & \sum_{k=0}^{\infty}
\frac{(1-r_m)^{n+k}}{n+k}\bigg[\Psi(n+k)-\Psi(n)\bigg].
\end{eqnarray}
Alternative expressions, useful for small $r_m$, are:
\begin{eqnarray}
s_1(n,1-r_m)&=& -\Big(\ln(r_m)+\Psi(n)+\gamma_E\Big) \,-\,
\sum_{k=1}^{\infty}
\frac{(-r_m)^k\,\Gamma(n)}{\Gamma(k+1)\,k\,\Gamma(n-k)}\\
\nonumber s_2(n,1-r_m)&=&
\frac{1}{2}\Big(\ln(r_m)+\Psi(n)+\gamma_E\Big)^2+\frac{1}{12}\pi^2
-\frac{1}{2}\Psi_1(n) \,+\sum_{k=1}^{\infty}
\frac{(-r_m)^k\,\Gamma(n)}{\Gamma(k+1)\,k^2\,\Gamma(n-k)}.
\end{eqnarray}

>From the definitions of $i_{0,1}$ and \eq{C_1_sing} it follows that
the first moment ($n=1$) of the singular part, which is relevant for
the total width, is simply:
\begin{eqnarray}
\label{C_1_sing_n_eq_1} K_1^{n=1\,\sing} (\lambda,x)&=&
-w_0(\lambda,x)\bigg[\frac12 \ln^2(r_m) +\frac74\ln(r_m) \bigg].
\end{eqnarray}
We note that $i_{0,1}^n(r_m)={\cal O}(r_m)$, so for any moment the
small--$r_m$ limit ($x\longrightarrow 1$) is particularly simple:
\[
K_1^{n\,\sing} (\lambda,x\longrightarrow 1) \longrightarrow
K_1^{n=1\,\sing} (\lambda,x \longrightarrow 1).
\]
For fixed $r_m$ the sums $s_{1,2}(n,1-r_m)$ in \eq{s12} vanish
exponentially at large $n$. Thus, in the final expression in Eqs.
(\ref{i0}) and (\ref{i1}) only the square brackets contains terms
that do not vanish at $n\longrightarrow \infty$. The logarithms of
$r_m$ appearing explicitly in \eq{C_1_sing} cancel against the ones
in $i_1$ and $i_0$, yielding
\begin{eqnarray}
\label{C_1_sing_explicit} K_1^{n\,\sing} (\lambda,x) &=&
-w_0(\lambda,x)\bigg\{ \frac12\left[
\frac{\pi^2}{6}-\Psi_1(n)+\left(\Psi(n)+\gamma_E\right)^2
\right]-\frac74(\Psi(n)+\gamma_E)\\ \nonumber && -s_1(n,1-r_m)
\ln(r_m) -s_2(n,1-r_m) -\frac74 s_1(n,1-r_m) \bigg\},
\end{eqnarray}
so the $r_m$ dependence of $K_1^{n\,\sing} (\lambda,x)$ appears only
though ${\cal O}(1/n)$ terms. The terms is the curly brackets in
\eq{C_1_sing_explicit} in the first and second lines can interpreted
as originating in the two following integrals, respectively,
\begin{eqnarray}
\label{C_1_sing_explicit_interpretation} K_1^{n\,\sing} (\lambda,x)
&=& -w_0(\lambda,x)\bigg\{ \int_0^1(1-r)^{n-1}\left(\frac{\ln
r}{r}+\frac74\frac{1}{r}\right)_{*} \\ \nonumber && \hspace*{100pt}-
\int_{r_m}^1(1-r)^{n-1}\left(\frac{\ln
r}{r}+\frac74\frac{1}{r}\right)
 \bigg\}.
\end{eqnarray}

\subsubsection*{Regular part}

Let us consider now the regular part of \eq{C1_reg_def}. The first
moment ($n=1$), which is useful in calculating the integrated width,
is fairly simple:
\begin{eqnarray}
\label{C1_reg_n_eq_1} K_1^{n=1\,\,\reg} (\lambda,x) &=& 6\,{\rm
Li}_2\left(1 - {\displaystyle \frac {1 - x}{\lambda }}
\right)\bigg[8\,\lambda  + 36\, \lambda ^{5} + 232\,\lambda ^{3} -
190\,\lambda ^{4} - 90\, \lambda ^{2} \\ \nonumber &&
\hspace*{-30pt} \mbox{} - \bigg(4 - 196\,\lambda ^{3} - 76\,\lambda
+ 225\,\lambda ^{ 2} + 40\,\lambda ^{4}\bigg)\,(  1 - x) \\
\nonumber && \hspace*{-30pt} \mbox{} + (34\,\lambda  - 33\,\lambda
^{2} + 8\,\lambda ^{3} - 8) \,(1 - x)^{2}\bigg]\mbox{} - \ln
\left({\displaystyle \frac {1 - x}{\lambda }} \right)\,\frac{1 -
x}{\lambda}\, \bigg[216\,\lambda ^{5} - 1140\,\lambda ^{4}
  \\ \nonumber && \hspace*{-30pt}
\mbox{}+ 84\,\lambda  + 1416\,\lambda ^{3} - 612\,\lambda ^{2} -
3\,\bigg( - 199\, \lambda ^{3} + 211\,\lambda ^{2} + 2 - 70\,\lambda
+ 44\,\lambda
 ^{4}\bigg)\,( 1 - x) \\ \nonumber && \hspace*{-30pt}
\mbox{} - 2\,( 1 - \lambda )\,\bigg(\lambda ^{2} - 2\,\lambda  +
2\bigg) \,( 1 -x)^{2}\bigg] \mbox{} + \frac{1}{6\,\lambda} \bigg[
540\,\pi ^{2}\,\lambda ^{3} - 216\,\pi ^{2}\,\lambda ^{ 6} -
1392\,\pi ^{2}\,\lambda ^{4}
 \\ \nonumber && \hspace*{-30pt}
\mbox{} + 1140\,\pi ^{2}\,\lambda ^{5} - 48\,\pi ^{2}\,\lambda ^{
2}\bigg] + (1 - x)\,\bigg[4\,\pi ^{2} - 516\,\lambda  - 196\,\pi
^{2}\, \lambda ^{3} - 76\,\pi ^{2}\,\lambda   \\ \nonumber &&
\hspace*{-30pt} \mbox{} + 225\,\pi ^{2}\,\lambda ^{2} + 216\,\lambda
^{4} - 1140\,\lambda ^{3} + 1365\,\lambda ^{2} + 60 + 40\,\pi
^{2}\,\lambda ^{4}\bigg] - \frac{(1 - x)^{2}}{2\,\lambda}\,\bigg[ -
624\,\lambda   \\ \nonumber && \hspace*{-30pt} \mbox{} + 39 +
16\,\pi ^{2}\,\lambda ^{4} + 1995\,\lambda ^{2} - 1812 \,\lambda
^{3} + 372\,\lambda ^{4} - 16\,\pi ^{2}\,\lambda  - 66 \,\pi
^{2}\,\lambda ^{3} + 68\,\pi ^{2}\,\lambda ^{2}\bigg]\\ \nonumber &&
- \frac{1}{3\,\lambda}\,( 1 - \lambda )\,\bigg[29\,\lambda ^{2}
 - 70\,\lambda  + 43\bigg].
\end{eqnarray}

To express a general $n$-th moment of the regular part, we write
\begin{equation}
\label{c_1_reg_sum} k_1^{n \,\reg}(\lambda,r,x)=\sum_{i=-4}^2
f_i(\lambda,x) (1-r)^i\ln (r)+\sum_{i=-3}^2
\tilde{f}_i(\lambda,x)(1-r)^i,
\end{equation}
where we consider $n$ complex with sufficiently large real part,
${\rm Re} (n)> 3$, to obtain convergence of the defining moment
integral in \eq{C1_moments} for individual terms in the sum, and
where
\begin{eqnarray}
\label{f} f_{-4}(\lambda,x)&=& \, - 36\,(\lambda  - 6)\,
( \lambda-1
 )^{2} \,( \lambda-1  + x)^{2}\\ \nonumber
f_{-3}(\lambda,x)&=&\, 36\,( \lambda-1 )\,(3\,\lambda ^{2}
 - 17\,\lambda  + 9)\,x^{2} + 36\,(7\,\lambda ^{2} - 40\,\lambda
 + 18)\,( \lambda-1 )^{2}\,x \\ \nonumber
&&\mbox{} + 36\,( \lambda-1 )^{3}\,(4\,\lambda ^{2} - 23\,
\lambda  + 9)
\\ \nonumber
f_{-2}(\lambda,x)&=&\,  - 6\,(19\,\lambda ^{3} - 102\,\lambda ^{2} -
34 + 112\,\lambda )\,x^{2} \\ \nonumber &&\mbox{} - 12\,( - 1 +
\lambda )\,(29\,\lambda ^{3} - 159\,\lambda
 ^{2} + 148\,\lambda  - 38)\,x - 4008\,\lambda ^{3} - 1608\,
\lambda \\
\nonumber &&+ 1812\,\lambda ^{4}  - 240\,\lambda ^{5} + 252 +
3792\,\lambda ^{2}
\\
\nonumber f_{-1}(\lambda,x)&=&\, 6\,(8\,\lambda ^{3} - 33\,\lambda
^{2}
 - 8 + 34\,\lambda )\,x^{2} + 6\,( - 212\,\lambda ^{3} + 291\,
\lambda ^{2} - 144\,\lambda  + 20 + 40\,\lambda ^{4})\,x \\
\nonumber &&\mbox{} + 2616\,\lambda ^{3} + 708\,\lambda  -
1380\,\lambda ^{4}
 + 216\,\lambda ^{5} - 72 - 2088\,\lambda ^{2}
\\
\nonumber f_{0}(\lambda,x)&=&\,  - 6\,( \lambda-1 )\,(\lambda
^{2}
 + 2 - 2\,\lambda )\,x^{2} - 6\,( - 58\,\lambda ^{3} - 34\,
\lambda  + 71\,\lambda ^{2} + 6 + 14\,\lambda ^{4})\,x \\ \nonumber
&&\mbox{} - 990\,\lambda ^{3} + 24 - 114\,\lambda ^{5} + 600\,
\lambda ^{4} - 228\,\lambda  + 708\,\lambda ^{2}
\\
\nonumber f_{1}(\lambda,x)&=&\, \,12\,( \lambda-1 )\,\lambda \,(
\lambda ^{2} + 2 - 2\,\lambda )\,x + 6\,( \lambda-1 )\, \lambda
\,(6\,\lambda ^{3} - 19\,\lambda ^{2} + 18\,\lambda  - 6)
\\
\nonumber f_{2}(\lambda,x)&=& \, - 6\,\lambda ^{2}\,( \lambda-1
) \,(\lambda ^{2} + 2 - 2\,\lambda )
\end{eqnarray}
and
\begin{eqnarray}
\label{tilde_f} \tilde{f}_{-3}(\lambda,x)&=&\,  - 36\,(\lambda  -
6)\,(  \lambda- 1 )^{2}\,x^{2} - 72\,(\lambda  - 6)\,( \lambda- 1
)^{3} \,x + 216 + 360\,\lambda ^{4}  \nonumber
\\
 &&\mbox{} - 36\,\lambda ^{5} + 1440\,\lambda ^{2} - 900\,\lambda  -
1080\,\lambda ^{3}
\\ \nonumber
\tilde{f}_{-2}(\lambda,x)&=&\, 18\,( \lambda - 1 )\,(5\,\lambda ^{2}
- 27\,\lambda  + 12)\,x^{2} + 108\,(2\,\lambda ^{2} - 11\, \lambda +
4)\,( \lambda- 1  )^{2}\,x \\  \nonumber && \mbox{} + 126\,\lambda
^{5} - 1080\,\lambda ^{4} + 2700\,\lambda ^{3} - 216 - 2880\,\lambda
^{2} + 1350\,\lambda
\\ \nonumber
\tilde{f}_{-1}(\lambda,x)&=&\,  - 6\,(12\,\lambda ^{3} - 58\,
\lambda ^{2} + 60\,\lambda  - 19)\,x^{2} \\
\nonumber && \mbox{} - 6\,(  \lambda - 1)\,(41\,\lambda ^{3} -
209\,\lambda ^{2} + 174\,\lambda  - 46)\,x + 162 - 1008\,\lambda  +
1302\,
\lambda ^{4}  \\
\nonumber && \mbox{} + 2472\,\lambda ^{2}- 2748\,\lambda ^{3} -
180\,\lambda ^{5}
\\ \nonumber
\tilde{f}_{0}(\lambda,x)&=&\, 3\,(  \lambda- 1 )\,(6\,\lambda
^{2} - 14\,\lambda  + 7)\,x^{2} \\
\nonumber && \mbox{} + 3\,(21 - 120\,\lambda  - 210\,\lambda ^{3} +
262\, \lambda ^{2} + 44\,\lambda ^{4})\,x - 42 + 342\,\lambda  -
804\,
\lambda ^{4} \\
\nonumber && \mbox{}  - 1053\,\lambda ^{2} + 1422\,\lambda ^{3} +
135\,\lambda ^{5}
\\ \nonumber
\tilde{f}_{1}(\lambda,x)&=&\, \, - 6\,\lambda \,(5\,\lambda  - 7)
\,(  \lambda- 1 )^{2}\,x - 63\,\lambda  + 261\,\lambda ^{4} +
252\,\lambda ^{2} - 396\,\lambda ^{3} - 54\,\lambda ^{5}
\\ \nonumber
\tilde{f}_{2}(\lambda,x)&=&\, \,3\,\lambda ^{2}\,(3\,\lambda  - 7
)\,( \lambda- 1  )^{2}
\end{eqnarray}
Despite appearance $k_1^{n\,\reg} (\lambda,x)$  is regular also for
$r\longrightarrow 1$ (recall that the phase space is
$r<(1-x)/\lambda<1$): there are cancellations between the would-be
singular terms in \eq{c_1_reg_sum}.

Performing the integration in \eq{C1_reg_def} we obtain the
following moments:
\begin{eqnarray}
\label{C_1_reg} K_1^{n\,\reg} (\lambda,x) &=& \sum_{i=-4}^{2}
f_i(\lambda,x) I_{n+i}(r_m)+\sum_{i=-3}^{2}
\tilde{f}_i(\lambda,x)J_{n+i}(r_m)
\end{eqnarray}
where we define:
\begin{eqnarray}
\label{Jn} J_n(r_m) &\equiv& \int_0^{r_m}dr
(1-r)^{n-1}=\frac{1}{n}\left(1-(1-r_m)^n\right),
\end{eqnarray}
and
\begin{eqnarray}
\label{In} I_n(r_m) &\equiv& \int_0^{r_m}dr (1-r)^{n-1}\ln{r}
=          -\frac{1}{n^2}-\frac{1}{n}(\Psi(n)+\gamma_E)+
           \sum_{k=1}^{\infty}\frac{(1-r_m)^{n+k}}{k(n+k)}
           \\ \nonumber
&=& -\frac{1}{n}\bigg[\frac{1}{n}\left(1-(1-r_m)^n\right)
+(1-r_m)^n\ln(r_m)+(\Psi(n)+\gamma_E)
+\sum_{k=0}^{\infty}\frac{(1-r_m)^{n+k}}{n+k}\bigg]\nonumber \\
\nonumber
           &=&
           \frac{1}{n}\bigg[i_1^n(r_m)+\left(1-(1-r_m)^n\right)
           \left(\ln(r_m)-\frac{1}{n}\right)\bigg],
\end{eqnarray}
where $i_1^n(r_m)$ is given in \eq{i1}. Note that $K_1^{n\,\reg}
(\lambda,x)$ has {\em  no} singularities\footnote{In \eq{In} there
is exact cancellation of singularities at all non-posituve integer
between the $\Psi$ term and the sum.} on the real $n$ axis, neither
at positive nor negative values of $n$. This follows directly form
\eq{C1_reg_def} as the upper limit of integration obeys $r_m<1$.
Note also that for fixed $n$ the functions $J_n(r_m)$ and $I_n(r_m)$
vanish in the $r_m\longrightarrow 0$ limit, and the leading order
approximation to these functions in this limit is $n$--independent:
\begin{eqnarray}
\label{JI_small_rm__first}
J_n(r_m)&\simeq& J_{n=1}(r_m)+{\cal O}(r_m^2); \qquad  J_{n=1}(r_m)=r_m\\ \nonumber
I_n(r_m)&\simeq& I_{n=1}(r_m)+{\cal O}(r_m^2); \qquad  I_{n=1}(r_m)=r_m (\ln(r_m)-1).
\end{eqnarray}

\section{Matching coefficients for the electron--energy integrated
width\label{sec:partially_integrated_matching}}

Here we compute the matching coefficients  $\overline{V}_r(\lambda,x>x_0)$ and
$\overline{\Delta K_1}^n(\lambda,x>x_0)$ in \eq{double_diff_matched}, by
integrating the virtual coefficients of Eqs. (\ref{V_and_D_r}) and (\ref{V_r_coef}) and the
real--emission moments of \eq{tilde_C_1_explicit}, respectively.

The virtual terms are:
\begin{equation}
\overline{V}_r(\lambda,x>x_0)=\bar{w}_0(\lambda,x_0)
+\frac{C_F\alpha_s(m)}{\pi}\bar{w}_1(\lambda,x_0)+\cdots,
\end{equation}
where the LO coefficient is
\begin{eqnarray}
\label{bar_w0} \bar{w}_0(\lambda,x_0)=\left\{\begin{array}{llr}
\bar{w}_0^a(\lambda)=&
2\,\lambda ^{2}\,(3-2\,\lambda ),\qquad& \lambda<1-x_0 \\ \\
\bar{w}_0^b(\lambda,x_0)=&4\,x_0^{3} + 6\,\left( - {\displaystyle
{3}}  + 2\lambda \right)\,x_0^{2} + 12\, (1 - \lambda)(2- \lambda
)\,x_0 \\&- 10 - 12\,\lambda ^{2} + 24\,\lambda, & \lambda\geq 1-x_0
\end{array} \right.
\end{eqnarray}
and the NLO coefficient is:
\begin{eqnarray}
\bar{w}_1(\lambda,x_0)=\left\{\begin{array}{lr}
\bar{w}_1^a(\lambda)  \qquad  \qquad & \lambda<1-x_0 \\ \\
\bar{w}_1^b(\lambda,x_0)\qquad\qquad & \lambda\geq 1-x_0
\end{array} \right.
\end{eqnarray}
with
\begin{eqnarray}
\label{bar_w1} \bar{w}_{1}^{a}(\lambda)&=& 2\lambda^2
(-3+2\lambda){\rm Li}_2(1-\lambda)
+\lambda^2(4\lambda-9) \ln(\lambda)+\frac{1}{6}\lambda^2(4\pi^2+15)(-3+2\lambda) \nonumber\\
\bar{w}_{1}^{b}(\lambda,x_0)&=& 2 (1-x_0) (2 x_0^2-7 x_0+6 x_0
\lambda-12 \lambda+6 \lambda^2+5) {\rm Li}_2(1-\lambda) \\ \nonumber
&& +(1-x_0) (-17 x_0+4 x_0^2-30 \lambda+12 \lambda^2+13+12 x_0
\lambda) \ln(\lambda) \\ \nonumber && +\frac{1}{6} (1-x_0) (4
\pi^2+15) (2 x_0^2-7 x_0+6 x_0 \lambda-12 \lambda+6 \lambda^2+5)
\end{eqnarray}
\eq{bar_w0} shows that the small $\lambda$ limit is power
suppressed. This property is crucial for the validity of the
perturbative treatment, as the effective {\em hard scale} in the
Sudakov factor is $\lambda m_b$.

The integration over $\widetilde{\Delta K}_1^{n} (\lambda,x)$ is
more involved as it depends on $x$ in a non-trivial way through
$r_m=(1-x)/\lambda$, which was the upper limit in the $r$
integration. It is convenient to arrange the polynomial dependence
on $x$ in powers of $r_m$, namely
\begin{eqnarray}
\label{coef_decomposition} \hspace*{-20pt} w_0(\lambda,x)\equiv
\sum_{j=0}^2 w_0^{(j)}(\lambda)\, r_m^j\,; \quad
f_i(\lambda,x)\equiv \sum_{j=0}^2 f^{(i,j)}(\lambda)\, r_m^j\,;
\quad \tilde{f}_i(\lambda,x)\equiv \sum_{j=0}^2
\tilde{f}^{(i,j)}(\lambda)\, r_m^j,
\end{eqnarray}
where the coefficients are given in Tables \ref{coef_table_w0} and
\ref{coef_table_f},
\begin{table}
\begin{center}
\begin{tabular}{|c|c|}
  \hline
 $w_0^{(0)}(\lambda)$  & $12\lambda^2(1-\lambda)$ \\
  \hline
 $w_0^{(1)}(\lambda)$  & $12\lambda^2(-1+2\lambda)$ \\
 \hline
 $w_0^{(2)}(\lambda)$  & $-12\lambda^3$ \\
  \hline
\end{tabular}
\caption{The Born level coefficient, $w_0(\lambda,x)$ of
\eq{V_r_coef}, decomposed according to
\eq{coef_decomposition}.\label{coef_table_w0}}
\end{center}
\end{table}
\begin{table}
\begin{center}
\begin{tabular}{|c|c|c|c|}
  \hline
  $i$ & $j$ & $f^{(i,j)}(\lambda)$ & $\tilde{f}^{(i,j)}(\lambda)$ \\
  \hline
-4&   0 & $-36 (\lambda-6)$ $(1-\lambda)^2 \lambda^3$ & 0 \\
-4 &  1 & $72 (\lambda-6)$ $(1-\lambda)^2 \lambda^3$ & 0 \\
-4 &  2 & $-36 (\lambda-6)$ $(1-\lambda)^2 \lambda^3$ & 0 \\
  \hline
-3 &  0 & $-36 \lambda^3 (1-\lambda) (4 \lambda^2-24 \lambda+15)$ & $-36 (\lambda-6) (1-\lambda)^2 \lambda^3$ \\
-3 &  1 & $36 \lambda^3 (1-\lambda) (7 \lambda^2-41 \lambda+24)$ & $72 (\lambda-6) (1-\lambda)^2 \lambda^3$ \\
-3 &  2 & $-36 \lambda^3 (1-\lambda) (3 \lambda^2-17 \lambda+9)$ & $-36 (\lambda-6) (1-\lambda)^2 \lambda^3$ \\
  \hline
-2 &  0 & $-6 \lambda^2 (8-120 \lambda+311 \lambda^2-244 \lambda^3+40 \lambda^4)$ & $-18 \lambda^3 (1-\lambda) (7 \lambda^2-41 \lambda+24)$ \\
-2 &  1 & $12 \lambda^2 (4+205 \lambda^2-74 \lambda-169 \lambda^3+29 \lambda^4)$ & $72 \lambda^3 (1-\lambda) (3 \lambda^2-17 \lambda+9)$ \\
-2 &  2 & $-6 \lambda^3 (-34+19 \lambda^3+112 \lambda-102 \lambda^2)$ & $-18 \lambda^3 (1-\lambda) (5 \lambda^2-27 \lambda+12)$ \\
  \hline
-1&  0 & $12 \lambda^2 (4-45 \lambda+116 \lambda^2-95 \lambda^3+18 \lambda^4)$ & $-6 \lambda^2 (8-87 \lambda+220 \lambda^2-176 \lambda^3+30 \lambda^4)$ \\
-1&  1 & $-6 \lambda^2 (4+225 \lambda^2-76 \lambda-196 \lambda^3+40 \lambda^4)$ & $6 \lambda^2 (8+267 \lambda^2-100 \lambda-226 \lambda^3+41 \lambda^4)$ \\
-1&  2 & $6 \lambda^3 (-8+8 \lambda^3+34 \lambda-33 \lambda^2)$ & $-6 \lambda^3 (-58 \lambda^2+60 \lambda-19+12 \lambda^3)$ \\
  \hline
0&  0 & $-6 \lambda^2 (8-50 \lambda+108 \lambda^2-86 \lambda^3+19 \lambda^4)$ & $3 \lambda^2 (15-109 \lambda+270 \lambda^2-224 \lambda^3+45 \lambda^4)$ \\
0&  1 & $6 \lambda^2 (2+65 \lambda^2-26 \lambda-56 \lambda^3+14 \lambda^4)$ & $-3 \lambda^2 (7+222 \lambda^2-78 \lambda-198 \lambda^3+44 \lambda^4)$ \\
0&  2 & $-6 \lambda^3 (-1+\lambda) (\lambda^2-2 \lambda+2)$ & $3 \lambda^3 (-1+\lambda) (6 \lambda^2-14 \lambda+7)$ \\
  \hline
1&  0 & $-6 \lambda^2 (1-\lambda) (6 \lambda^3-17 \lambda^2+14 \lambda-2)$ & $3 \lambda^2 (1-\lambda) (-1+2 \lambda) (9 \lambda^2-25 \lambda+7)$ \\
1&  1 & $12 \lambda^3 (1-\lambda) (\lambda^2-2 \lambda+2)$ & $6 \lambda^3 (1-\lambda)^2 (5 \lambda-7)$ \\
1&  2 & 0 & 0 \\
  \hline
2&  0 & $6 \lambda^3 (1-\lambda) (\lambda^2-2 \lambda+2)$ & $3 \lambda^3 (3 \lambda-7) (1-\lambda)^2$ \\
2&  1 & 0 & 0 \\
2&  2 & 0 & 0 \\
  \hline
\end{tabular}
\caption{The NLO matching coefficients, $f_i(\lambda,x)$ and
$\tilde{f}_{i}(\lambda,x)$ of \eq{f} and \eq{tilde_f}, respectively,
decomposed according to
\eq{coef_decomposition}.\label{coef_table_f}}
\end{center}
\end{table}
and change variables from $x$ to
\[
v\equiv 1-r_m=1-(1-x)/\lambda.
\]
One then obtains:
\begin{eqnarray}
\label{overline_Delta_C_1_n} \hspace*{-20pt}\overline{\Delta
K_1}^n(\lambda,x>x_0) \equiv
\int_{\max\{x_0,1-\lambda\}}^1\hspace*{-20pt} dx  \widetilde{\Delta
K}_1^{n} (\lambda,x)
 = \lambda\int_{\max\{v_0,0\}}
 ^1 \hspace*{-20pt}dv  \widetilde{\Delta K}_1^{n} (\lambda,x=1-\lambda(1-v))
\end{eqnarray}
where $v_0\equiv v_0(\lambda,x_0)\equiv 1-(1-x_0)/\lambda$. The
integration in \eq{overline_Delta_C_1_n} can be done analytically.
Let us write the result as
\begin{eqnarray}
&&\overline{\Delta K_1}^n(\lambda,x>x_0)= \left\{\begin{array}{lr}
{\cal C}_1^a(n,\lambda)= {\cal C}_1^b(n,\lambda,0)\qquad  \qquad & \lambda<1-x_0\nonumber \\ \\
{\cal C}_1^b(n,\lambda,v_0(\lambda,x_0))={\cal C}_1^{\sing\, b}
+{\cal C}_1^{\reg\, b} \qquad\qquad & \lambda\geq 1-x_0
\end{array} \right. .
\end{eqnarray}
The expressions for the first moment ($n=1$) are simple (see
\eq{tilde_C_1_explicit_n_eq_1}):
\begin{eqnarray}
{\cal C}_1^{\sing\, b}(n=1,\lambda,v_0)&=& \left(-3 \lambda^2+3
\lambda^2 v_0^2-2 \lambda^3 v_0^3+2 \lambda^3\right) \ln^2(1-v_0)
\\ \nonumber && \hspace*{-40pt}
+\left((4 \lambda^3-6 \lambda^2) v_0 +\left(2 \lambda^3+\frac{15}{2}
\lambda^2\right) v_0^2-\frac{17}{3} \lambda^3 v_0^3-\frac{3}{2}
\lambda^2-\frac{1}{3} \lambda^3\right) \ln(1-v_0)
\\ \nonumber && \hspace*{-40pt}
+\frac{17}{9} \lambda^3 v_0^3+\left(\frac{11}{6}
\lambda^3-\frac{15}{4} \lambda^2\right) v_0^2 +\left(-\frac{3}{2}
\lambda^2-\frac{1}{3} \lambda^3\right) v_0+\frac{21}{4}
\lambda^2-\frac{61}{18} \lambda^3\,,
\\ \nonumber
{\cal C}_1^{\reg\, b}(n=1,\lambda,v_0)&=& - \lambda ^{2}\,v_0\bigg(
- 16\,\lambda \, v_0^{2} - 66\,\lambda ^{3}\,v_0^{2} + 16\,\lambda
 ^{4}\,v_0^{2} + 68\,\lambda ^{2}\,v_0^{2} - 180
\,\lambda \,v_0  \\ \nonumber && \hspace*{-40pt}
  + 72\,\lambda ^{4}\,v_0 - 390\,\lambda ^{3}\,v_0 + 12\,v_0 + 471
\,\lambda ^{2}\,v_0 + 24\,\lambda ^{4} + 246\,\lambda ^{2 } -
132\,\lambda  + 24
\\ \nonumber && \hspace*{-40pt}
 - 162\,\lambda ^{3}\bigg) {\rm Li}_2(v_0)
+ {\displaystyle \frac {1}{6}} \lambda
^{2}\,(1-v_0)\bigg(12\,v_0^{3} \,\lambda ^{2} - 9\,v_0^{3}\,\lambda
^{3} + 3\,v_0^{3}\,\lambda ^{4} - 6\,v_0^{3}\,\lambda  \\ \nonumber
&& \hspace*{-40pt}
   + 287\,\lambda ^{4}\,v_0^{2} - 1299\,\lambda ^{3}
\,v_0^{2} + 1366\,\lambda ^{2}\,v_0^{2} - 434\, \lambda \,v_0^{2} +
12\,v_0^{2} + 377\,\lambda ^{ 4}\,v_0 \\ \nonumber &&
\hspace*{-40pt}+ 3301\,\lambda ^{2}\,v_0 + 264\,v_0 -2361\,\lambda
^{3}\,v_0 - 1586\,\lambda \,v_0 + 5 \,\lambda ^{4} + 31\,\lambda
^{2} + 58\,\lambda  - 39\,\lambda ^{3}
 \\ \nonumber && \hspace*{-40pt}
 - 60\bigg)\ln(1 - v_0) +{\displaystyle \frac{v_0^{4}}{24} {\lambda^{3}\,( \lambda-1 )\,(61\,
\lambda ^{2} - 146\,\lambda  + 92)\,}}  + \frac{v_0^{3}}{18} \lambda
^{2}\bigg(1229\,\lambda ^{4}
\\ \nonumber && \hspace*{-40pt}
- 6141\,\lambda ^{3} + 6673\,\lambda ^{2}  - 2048\,\lambda   + 129 +
\pi^2(48\,\lambda ^{4}- 198\,\lambda ^{3}+ 204\,\lambda ^{2}-
48\,\lambda) \bigg)
 \\ \nonumber && \hspace*{-40pt}
 + \frac{v_0^{2}}{12} \lambda ^{2}\bigg( - 385\,\lambda ^{4} +
144\,\pi ^{2}\,\lambda ^{4} + 1077\,\lambda ^{3} - 780\,\pi ^{2}
\,\lambda ^{3} + 199\,\lambda ^{2} + 942\,\pi ^{2}\,\lambda ^{2}
 \\ \nonumber && \hspace*{-40pt}
  - 1196\,\lambda  - 360\,\pi ^{2}\,\lambda  + 390 + 24\,
\pi ^{2}\bigg)  + \frac{v_0}{6} \lambda ^{2}\bigg( - 233\,\lambda
 ^{4} + 24\,\pi ^{2}\,\lambda ^{4} + 1563\,\lambda ^{3} \\ \nonumber && \hspace*{-40pt}
  - 162\,\pi ^{2}\,\lambda ^{3} - 2400\,\lambda ^{2} + 246
\,\pi ^{2}\,\lambda ^{2} - 132\,\pi ^{2}\,\lambda  + 1368\, \lambda
- 303 + 24\,\pi ^{2}\bigg) \\ \nonumber && \hspace*{-40pt} +\lambda
^{2}\,\bigg({\displaystyle \frac {7}{72}} \,\lambda ^{4} -
{\displaystyle \frac {11}{24}} \,\lambda ^{3} + {\displaystyle \frac
{25}{9}} \,\lambda ^{2} - {\displaystyle \frac {193}{18}} \,\lambda
+ {\displaystyle \frac {65}{6}} \bigg),
\end{eqnarray}
and
\begin{eqnarray}
\!\!\!\!\! {\cal C}_1^a(n=1,\lambda)&=&
\lambda^2\left(\frac{21}{4}-\frac{61}{18}\lambda\right)\,+\, \lambda
^{2}\,\bigg({\displaystyle \frac {7}{72}} \,\lambda ^{4} -
{\displaystyle \frac {11}{24}} \,\lambda ^{3} + {\displaystyle \frac
{25}{9}} \,\lambda ^{2} - {\displaystyle \frac {193}{18}} \,\lambda
+ {\displaystyle \frac {65}{6}} \bigg),
\end{eqnarray}
while for general complex $n$ (with ${\rm Re}(n)>3$ i.e. ${\rm
Re}(n+i)>-1$, so that convergence of the moment integral for
individual terms in the sum is guaranteed) ${\cal
C}_1^b(n,\lambda,v_0(\lambda,x_0))$ can be written in terms of a few
known integrals:
\begin{eqnarray}
&&{\cal C}_1^b(n,\lambda,v_0(\lambda,x_0))= \sum_{j=0}^2
\,\bigg[\left(K_{n,j}(v_0)+\frac{7}{4} L_{n,j}(v_0)
+M_{n,j}(v_0)\right)\times w_0^{(j)}(\lambda)\nonumber\\
\hspace*{20pt}&&+ \sum_{i=-4}^{2} f^{(i,j)}(\lambda) \frac{1}{n+i}
\bigg(-\left(\Psi(n+i+1)+\gamma_E\right) P_{j}(v_0) +\frac{1}{n+i}
Q_{n+i,j}(v_0) \\ \nonumber && \hspace*{20pt} - R_{n+i,j}(v_0) -
L_{n+i,j}(v_0) \bigg)
+\sum_{i=-3}^{2} \tilde{f}^{(i,j)}(\lambda) \frac{1}{n+i}
\bigg(P_{j}(v_0)-Q_{n+i,j}(v_0)\bigg) \bigg].
\end{eqnarray}
All these integrals can be expressed as power series in $v_0$, which
is convenient for numerical evaluation:
\begin{eqnarray}
P_{j}(v_0) &\equiv& \int_{v_0}^1 dv  (1-v)^j  \,=\,
\frac{(1-v_0)^{j+1}}{j+1}, \\\nonumber \bar{Q}_{n,j}(v_0) &\equiv&
\int_{0}^{v0} dv  (1-v)^j v^n = \sum_{k=0}^{\infty}
 (1-v_0)^{j+1} v_0^{n+k+1} \frac{\Gamma(n+1)\Gamma(k+j+n+2)}{\Gamma(j+n+2)\Gamma(k+n+2)}, \\\nonumber
Q_{n,j}(v_0) &\equiv& \int_{v_0}^1 dv  (1-v)^j v^n =
\frac{\Gamma(j+1)\Gamma(n+1)}{\Gamma(j+n+2)}-\bar{Q}_{n,j}(v_0),
\\\nonumber R_{n,j}(v_0) &\equiv& \int_{v_0}^1 dv  (1-v)^j v^n \ln
(1-v) \,=\,\frac{d}{dj} Q_{n,j}(v_0), \\\nonumber L_{n,j}(v_0)
&\equiv& \int_{v_0}^1 dv  (1-v)^j v^n
\sum_{k=0}^{\infty}\frac{v^k}{n+k}
\,=\,\frac{\Gamma(n)\Gamma(j+1)}{(j+1)\Gamma(n+j+1)} \\\nonumber
&&-\frac{(1-v_0)^{j+1} v_0^{n}}{j+1} \sum_{k=0}^{\infty} v_0^k
\left(\frac{\Gamma(n)
\Gamma(k+1+j+n)}{\Gamma(j+1+n)\Gamma(k+n+1)}-\frac{1}{n+k}\right),
\\\nonumber K_{n,j}(v_0) &\equiv& \int_{v_0}^1 dv  (1-v)^j v^n \ln
(1-v) \sum_{k=0}^{\infty}\frac{v^k}{n+k}\,=\,\frac{d}{dj}
L_{n,j}(v_0)\\\nonumber M_{n,j}(v_0) &\equiv& \int_{v_0}^1 dv
(1-v)^j v^n
\sum_{k=0}^{\infty}\frac{v^k}{n+k}\left(\Psi(n+k)-\Psi(n)\right)\\\nonumber&=&
\frac{\Gamma(n)\Gamma(j+1)}{\Gamma(1+n+j)(1+j)^2}-\sum_{k=0}^{\infty}
\frac{\bar{Q}_{n+k,j}(v_0)}{n+k}\left(\Psi(n+k)-\Psi(n)\right),
\end{eqnarray}
These series converge within the physical region, i.e. $0\leq
v_0<1$, and complex $n$ with ${\rm Re}(n)>3$ (which can have large
imaginary part). The series expansions of $R_{n,j}(v_0)$ and
$K_{n,j}(v_0)$ can be readily obtained by differentiating the
expressions for $Q_{n,j}(v_0)$ and $L_{n,j}(v_0)$, respectively,
with respect to $j$. For integer values of~$j$, $M_{n,j}(v_0)$
becomes a single infinite sum since $\bar{Q}_{n,j}(v_0)$ reduces
then to simple algebraic expressions:
\begin{eqnarray}
\bar{Q}_{n,0} &=& v^{1+n}\frac{1}{1+n}\\ \nonumber \bar{Q}_{n,1} &=&
v^{1+n}\left[\frac{1}{1+n}-\frac{v}{2+n}\right] \\ \nonumber
\bar{Q}_{n,2} &=&
v^{1+n}\left[\frac{v^2}{3+n}-\frac{2v}{2+n}+\frac{1}{1+n}\right].
\end{eqnarray}

In contrast to ${\cal C}_1^b(n,\lambda,v_0)$, the final expression
for ${\cal C}_1^a(n,\lambda)= {\cal C}_1^b(n,\lambda,0)$ takes a
simple form:
\begin{eqnarray}
{\cal C}_1^a(n,\lambda) &=& \lambda^2 (\Psi(n+4)+\gamma_E)
\bigg[\frac{18-40 \lambda-12 \lambda^3+2 \lambda^4+35
\lambda^2}{n+1} +\frac{-6+4 \lambda}{n}\\ \nonumber && +\frac{-6+26
\lambda+18 \lambda^3-4 \lambda^4-35 \lambda^2} {2+n} +2 \lambda
\frac{-3 \lambda^2+4 \lambda-2+\lambda^3}{3+n}\bigg]\\ \nonumber &&
+\lambda^2 \bigg[\frac{1}{2} \frac{114 \lambda-105 \lambda^2+36
\lambda^3-6 \lambda^4-45}{n+1} +\frac{1}{6} \frac{-86
\lambda+129}{n}\\ \nonumber && +\frac{1}{2} \frac{-66 \lambda+80
\lambda^2-44 \lambda^3+12 \lambda^4+21}{2+n} +\frac{1}{6} \frac{2
\lambda-15 \lambda^2+24 \lambda^3-18 \lambda^4+6}{3+n}\bigg].
\end{eqnarray}

It follows, in particular, that the resummed width, {\em integrated
with respect to the lepton energy} (no cut, i.e. $x_0=0$), is:
\begin{eqnarray}
\label{double_diff_r_lambda_no_x0_cut} &&\hspace*{-10pt}
\frac{1}{\Gamma_0} \frac{d\Gamma(\lambda,r)}{d\lambda dr}
=\int_{c-i\infty}^{c+i\infty} \frac{d n}{2\pi i} \,\frac{d\Gamma_{n}
(\lambda)}{d\lambda} \, \left(1-{r}\right)^{-n}
\end{eqnarray}
with
\begin{eqnarray}
\label{double_diff_matched_no_x0_cut} &&\left.\frac{d\Gamma_{n}
(\lambda)}{d\lambda}\right\vert_{\rm matched} \equiv
\int_{1-\lambda}^1dx \,\left.\frac{d\Gamma_{n} (\lambda,x)}{d\lambda
dx}\right\vert_{\rm matched}\\\nonumber &&\hspace*{30pt}=
\left[\bar{w}_0^{a}(\lambda) \,+ \left(\bar{w}_1^{a}(\lambda)+{\cal
C}_1^a(n,\lambda)\right) \frac{C_F\alpha_s(m_b)}{\pi} + {\cal
O}(\alpha_s^2) \right] \times \\ \nonumber &&\hspace*{50pt}
\exp\bigg\{ \frac{C_F}{\beta_0}\int_0^{\infty}\frac{du}{u} \, T(u)
\left(\frac{\Lambda^2}{m_b^2\lambda^2}\right)^u
 \, \bigg[B_{\cal
S}(u)\Gamma(-2u)\left(\frac{\Gamma(n)}{\Gamma(n-2u)}-\frac{1}{\Gamma(1-2u)}\right)
\\ \nonumber &&\hspace*{100pt}
 -B_{\cal
J}(u)\Gamma(-u)\left(\frac{\Gamma(n)}{\Gamma(n-u)}-\frac{1}{\Gamma(1-u)}\right)
\bigg] \bigg\}.
\end{eqnarray}
where $\bar{w}_{0,1}^{a}(\lambda)$ are given in Eqs. (\ref{bar_w0})
and (\ref{bar_w1}).

\section{Phase--space integrals for partially integrated width\label{sec:Phase_space}}

\subsection{Partially integrated width in hadronic variables\label{sec:diff_and_int}}

We have seen that the resummed moments ${d\Gamma_n(\lambda,x)}/{d\lambda dx}$ of
\eq{Gamma_n_lnR_matched} (or \eq{Gamma_n_lnR_matched_w_constant})
where $n$ is the Mellin conjugate to $r= (P^+ -
\bar{\Lambda})/(P^--\bar{\Lambda})$ can be used to compute the
differential width using~\eq{Pplus_diff}. Similarly, the partially integrated
width with a cut $P^+ < P_{\max}^+$ can be computed by
\begin{eqnarray}
\label{Pplus_int}
\frac{1}{\Gamma_0}\frac{d\Gamma(P^+<P_{\max}^+,\,P^-,\,E_l)}{dP^-\,dE_l
}&=& \frac{2}{m_b^2}\,\int_{c-i\infty}^{c+i\infty}\frac{dn} {2\pi
i(n-1)}
\left(1-\frac{\bar{\Lambda}}{P^-}\right)^{n-1}\times \\
\nonumber && \hspace*{30pt}
\left(1-\frac{P_{\max}^+}{P^-}\right)^{1-n}
\frac{d\Gamma_n(\lambda=\frac{P^--\bar{\Lambda}}{m_b},\,x=\frac{2E_l}{m_b})}{d\lambda
dx}.
\end{eqnarray}
This relation is useful in computing partially integrated widths
with a variety of experimental cuts by performing integration over
$E_l$ and $P^-$ as well as the inverse Mellin transform and the
Principal Values Borel integration numerically.

The expression for the width with a cut on the invariant mass, in
addition to a cut on the lepton energy, takes the form:
\begin{eqnarray}
\label{diff_M_X_cut}
&&\frac{\Gamma(P^+P^-<M_X^2,E_l>E_0)}{\Gamma_0}= \int_{0}^{M_B/2}
dE_l \,\,\theta(E_l>E_0)\, \int_{M_B-2E_l}^{M_B} dP^-\,\\ \nonumber
&&\hspace*{40pt}\bigg[ \theta(M_B-2E_l<M_X^2/P^-)\,
\frac{1}{\Gamma_0}\frac{d\Gamma(P^+<M_B-2E_l,\,P^-,\,E_l)}{dP^-\,dE_l
}
\\ \nonumber &&\hspace*{65pt} +  \theta(M_X^2/P^-<M_B-2E_l)\,
\frac{1}{\Gamma_0}\frac{d\Gamma(P^+<M_X^2/P^-,\,P^-,\,E_l)}{dP^-\,dE_l
}\bigg],
\end{eqnarray}
so \eq{Pplus_int} is used twice, in the latter term $P_{\max}^+$ is
set by the external invariant mass cut value $M_X^2/P^-$, while in
the former by the actual phase--space limit $M_B-2E_l$. Note however
that the integrated distribution over the entire $P^+$ phase space
($P^+<M_B-2E_l$) simply corresponds to the first moment,
\begin{eqnarray}
\frac{1}{\Gamma_0}\frac{d\Gamma(P^+<M_B-2E_l,\,P^-,\,E_l)}{dP^-\,dE_l
}\,=\,\frac{2}{m_b^2}
\frac{d\Gamma_{n=1}(\lambda=\frac{P^--\bar{\Lambda}}{m_b},\,x=\frac{2E_l}{m_b})}{d\lambda\,dx
},
\end{eqnarray}
and therefore need not be computed by an inverse Mellin integral;
such is necessary of course for the term where the upper limit on
$P^+$ is set by $M_X^2/P^-$.

When computing the phase--space integrals in \eq{diff_M_X_cut} one
should distinguish between three cases: for cuts where
$M_X^2>M_B^2-2E_0M_B$ (in Fig.~\ref{fig:P_minus_El_ps} this would
correspond to $E_l^{(F)}<E_0$) the hadronic invariant--mass cut is
irrelevant and
\begin{eqnarray}
\label{large_S_cut}
\hspace*{-23pt}\frac{\Gamma(P^+P^-<M_X^2,E_l>E_0)}{\Gamma_0}&=&\!\!
\int_{E_0}^{M_B/2}\!\! \!\!
dE_l \int_{M_B-2E_l}^{M_B} \!\!\!\!\!\!\!\!dP^-\,
\frac{1}{\Gamma_0}\frac{d\Gamma(P^+<M_B-2E_l,\,P^-,\,E_l)}{dP^-\,dE_l
}.
\end{eqnarray}
For cuts with $(M_B-2E_0)^2<M_X^2<M_B^2-2E_0M_B$ (in
Fig.~\ref{fig:P_minus_El_ps} this would correspond to
$E_l^{(C)}<E_0<E_l^{(F)}$):
\begin{eqnarray}
\label{intermediate_S_cut} &&\nonumber
\frac{\Gamma(P^+P^-<M_X^2,E_l>E_0)}{\Gamma_0}=
\int_{E_0}^{\frac{1}{2}\left(M_B-{M_X^2}/{M_B}\right)}\!\! \!\!dE_l
\\ \nonumber
&& \hspace*{30pt}
\bigg\{\int_{M_B-2E_l}^{M_X^2/(M_B-2E_l)} \!\!\!\!dP^-\,
\frac{1}{\Gamma_0}\frac{d\Gamma(P^+<M_B-2E_l,\,P^-,\,E_l)}{dP^-\,dE_l
} \\ \nonumber
&&\hspace*{30pt}\,\,\,\,+\,\int_{M_X^2/(M_B-2E_l)}^{M_B} \!\!\!\!dP^-\,
\frac{1}{\Gamma_0}\frac{d\Gamma(P^+<M_X^2/P^-,\,P^-,\,E_l)}{dP^-\,dE_l }\bigg\} \\
&&\hspace*{30pt}+
\int_{\frac{1}{2}\left(M_B-{M_X^2}/{M_B}\right)}^{M_B/2}\!\! \!\!
dE_l \int_{M_B-2E_l}^{M_B} \!\!\!\!dP^-\,
\frac{1}{\Gamma_0}\frac{d\Gamma(P^+<M_B-2E_l,\,P^-,\,E_l)}{dP^-\,dE_l
}.
\end{eqnarray}
Finally, for cuts $M_X^2<(M_B-2E_0)^2$ as in
Fig.~\ref{fig:P_minus_El_ps} where $E_0<E_l^{(C)}<E_l^{(F)}$:
\begin{eqnarray}
\label{small_S_cut} &&\nonumber
\frac{\Gamma(P^+P^-<M_X^2,E_l>E_0)}{\Gamma_0}=
\int_{E_0}^{\frac{1}{2}(M_B-M_X)}\!\! \!\!dE_l
\int_{M_B-2E_l}^{M_B} \!\!\!\!dP^-\,
\frac{1}{\Gamma_0}\frac{d\Gamma(P^+<M_X^2/P^-,\,P^-,\,E_l)}{dP^-\,dE_l
}
\\ \nonumber
&&\hspace*{30pt}+\int_{\frac{1}{2}(M_B-M_X)}^{\frac{1}{2}\left(M_B-M_X^2/{M_B}\right)}\!\!
\!\!dE_l
\bigg\{\int_{M_B-2E_l}^{M_X^2/(M_B-2E_l)} \!\!\!\!dP^-\,
\frac{1}{\Gamma_0}\frac{d\Gamma(P^+<M_B-2E_l,\,P^-,\,E_l)}{dP^-\,dE_l
} \\ \nonumber
&&\hspace*{30pt}+\,\int_{M_X^2/(M_B-2E_l)}^{M_B} \!\!\!\!dP^-\,
\frac{1}{\Gamma_0}\frac{d\Gamma(P^+<M_X^2/P^-,\,P^-,\,E_l)}{dP^-\,dE_l }\bigg\} \\
&&\hspace*{30pt}+
\int_{\frac{1}{2}\left(M_B-{M_X^2}/{M_B}\right)}^{M_B/2}\!\! \!\!
dE_l \int_{M_B-2E_l}^{M_B} \!\!\!\!dP^-\,
\frac{1}{\Gamma_0}\frac{d\Gamma(P^+<M_B-2E_l,\,P^-,\,E_l)}{dP^-\,dE_l
}.
\end{eqnarray}
The last expression, corresponding to a stringent
(charm--discriminating) hadronic--mass cut and a mild lepton--energy
cut, is most useful for extracting $\left\vert V_{ub}\right\vert$.

\subsection{Distributions and partial widths with lepton energy cut~\label{sec:int_lep_energy}}

The analytic integration over the lepton energy performed in
Sec.~\ref{sec:Analytic_lepton_energy_cut} and
Appendix~\ref{sec:partially_integrated_matching} can be used as an
alternative to the numerical integration over $E_l$. Starting with
the resummed distribution of \eq{double_diff_r_lambda} with $x>x_0$
and converting to hadronic lightcone  momenta using \eq{P^+} we
obtain the following expression for the double differential
distribution:
\begin{eqnarray}
\label{double_diff_r_H_lambda_P} \hspace*{-10pt} \frac{1}{\Gamma_0}
\frac{d\Gamma(P^-,P^+,E_l>E_0)}{dP^- dP^+}
&=&\frac{1}{P^-}\,\int_{c-i\infty}^{c+i\infty} \frac{d n}{2\pi i}
\,\left(1-\frac{\bar{\Lambda}}{P^-}\right)^{n-1}
\left(1-\frac{P^+}{P^-}\right)^{-n}\!\times \\ \nonumber &&
\hspace*{100pt} \frac{1}{m_b}\frac{d\Gamma_{n}
(\lambda=\frac{P^--\bar{\Lambda}}{m_b},\,x>x_0=\frac{2E_0}{m_b})}{d\lambda}.
\end{eqnarray}
The corresponding integrated width with a cut on $P^{+}$ is:
\begin{eqnarray}
\label{int_r_H_lambda_P} \hspace*{-10pt} \frac{1}{\Gamma_0}
\frac{d\Gamma(P^-,P^+<P_{\max}^{+},E_l>E_0)}{dP^-}
&=&\int_{c-i\infty}^{c+i\infty} \frac{d n}{2\pi i (n-1)}
\,\left(1-\frac{\bar{\Lambda}}{P^-}\right)^{n-1}
\left(1-\frac{P_{\max}^+}{P^-}\right)^{1-n}\!\!\!\!\!\!\times
\nonumber \\  && \hspace*{40pt} \frac{1}{m_b}\frac{d\Gamma_{n}
(\lambda=\frac{P^--\bar{\Lambda}}{m_b},\,x>x_0=\frac{2E_0}{m_b})}{d\lambda}.
\end{eqnarray}
In order to compute the integrated width with an experimental
$P_{\max}^{+}$ cut one has to sum over two contributions, one from
the small $P^-$ region, where the experimental cut as well as the
resummation are irrelevant, and one over the more important region
of larger $P^-$ where they are relevant. The result is:
\begin{eqnarray}
\label{int_M_X_cut_P_plus} \frac{\Gamma(P^+<P_{\max}^{+},
E_l>E_0)}{\Gamma_0}&=&\int_0^{P_{\max}^{+}} dP^-\,
\frac{d\Gamma_{n=1}(\lambda=\frac{P^--\bar{\Lambda}}{m_b},\,x>\frac{2E_0}{m_b})}{m_b
d\lambda }
\\ \nonumber
&+&\int_{P_{\max}^{+}}^{M_B} dP^- \frac{1}{\Gamma_0}
\frac{d\Gamma(P^-,P^+<P_{\max}^{+},E_l>E_0)}{dP^-},
\end{eqnarray}
where in the second term \eq{int_r_H_lambda_P} is used. Recall that
the matching term in $\frac{d\Gamma_{n} (\lambda,x>x_0)}{d\lambda}$
has different analytic expressions depending on the sign of
$M_B-2E_0-P^{-}$, see \eq{double_diff_matched} and below.

The double differential width of \eq{double_diff_r_H_lambda_P} can
be readily converted into other kinematic variables, for example,
the hadronic and leptonic invariant--mass distribution with lepton
energy cut is:
\begin{eqnarray}
\label{double_diff_sH_q2} \frac{1}{\Gamma_0}
\frac{d\Gamma(M_X^2,q^2,E_l>E_0)}{dM_X^2 dq^2} &=&\int dP^+ dP^-\,
\int_{c-i\infty}^{c+i\infty} \frac{d n}{2\pi i}
\,\frac{\left(1-\frac{\bar{\Lambda}}{P^-}\right)^{n-1}}{P^-}\,
\left(1-\frac{P^+}{P^-}\right)^{-n}\,\times \nonumber \\
&& \hspace*{-80pt} \frac{d\Gamma_{n}
(\lambda=\frac{P^--\bar{\Lambda}}{m_b},\,x>\frac{2E_0}{m_b})}{m_b\,d\lambda}
 \delta (M_X^2-P^+P^-)\,\delta (q^2-(M_B-P^+)(M_B-P^-))\nonumber \\
 &=&
 \int_{c-i\infty}^{c+i\infty}
\frac{d n}{2\pi i}
\,\frac{\left(1-\frac{\bar{\Lambda}}{P^-}\right)^{n-1}}{(P^-)^2-M_X^2}\,
\left(1-\frac{M_X^2}{(P^-)^2}\right)^{-n}
 \nonumber \\
&& \left. \frac{d\Gamma_{n}
(\lambda=\frac{P^--\bar{\Lambda}}{m_b},\,x>\frac{2E_0}{m_b})}{M_B\,
m_b\,d\lambda}\right\vert_{\begin{array}{l}
P^-=E_j+\sqrt{E_j^2-M_X^2}\\
E_j=(M_B^2-M_X^2-q^2)/(2M_B)\end{array}}
\end{eqnarray}

\eq{int_r_H_lambda_P} can also be used to obtain the partial width
with a cut on the hadronic invariant mass $M_X^2$, i.e.
\begin{eqnarray}
\label{int_M_X_cut} \frac{\Gamma(P^- P^+<M_X^2,
E_l>E_0)}{\Gamma_0}&=&\int_0^{M_X} dP^-\,
\frac{d\Gamma_{n=1}(\lambda=\frac{P^--\bar{\Lambda}}{m_b},\,x>\frac{2E_0}{m_b})}{m_b
d\lambda }
\\ \nonumber
&+&\int_{M_X}^{M_B} dP^- \frac{1}{\Gamma_0}
\frac{d\Gamma(P^-,P^+<M_X^2/P^-,E_l>E_0)}{dP^-}
\end{eqnarray}
where in the second term \eq{int_r_H_lambda_P} is used.

\end{document}